\documentclass[12pt]{article}
\usepackage[english]{babel}
\usepackage{amssymb}
\usepackage{amsmath}
\usepackage{subfig}
\usepackage{epsfig}
\usepackage{graphicx}
\usepackage{color}
\input{psfig.tex}

\long\def\comment#1{}

\newfont{\bbb}{msbm10 scaled 700}

\newfont{\bb}{msbm10 scaled 1100}

\newcommand{\PP}{\mbox{\bb P}}
\newcommand{\RR}{\mbox{\bb R}}

\newcommand{\FF}{\mbox{\bb F}}

\newcommand{\EE}{\mbox{\bb E}}

\newcommand{\bv}{{\bf b}}
\newcommand{\cv}{{\bf c}}

\newcommand{\pv}{{\bf p}}

\newcommand{\sv}{{\bf s}}

\newcommand{\uv}{{\bf u}}

\newcommand{\vv}{{\bf v}}
\newcommand{\xv}{{\bf x}}
\newcommand{\yv}{{\bf y}}
\newcommand{\zv}{{\bf z}}

\newcommand{\onev}{{\bf 1}}

\newcommand{\Am}{{\bf A}}
\newcommand{\Bm}{{\bf B}}

\newcommand{\Fm}{{\bf F}}
\newcommand{\Gm}{{\bf G}}
\newcommand{\Hm}{{\bf H}}
\newcommand{\Id}{{\bf I}}

\newcommand{\Pm}{{\bf P}}

\newcommand{\Rm}{{\bf R}}

\newcommand{\Ac}{{\cal A}}

\newcommand{\Cc}{{\cal C}}

\newcommand{\Hc}{{\cal H}}

\newcommand{\Kc}{{\cal K}}

\newcommand{\Mc}{{\cal M}}

\newcommand{\Qc}{{\cal Q}}

\newcommand{\Sc}{{\cal S}}

\newcommand{\Wc}{{\cal W}}

\newcommand{\Yc}{{\cal Y}}

\newcommand{\tauv}{\hbox{\boldmath$\tau$}}

\newcommand{\Pim}{\hbox{\boldmath$\Pi$}}

\renewcommand{\arg}{{\hbox{arg}}}

\newcommand{\eqdef}{\stackrel{\Delta}{=}}

\title{A Practical Approach to Lossy Joint Source-Channel Coding}
\author{Maria Fresia$^\dagger$ and Giuseppe Caire$^\star$}

\begin{document}
\maketitle

\vskip 4cm
\noindent
$\dagger$ Maria Fresia, Communications and Remote Sensing Lab.
Universit\'e catholique de Louvain, Belgium (E-mail: {\tt fresia@tele.ucl.ac.be}) \\[12pt]
$\star$ Giuseppe Caire, Ming Hsieh Dept. of Electrical Engineering,  University of Southern California,
Los Angeles, CA, USA (E-mail: {\tt caire@usc.edu}).

\newpage
\clearpage

\begin{abstract}
This work is devoted to {\em practical} joint source channel coding.
Although the proposed approach has more general scope,
for the sake of clarity we focus on a specific application example, namely,
the transmission of digital images over noisy binary-input output-symmetric channels.
The basic building blocks of most state-of the art source coders are: 1) a linear transformation; 
2) scalar quantization of the transform coefficients; 3)
probability modeling of the sequence of quantization indices; 4) an entropy coding stage. 
We identify the weakness of the conventional separated source-channel coding approach in the
catastrophic behavior of the entropy coding stage. Hence, we replace this stage with {\em linear} coding,
that maps directly the sequence of redundant quantizer output symbols into a channel codeword.
We show that this approach does not entail any loss of optimality in the asymptotic regime of large block length.
However, in the practical regime of finite block length and low decoding complexity our approach yields
very significant improvements. Furthermore, our scheme allows to retain the transform, 
quantization and probability modeling of current state-of the art source coders, that are carefully matched to the 
features of specific classes of sources. In our working example, we make use of ``bit-planes'' and ``contexts'' model defined by
the JPEG2000 standard and we re-interpret the underlying probability model as a sequence of conditionally Markov sources. 
The Markov structure allows to derive a simple successive coding and decoding scheme, where the latter is based on
iterative Belief Propagation. We provide a construction example of the proposed scheme based on punctured Turbo Codes 
and we demonstrate the gain over a conventional separated scheme by running extensive numerical experiments 
on test images.
\end{abstract}

\newpage
\clearpage

\section{Introduction} \label{sec:Intro}

Shannon's source-channel {\em separation principle} \cite{CoTh} states that, in the limit of large blocklength and for a large class 
of communication setups, the optimal performance can be approached by independently designing the source coding
and the channel coding schemes. 
Driven by the separation principle, modern communication systems have been developed according to a rather
rigid layered architecture \cite{kurose2003cnt}: the source coding functions are essentially
relegated to the application layer, at the top of the protocol stack, while the channel coding functions are located 
at the link and physical layers, at the bottom of the protocol stack. 
While a separated (layered) architecture 
has the unquestioned advantage of modular system design, allowing the convergence of a great variety of services on a common data network 
infrastructure\footnote{The modern evolution of Internet, offering telephone, video, multimedia streaming and data on the same
wired and wireless infrastructure is the paramount example of this trend.}, there are cases where
a Joint Source-Channel Coding (JSCC) approach is called for. 
On one hand, there exist several relevant multi-terminal settings where the separated approach is 
known to be suboptimal \cite{CoTh}. 
On the other hand, even in standard point to point channels where Separated Source-Channel Coding (SSCC) is asymptotically optimal, 
the use of independently designed source and channel codes may result in poor performance in the practical 
non-asymptotic regime of finite block length and low complexity encoding/decoding. 
This paper is concerned with this second case.

As a typical example, consider the transmission of digital images on a wireless 
channel.\footnote{This has become a killer application in cellular systems thanks to the widespread use of camera-equipped mobile phones.}
Present systems treat the compressed image as a data packet that must be delivered 
error-free to the destination, despite the fact that, differently from other kind of data, 
an image may be represented at the destination within some tolerable distortion. 
In a system based on separation, the distortion is introduced uniquely by  
source coding and the underlying link and physical layers struggle to deliver the source-encoded bits error-free, 
by using a combination of channel coding and retransmissions. 
Another typical example is represented by terrestrial or satellite Digital Video Broadcasting \cite{DVB_T,DVB_S}.
Here retransmissions cannot be used. Therefore, the channel coding scheme is designed to achieve a very strict BER target of $10^{-10}$ 
or below. In both these examples, a very demanding performance requirement is imposed on the physical  
layer. This is due to the fact that the source coding scheme, that was designed assuming an error-free channel, 
exhibits a {\em catastrophic} behavior with respect to channel errors: even a small fraction of bits in error at the source decoder 
input produces a very large distortion of the reconstructed source. 

The purpose of this paper is to outline a new {\em pragmatic} approach
to the design of JSCC schemes. 
Instead of focusing on idealized sources, channels and distortion measures as in most information-theoretic literature on 
JSCC (e.g., see \cite{gastpar} and references therein), 
we start from the typical structure of state-of the art source 
coders, as outlined in \cite{ortega-spmag98}. 
Although the focus of this paper is on {\em practical system design}, our approach is based on two key information theoretic
results: a) linear codes achieve the Shannon limit of $C/\overline{H}$ source symbols per channel use in order to transmit 
an arbitrary source with {\em sup-entropy rate} $\overline{H}$ over a symmetric memoryless channel with capacity $C$ 
(see Theorem 1 in the following); b) under quadratic distortion, the concatenation of (dithered) scalar quantization with entropy coding 
achieves the rate-distortion function of a stationary ergodic source within bounded rate penalty \cite{ziv-IT85,zamir-feder}. 

In our view, result (b) is the theoretical foundation of most state-of the art source ``transform'' coders, that are based on a linear 
transformation in order to project the source onto a convenient basis, followed by scalar quantization of the transform coefficients and 
entropy coding of the quantization indices. The latter is based on a carefully designed probability model matched to the class of sources 
to be encoded (see \cite{ortega-spmag98} and Section \ref{sec:jpeg2000-concept}). The catastropic behavior said above is due to the 
presence of the entropy coding stage, that is typically implemented by using standard data compression algortihms such as adaptive 
arithmetic coding or Huffman coding \cite{ortega-spmag98,TaMa,CoTh}. Fortunately, thanks to result (a), the conventional 
entropy coding stage can be replaced by a linear non-catastrophic encoder that maps directly the redundant symbol sequence output 
by the quantizer into a channel codeword (see Section \ref{sec:proposed-approach}).

In order to illustrate our ideas, we use as running example the transmission of digital images
over a Binary-Input Output-Symmetric (BIOS) channel and use JPEG2000 \cite{TaMa} as our baseline source coder.
Since the sequence of quantization indices is not binary in general, we use a successive encoding and ``onion-peeling'' decoding architecture:
the chain rule decomposition of entropy ensure asymptotic optimality of this approach, that has the advantage of making use of
binary linear codes, whose design is well understood.
As argued in Section \ref{sec:Conclusions}, our approach can be readily extended to a variety of channels and sources.

As far as low-complexity decoding is concerned, in Section \ref{sec:Encoder} we reduce the probability model defined by JPEG2000 
to a Markovian model, that admits a very simple trellis {\em Factor Graph}. This yields an efficient joint source-channel 
iterative decoding scheme based on Belief Propagation (BP) \cite{Ksch,RiU01}.
In Section \ref{sec:Decoder} we discuss the design of the linear coding stage. 
In this paper we consider the use of punctured Turbo Codes, although other families of linear codes can be easily used instead.
Numerical experiments that demonstrate the advantage of the proposed scheme are presented in Section \ref{sec:Results}.

{\bf Brief literature survey.}
JSCC is a very vast subject and covering it is well beyond the scope of this paper.
Here, we focus only on classical and recent works directly related to our approach.

The fact that linear codes with syndrome (linear) encoding achieve entropy for discrete sources is well-known 
(see \cite{CK} and references therein). This result was recently extended to arbitrary sources (also non-stationary non-ergodic)
in \cite{CaShVe_FnT}. We make use of this result to prove our Theorem 1, which is indeed a simple corollary.
The optimality of linear codes for (almost) lossless fixed-to-fixed length source coding, together with the advances in channel coding
that followed the discovery of Turbo Codes \cite{BerGlav} and the re-discovery of LDPC codes \cite{RiU01}, 
spurred an impressive amount of 
work aimed at using these families of codes with low-complexity BP decoding for data compression 
(see for example 
\cite{aaron-girod,garcia-frias-zhao,bajcsy-mitran,Hagenauer04,Hagenauer05,CaShVe_paris,CaShVe_brest,CaShVe_dimacs1,CaShVe_dimacs2,CaShVe_FnT}).
While a channel coding approach is needed for Slepian-Wolf separated coding of correlated sources, 
it is quite immediate to verify that it is not competitive 
with standard fixed-to-variable length entropy coding
both in terms of complexity and in terms of performance in the standard lossless data compression setting.
The only exception is, perhaps, the algorithm devised in \cite{CaShVe_paris,CaShVe_brest,CaShVe_dimacs1,CaShVe_dimacs2,CaShVe_FnT}
based on closed-loop ``doping'', that allows the decoder to achieve zero-error by allowing for some small variability in the encoding length.
However, when transmitting over a noisy channel closed-loop doping cannot be applied in a straightforward manner since the 
encoder cannot replicate exactly the decoding process. 

On the other hand, it is well-known that linear codes with {\em linear} encoding are bounded away from the rate-distortion 
function \cite{ancheta-IT,ancheta-thesis,massey78}. Linear codes in the lossy source coding setting have been proposed as a structured way 
to construct the quantization codebook. However, the encoder must be non-linear, and typically involves a high complexity.
Classical results on this topic are \cite{gallager,jelinek,hellman,viterbi-omura,koshelev} and 
more recent results can be found, for example, in \cite{wainwright,mezard,yedidia}.

We would like to stress the fact that our approach is {\em very different} from the above works: we do make use of linear codes 
and {\em linear} encoding. However, linear encoding is applied to the output of a scalar quantizer in order to map the (redundant) quantization indices
onto channel symbols. The scheme is not limited by Ancheta's negative result \cite{ancheta-IT,massey78} because quantization 
is a non-linear operation. However, the scheme is also not limited by the lossless or almost lossless 
requirement because we are in a lossy setting.
Furthermore, we take advantage of the very low complexity of scalar quantization and linear encoding.
In simple terms, we take the best of both non-linear and linear encoding approaches without paying a high price 
in terms of performance.

Another set of relevant related works deals with JSCC for redundant data, 
with a BER (Hamming distortion) performance criterion.
This has been recently pursued, for example, in \cite{ZhuAlaMitr, HAgWa06}.  Our use of punctured Turbo Codes
for the linear coding stage is largely inspired by \cite{ZhuAlaMitr}, with some differences that shall be pointed out in Section
\ref{sec:Decoder}. 
At the decoder we do make use of BP iterative decoding taking the source statistics into account.
This approach, generally referred to as joint source-channel {\em decoding}, 
or source-aided decoding, is treated in a very large number of works, as for example
\cite{vary,Hagen,gortz01,scanavino-grangetto,carlach05,lamy02,lamy05}.
Several works do not consider eliminating the entropy coding stage in the encoder as done here,
but make use of an iterative decoder that exploits some structure of the entropy code (Huffman, as in \cite{carlach05,lamy05} or
arithmetic, as in \cite{scanavino-grangetto}) in order to mitigate the catastrophic 
effect of residual bit errors after channel decoding on the entropy code inverse (decompression).
Unfortunately, classical entropy coding algorithms do not lend themselves easily to soft-input soft-output
BP decoding. Therefore these iterative scheme have typically high complexity and often not so exciting 
performance.

As a final remark, we would like to mention that several works considered a milder form of 
JSCC based on the optimization of the error protection (redundancy) of channel coding in order to optimize 
the end-to-end distortion performance of some standard {\em embedded} source coding scheme. 
This idea, which is directly related to the concept of 
{\em unequal error protection}, appears for example in \cite{Nosra, Chande, ChFa, Stan, Stankov, Hama05, maria_icc07} and it is briefly
discussed in Section \ref{sec:separated-scheme}. We would like to mention that our approach does not make use of
unequal error protection on the source data, with the exception of the model probability parameters 
that must be received error-free. This is similar to the {\em header} high-protection required by standard 
separated source-channel coding, and involves only a small fraction, vanishing in the limit of large block length, 
of the overall source length.

\section{Main ideas} \label{sec:Main-ideas}

This section is devoted to the main ideas driving the proposed JSCC scheme. 
The presentation is kept as general as possible.
A more detailed presentation of the encoder and decoder design for the specific example of 
digital images on BIOS channels is provided in Sections \ref{sec:Encoder} and \ref{sec:Decoder}, respectively.

\subsection{A typical source transform coding architecture} \label{sec:jpeg2000-concept}

\begin{figure}[thpb]
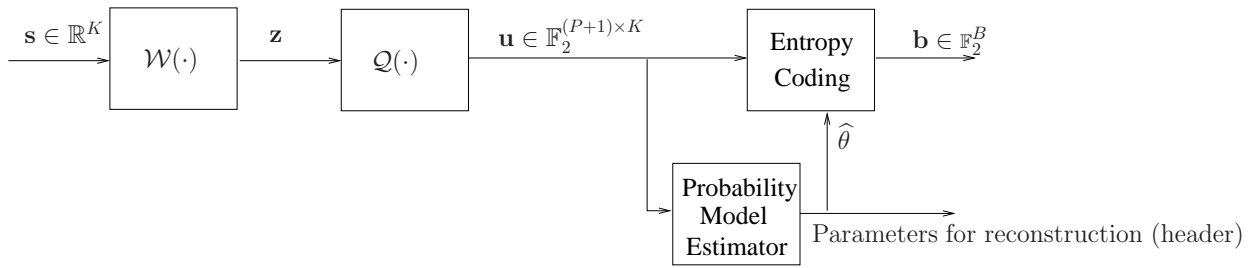

\centerline{\resizebox{1.0\textwidth}{!}{\input jpeg2000encoder.pstex_t}}
\caption{A JPEG2000-like source encoder. $\widehat{\theta}$ denotes the estimated probability model 
assumed by the entropy coding stage.}
\label{source-encoder}
\end{figure}

We consider a general transform source coding architecture illustrated by the block diagram of Fig. \ref{source-encoder} 
(see \cite{ortega-spmag98} and references therein), inspired by JPEG2000 \cite{TaMa}.

The source sequence $\sv \in \RR^K$ is transformed by the linear map
$\Wc : \RR^K \rightarrow \RR^K$ (e.g., a wavelet transform). 
The sequence of transform coefficients $\zv = (z_1,\ldots,z_K) = \Wc(\sv)$
is quantized by applying componentwise the scalar quantizers $\Qc_k : \RR \rightarrow \FF_2^{P+1}$,
for $k = 1,\ldots,K$, such that $u_k = \Qc_k(z_k)$.\footnote{The quantizers $\Qc_k$ may differ by their quantization regions and reconstruction points. 
For example, in the scheme considered here the quantizers $\Qc_k$
have a different dynamic range depending on which subband of the Discrete Wavelet Transform $\Wc$ the symbol $z_k$ belongs to.}
We denote by $P$ the number of quantization bits used to represent the
magnitude of sample $z_k$, and use one additional bit to represent the sign of $z_k$. Notice that this binary representation for
the quantization indices does not involve any loss of generality.
We let $\uv = \Qc(\zv) = (u_1,\ldots,u_K)$ denote the sequence of quantization indices, and
let $\{u_{p,k}: p = 0,\ldots,P\}$ denote the ``bits'' forming $u_k$.
We think of $\uv$ as a two-dimensional $(P+1)\times K$ binary array, where the rows
$\uv^{(p)} = (u_{p,1}, \ldots, u_{p,K})$, for $p \in \{0,\ldots,P\}$,
are referred to as the $p$-th ``bit-planes''.
The bit-plane $p = 0$ contains the sign bit, and the bit-planes $p=1,\ldots,P$ contain the magnitude bits, where
$p=1$ corresponds to the least significant bit and $p = P$ corresponds to the most significant bit.

The sequence of quantization indices $\uv$ is generally redundant.
Therefore, $\uv$ must be further compressed by a combination of {\em decimation} and {\em entropy coding}.
By ``decimation'' we mean discarding some segments of the bit-planes: only the segments of $\uv$
that are not discarded are effectively entropy-encoded and contribute to the encoder output $\bv \in \FF_2^B$.
For the sake of exposition simplicity, in this work we assume that no decimation 
is performed, although the case of decimation can be easily handled.

The entropy encoder is based on a probability parametric model, targeted to the specific class of 
sources (see \cite{ortega-spmag98} and references therein).
In the sequel, we assume the following probability model for our source: let $\{\PP^{(K)}_\theta(\cdot) : \theta \in \Theta, K=1,2,\ldots\}$ denote a family of processes, defined by the sequence of $K$ dimensional 
probability distributions\footnote{In the interest of notation simplicity, in this paper we use the symbol 
$\PP$ to denote probability with respect to the appropriately defined joint probability space.
We do not distinguish between the random variables and the arguments of the probability density or mass function,
since it is clear from the context.} 
on $\FF_2^{(P+1)\times K}$ parameterized by $\theta$. 
For the time being, we assume that for each $\theta \in \Theta$, the corresponding process 
is a stationary and ergodic, and denote its entropy rate by $H_\theta(U)$. 
This assumption shall be revisited in Section \ref{sec:Encoder}
when we discuss more specifically the probability model underlying JPEG2000.
Furthermore, we assume that the probability model is {\em matched}, that is,
$\uv \sim \PP^{(K)}_\theta$ for some $\theta \in \Theta$. A model mismatch would involve additional rate penalty. 
However, discussing mismatch in this context is rather pointless since the {\em actual} statistics of real-life 
sources such as images is essentially unknown.

The modeler estimates the probability parameter
$\widehat{\theta}$ from the current realization of $\uv$ and encodes losslessly 
the $p$-th bit-plane using about
\begin{equation} \label{outputlength}
B_p = - \log_2 \PP^{(K)}_{\widehat{\theta}}(\uv^{(p)}|\uv^{(p+1)},\ldots,\uv^{(P)}) 
\end{equation}
bits. As we will see in Section \ref{sec:Encoder}, in our working example the probability model is 
conditionally Markov with fixed state transition diagram, where the conditioning on the $p$-th bit-plane 
is due to the upper bit-planes $p+1,\ldots,P$.
Then, the model parameter $\theta$ consists of the collection of transition matrices defining 
these Markov chains, the elements of which can be easily Maximum-Likelihood (ML) estimated 
by counting the empirical frequency of symbols corresponding to each state transition.

In source coding schemes such as JPEG2000, the model parameter estimation and the entropy coding is performed simultaneously, 
by sequential estimation of the state transition probabilities with a Krichevsky-Trofimov (KT) probability estimator \cite{Trofimov} together 
with arithmetic coding \cite{Witten87, TaMa}. In Fig.~\ref{source-encoder} we represent the probability model estimation 
and the entropy  coding as two separate blocks for conceptual simplicity and because in the proposed scheme 
(see Section \ref{sec:proposed-approach}) these two functions are indeed performed separately.

The output length of the source encoder is given by
$B = \sum_{p=0}^{P} B_p$. It follows that the source coding rate is given by
\[ R_s = \frac{B}{K} \;\;\; \hbox{bit/source sample} \]
This corresponds to some target distortion level.
In the case of no decimation, distortion is due uniquely
to the quantization error. We shall refer to this distortion value, denoted by $D_{\Qc}$, 
as the {\em quantization distortion}. Of course, a more flexible tradeoff between
distortion and rate can be achieved by considering decimation.

\subsection{Separated approach and Shannon limit} \label{sec:separated-scheme}

We consider the problem of transmitting the source sequence $\sv$
over a stationary memoryless BIOS channel \cite{RiU01} 
with capacity $C$ bit/channel use.

In a classical SSCC approach, the compressed bit sequence $\bv$ produced by the source encoder is mapped into a channel 
codeword $\xv \in \FF_2^N$ by a channel encoder of rate $R_c = B/N$.
The resulting encoding efficiency $\eta$ (measured in source samples per channel
use) is given by
\begin{equation} \label{eta}
\eta = \frac{K}{N} = \frac{R_c}{R_s}
\end{equation}
In the limit of large $K$, assuming stationarity and ergodicity of the source and a ML probability estimator 
such that $\widehat{\theta} \rightarrow \theta$, we have that $B \rightarrow K H_\theta(U)$.
Furthermore, when $\Theta$ is an $M$-dimensional compact set, 
Rissanen's bound \cite{Rissanen1984Universal} ensures that the model representation 
redundancy grows only as $\frac{M}{2} \log K$, and therefore communicating the probability model parameter 
$\widehat{\theta}$ as side 
information to the decoder  costs asymptotically a vanishing rate penalty (more details on the parameter representation 
are given in Section \ref{sec:Encoder}).

From what said above and Shannon source and channel coding theorems \cite{CoTh} it 
follows that, for arbitrary $\delta,\epsilon > 0$ and sufficiently large $K$, there exist 
pairs of source and channel codes achieving efficiency
\begin{equation} \label{etaopt}
\eta = \frac{C}{H_\theta(U)} - \delta
\end{equation}
with error probability $\PP(\widehat{\bv} \neq \bv) \leq \epsilon$,
where $\widehat{\bv}$ denotes the channel-decoded sequence of information bits.

The point with coordinates $(C/H_\theta(U),D_\Qc)$ on the
efficiency-distortion plane corresponds to the best possible performance for a 
JSCC with fixed quantizer $\Qc$ and source model parameter $\theta$. 
This point shall be referred to as the {\em Shannon limit} for our system.
Notice that for a source with rate distortion function $R(D)$, the best possible efficiency 
is given by $C/R(D_\Qc) \geq C/H_\theta(U)$. 
Nevertheless, for complicated sources such as images the rate distortion function is not generally known.
Furthermore, scalar quantization followed by entropy coding is near-optimal over a wide range of rates
a wide class of sources \cite{gish-pierce,ziv-IT85,zamir-feder,linder-IT00,marco-neuhoff-IT06}.
Following the pragmatic approach advocated in \cite{ortega-spmag98}, we say that the rate-distortion point 
$(H_\theta(U),D_\Qc)$ is a point on the source encoder {\em operational rate-distortion curve}.

As already argued in Section \ref{sec:Intro}, it is well-known that 
for practical (finite) source block lengths and channel encoding/decoding complexity the SSCC scheme obtained by the 
concatenation of the transform coder described in Section \ref{sec:jpeg2000-concept} with a channel code might perform quite poorly. 
It is well-known that entropy coding has a {\em catastrophic} behavior: its inverse function is ill-conditioned. 
A small number of bit-errors in the channel-decoded bit sequence $\widehat{\bv}$ generates a large number of symbol errors in
the entropy-decoded quantization index sequence $\widehat{\uv}$, and eventually a large distortion in the
reconstructed sequence $\widehat{\sv} = \Wc^{-1}(\Qc^{-1}(\widehat{\uv}))$.
\footnote{With some abuse of
notation, we denote by $\Qc^{-1}$ the reconstruction mapping of
the quantizer, i.e., $\Qc^{-1}(\uv)$ denotes the representation point $\widehat{\sv}$ of the quantization bin
indexed by $\uv$.}

The catastrophic behavior of the source encoder can be partially
mitigated by imposing an {\em embedded} structure. An encoder is
said to be embedded if for any $B' < B$, the output sequence
$\bv'$ produced for output length $B'$ is the prefix of sequence
$\bv$ produced for output length $B$. With an embedded source
encoder, all the bits in $\widehat{\bv}$ from the beginning to the
first occurred bit-error can be used for reconstruction, while all
the rest must be discarded. Based on this idea, several works (see
for example \cite{Nosra, Chande, ChFa, Stan, Stankov, Hama05, maria_icc07}) have
addressed the optimization of channel coding
redundancy in order to maximize the average number of correctly
received bits or minimize the average distortion
before the occurrence of the first bit-error. 
In this paper we take a different approach, 
outlined in the next section.

\subsection{The proposed approach} \label{sec:proposed-approach}

\begin{figure}[thpb]
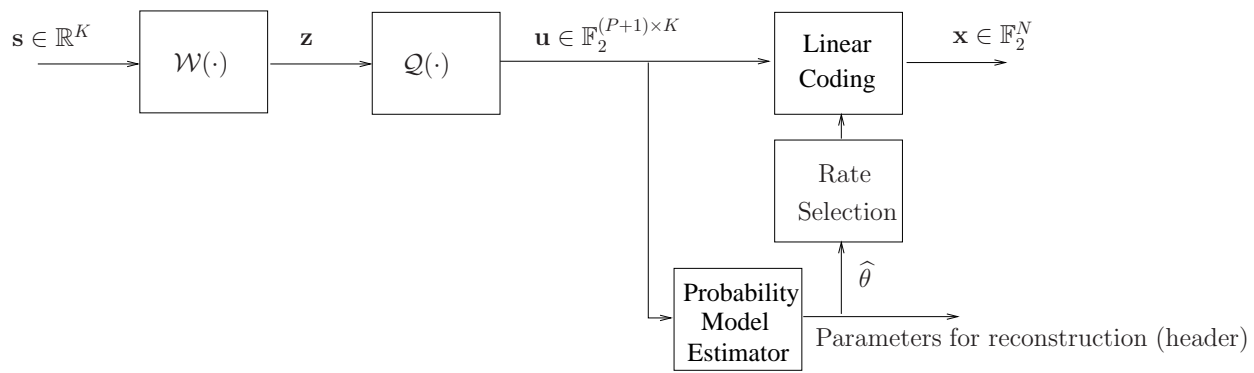

\centerline{\resizebox{1.0\textwidth}{!}{\input joint.pstex_t}}
\caption{Proposed joint source-channel coding scheme. The
estimated probability model parameters $\widehat{\theta}$
are separately transmitted as side information.}
\label{joint}
\end{figure}

Instead of {\em mitigating} the catastrophic behavior of the entropy coding stage,
we avoid it by taking a JSCC approach:
we merge entropy coding and channel coding into a single non-catastrophic
encoding operation, that maps the redundant sequence $\uv$ directly into the channel
codeword $\xv$ (see Fig.~\ref{joint}).

Since linear codes achieve the capacity of BIOS channels \cite{Csizar81} we shall
consider a linear map $\uv \mapsto \xv$. Furthermore, since {\em binary} linear codes are particularly 
simple and well understood, we shall implement this linear mapping in layers, bit-plane by
bit-plane. In particular, we consider $P+1$ linear codes
$\Cc_0,\ldots,\Cc_P$ with block length $N_0,\ldots,N_{P}$ and
generator matrices $\Gm_0,\ldots,\Gm_P$. We obtain the codeword
$\xv$ as the concatenation of $\xv^{(0)}, \ldots,\xv^{(P)}$, where
\begin{equation} \label{linear-coding}
\xv^{(p)} = \uv^{(p)} \Gm_p
\end{equation}
Three questions naturally arise at this point:
1) Suppose that for each $\theta$ we can pick the best encoding matrices, can we approach the Shannon limit?
2) Is it necessary to ``fine tuning'' the encoding matrices $\{\Gm_p\}$ for each set of source parameters $\theta \in \Theta$? 
3) Can we find a low-complexity decoder for the JSCC scheme?

Questions 1 and 2 are addressed simultaneously by Theorem 1 here below and by the comment that follows.
Question 3 is addressed in Section \ref{sec:Decoder}, where the Markov structure of the source probability model
is exploited in conjunction with the structure of the binary linear codes in order to obtain a low-complexity 
iterative joint source-channel decoder based on BP \cite{Ksch,RiU01,BerGlav}.

The asymptotic goodness of our scheme is supported by the following result,
that is an immediate corollary of the optimality of linear codes for both
lossless compression \cite{Csizar81,CaShVe_dimacs1,CaShVe_FnT} and for achieving
capacity of BIOS channels (see \cite{Csizar81} and references therein) 
and of the fact that the concatenation of two linear maps is a linear map. 

{\bf Theorem 1.} Consider a binary source $V$ defined by the
sequence of $K$-dimensional joint probability distributions
$\{P^{(K)}_V(\vv): K = 1,2,\ldots\}$ over $\FF_2^K$. Define the
sup-entropy rate $\overline{H}(V)$ of $V$ as \cite{verdu-han,han-book}
the limsup in probability of the sequence of random variables
\[ -\frac{1}{K} \log_2 P^{(K)}_V(\vv) \]
that is, the infimum of all values $h$ for which
\begin{equation} \label{sup-entropy}
\PP \left ( -\frac{1}{K} \log_2 P^{(K)}_V(\vv) \geq h \right ) \rightarrow 0
\end{equation}
as $K \rightarrow \infty$.
Consider a system that, for each length $K$, maps source sequences $\vv$ into binary codewords
$\cv = \vv\Gm$ of length $N$, and transmits $\cv$ over a BIOS channel with capacity $C$.
Let $\yv$ denote the channel output and $\psi : \yv \mapsto \widehat{\vv}$ be a 
suitable decoder.

Then, for any $\delta,\epsilon > 0$  and sufficiently large $K$
there exists a $K \times N$ matrix $\Gm$ and a decoder
$\psi$ such that $\PP(\widehat{\vv} \neq \vv) \leq \epsilon$ and $K/N \geq C/\overline{H}(V) - \delta$. \\[12pt]
{\bf Proof.} See Appendix \ref{app:proof}. \hfill $\square$ \\[12pt]

We argue that Theorem 1 has an important consequence for the joint source-channel coding 
of a family of sources $\Theta$ and a fixed BIOS channel. 
In fact, for each value $H$ there exists {\em one sequence} of encoding matrices of increasing block length $K$ 
and efficiency arbitrarily close to $C/H$ such that the decoding error probability vanishes {\em for all} 
the source statistics with parameters $\theta \in \Theta$ such that $H_\theta(U) = H$.
This fact is seen by considering  a mixed source \cite{han-book} $V$ defined by 
the family of distributions $\{\PP_\theta : \theta \in \Hc\}$, where $\Hc \subseteq \Theta$ is the set of parameters for which
$H_\theta(U) = H$ and where $\theta$ has a uniform prior probability over $\Hc$. 
It is easily seen that in this case $\overline{H}(V) = H$. Therefore, by Theorem 1, there exists a sequence of coding matrices of 
increasing block length such that $\PP_\theta(\widehat{\vv} \neq \vv) \leq \epsilon$ and $K/N \geq C/H - \delta$.

As far as implementation is concerned, this implies that we need only
to design one set of coding matrices $\{\Gm_p\}$ for each value $H$ of the source entropy rate. 
If ``tuning'' of the codes were needed for each particular $\theta \in \Theta$, the scheme would be impractical.
In fact, this would require a very large set of codes that do not differ only by their rate, but also by their structure (e.g., 
by their generator polynomials in the case of TCs, or degree distributions in the case of LDPCs). 
On the contrary, families of codes with different rate can be obtained by progressive puncturing of a single
``mother'' code, in a very convenient way for implementation. This is in fact the approach taken in the code design 
of Section \ref{sec:Decoder}: we define a fine quantization grid on the interval $[0,1]$ 
of possible values of the bit-plane empirical entropy rate and design a coding matrix $\Gm_p$ 
for each quantized rate value.  Then, when encoding the $p$-th bit-plane, we compute the conditional empirical 
entropy rate
\begin{equation} \label{empirical-entropy} 
\widehat{H}(U_p|U_{p+1},\ldots,U_P) = - \frac{1}{K} \log_2 \PP^{(K)}_{\widehat{\theta}}(\uv^{(p)}|\uv^{(p+1)},\ldots,\uv^{(P)})
\end{equation}
and choose the corresponding (pre-designed) encoding matrix.

A similar approach is followed in \cite{CaShVe_paris,CaShVe_brest,CaShVe_dimacs1,CaShVe_FnT} 
where linear block codes are used for universal lossless compression. 
Another possibility consists of using {\em rateless codes} \cite{raptor,luby,CaShVe_dimacs2}, in order to be able 
to generate an arbitrary amount of coded symbols and therefore a continuum of rates. 
This approach was followed in the context of universal lossless compression in 
\cite{CaShVe_dimacs2} and it is briefly discussed in Section \ref{sec:Conclusions} as a possible extension 
of the present work.

For an arbitrarily fine rate quantization grid, we let the codeword block length of the $p$-th bit-plane be given by 
\[ N_p = K  \left ( \frac{\widehat{H}(U_p|U_{p+1},\ldots,U_P)}{C} + \delta_p \right ) \]
where $\delta_p > 0$ is a small rate margin. The achieved coding efficiency is given by 
\begin{eqnarray} \label{eta-new}
\eta & = & \frac{K}{\sum_{p=0}^P N_p} \nonumber \\
& = & \frac{1}{\sum_{p=0}^P \frac{\widehat{H}(U_p|U_{p+1},\ldots,U_P)}{C} + \delta_p} \nonumber \\
& = & \frac{C}{\widehat{H}(U)} - \delta
\end{eqnarray}
where $\delta$ is a small positive quantity that vanishes as all 
$\delta_p \rightarrow 0$ and $\widehat{H}(U) = - \frac{1}{K} \log_2 \PP^{(K)}_{\widehat{\theta}}(\uv)$ is the empirical 
entropy rate of $\uv$. Since $\widehat{H}(U) \rightarrow H_\theta(U)$, it follows that for asymptotically large $K$ 
the proposed method approaches the operational Shannon limit defined in Section \ref{sec:separated-scheme}.

The advantage of JSCC over SSCC becomes clear in the non-asymptotic regime of moderate $K$ and
practical low complexity channel coding and decoding.
In fact, the design of non-catastrophic linear encoders is a very
well-known and well understood topic in coding theory \cite{McEliece}.
In particular, powerful channel coding families
such as Turbo Codes (TC) \cite{BerGlav}, Low-Density Parity-Check
(LDPC) codes \cite{RiU01, Gallager62, Gallager63, mncEL2} and
Irregular Repeat-Accumulate (IRA) codes \cite{Jin, Roumy} can
easily achieve post-decoding BER between $10^{-3}$ and $10^{-6}$
provided that their rate is below a certain threshold, that is
close to $C/H_\theta(U)$ even for moderate information block length $K$.
While, as said before, such values of BER would produce very large
distortion in the presence of the entropy coding stage, in the
proposed system the BER affects directly the bit-plane components $u_{p,k}$.
It is clear that some bits in error in the bit-planes yield a small output
distortion, since the inverse transform $\Wc^{-1}$ is linear,
unitary or close to unitary, and hence well-conditioned. 

As we will see in Section \ref{sec:Decoder}, the proposed decoder based on BP \cite{Ksch,RiU01,BerGlav}
computes efficiently an accurate approximation of the symbol-by-symbol posterior probabilities
\begin{equation} \label{posterior-marginals}
\left \{\PP(u_{p,k} | \yv) : p = 0,\ldots,P, \;\; k = 1,\ldots,K \right \}
\end{equation}
where $\yv$ is the channel output corresponding to the transmission of $\xv$.

As far as the source reconstruction is concerned, we let (without loss of generality) the $k$-th scalar quantizer 
map $u_k = Q_k(z)$ be
\begin{eqnarray} \label{Qk}
u_{0,k} & = & \left \{ \begin{array}{ll}
0 & z \geq 0 \\
1 & z < 0 \end{array} \right . \nonumber \\
(u_{1,k},\ldots,u_{P,k}) & = & \arg \min_{\vv \in \FF_2^P} \; \left | \frac{|z|}{\Delta_k} - \sum_{p=1}^P v_p 2^{p-1} \right |
\end{eqnarray}
where the parameter $\Delta_k$ determines the dynamic range of the quantizer for the $k$-th transform coefficient.
The corresponding reconstruction function is given by
\begin{equation} \label{Qinverse}
Q_k^{-1}(u_k) = (-1)^{u_{0,k}} \frac{\Delta_k}{2} \sum_{p=1}^{P} u_{p,k} 2^{p}
\end{equation}
Given the symbol-by-symbol posterior probabilities (\ref{posterior-marginals}) produced by the BP decoder,
both hard and soft reconstruction are possible. In the first case, symbol-by-symbol hard decisions
$\widehat{u}_{p,k} = \; \arg\max_{v \in \FF_2} \PP(u_{p,k}=v | \yv)$ are used in (\ref{Qinverse}) to generate
an estimate $\widetilde{z}_k$ of the $k$-th transform coefficient. In the second case,
the MMSE (conditional mean) estimator $\widetilde{z}_k = \EE[z_k|\yv]$ is computed.
Assuming zero-mean quantization noise statistically independent of the channel output $\yv$, this takes on the
appealing simple form\footnote{Of course, this is only an approximation if the BP decoder produces approximations of the true
symbol-by-symbol a posteriori probabilities (\ref{posterior-marginals}).}
\begin{equation} \label{soft-estimate1}
\widetilde{z}_k = \frac{\Delta_k}{2} \tanh\left (\frac{\lambda_{0,k}}{2} \right ) \sum_{p=1}^{P}
\frac{2^p}{1 + e^{\lambda_{p,k}}}
\end{equation}
where we define the a posteriori log-likelihood ratio (LLR) for symbol $u_{p,k}$ as
\begin{equation} \label{LLR}
\lambda_{p,k} = \log \frac{P(u_{p,k}=0 | \yv)}{P(u_{p,k}=1 | \yv)}
\end{equation}
After producing the sequence $\widetilde{\zv} = (\widetilde{z}_1,\ldots,\widetilde{z}_K)$,
the source is reconstructed by applying the inverse linear transform $\widetilde{\sv} = \Wc^{-1}(\widetilde{\zv})$.
The soft reconstruction approach defined in (\ref{soft-estimate1}) is sometimes referred to as ``soft-bit''
reconstruction in the literature \cite{fing97,fing01}.

\section{Probability model, estimation and lossless compression} \label{sec:Encoder}

From this section to the end of the paper we illustrate in more details an implementation 
example based on JPEG2000.
Despite the loss of generality, by developing this example we hope to corroborate the claims made
in Section \ref{sec:proposed-approach} and gain in clarity. 
Generalizations are discussed in Section \ref{sec:Conclusions}.

\begin{figure}[thpb]
\centerline{
\input{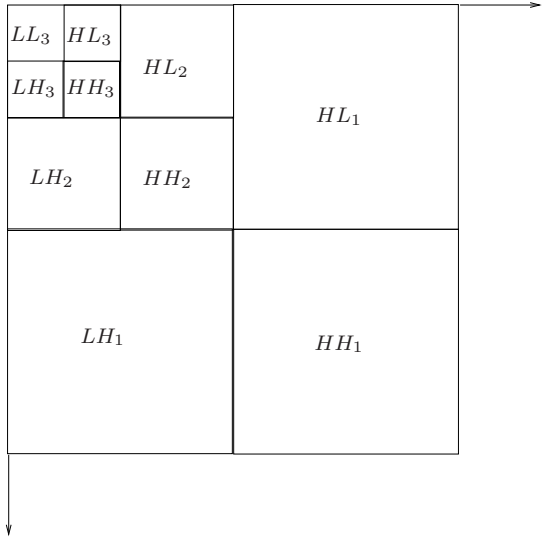}}
\caption{Subband decomposition in the DWT transform with $D=3$ levels.}
\label{DWT}
\end{figure}

In JPEG2000 the linear transform $\Wc$ in the block diagram of Fig.~\ref{source-encoder}
is a Discrete Wavelets transform (DWT) \cite{AnBa,LeGall,Swe}. This determines the subband structure shown
in Fig. \ref{DWT}. Let us consider a squared gray scale image of dimension $n \times n$ pixels.
After $D$ stages of DWT, the transform coefficients are partitioned into 
$3D + 1$ squared subbands (named $LL_D$ and $HL_d, LH_d$ and
$HH_d$, for $d = 1,\ldots,D$) of dimensions $\frac{n}{2^d} \times
\frac{n}{2^d}$, where $d$ is the decomposition step and where 
``L'' and ``H'' stand for {\em Low} and {\em High} frequency components, 
respectively.

The transform coefficients $z_k$ are quantized as in (\ref{Qk}).
The parameter $\Delta_k$ of the $k$-th quantizer is set to a fixed value for all positions $k$
in the same subband, and depends on the average energy content of the subband \cite{TaMa}.
These parameters are sent separately to the decoder for reconstruction.
Each quantized subband is partitioned into squared
blocks called ``code-blocks'',  that are independently entropy-encoded.
The typical size of code-blocks is $32 \times 32$ or $64 \times 64$.
This partitioning is done for the sake of decimation (discarding some code-blocks)
and in order to avoid error propagation across code-blocks at the reconstruction side:
if an error occurs in the channel-decoded stream, the error propagation will be limited
inside a code-block. The probability model is estimated {\em locally} on each code-block, 
thus allowing a better matching of the model used for entropy coding 
with the local statistics of the quantization indices. 
In JPEG2000 this is obtained by resetting the KT probability estimator at the beginning of each code-block.
Since in our system we do not have such error propagation problems and, for simplicity, we do not consider decimation,
we shall not use a rigid partitioning into code-blocks. Nevertheless, we shall consider the matching of the probability model parameters
to the local statistics, depending on the bit-plane index $p$ and subband index $d$. 
The binary data corresponding to the bit-plane/subband index pair $(p,d)$ will be referred to in the following as
the $(p,d)$ ``segment''.

\begin{figure}[thpb]
\centerline{
\input{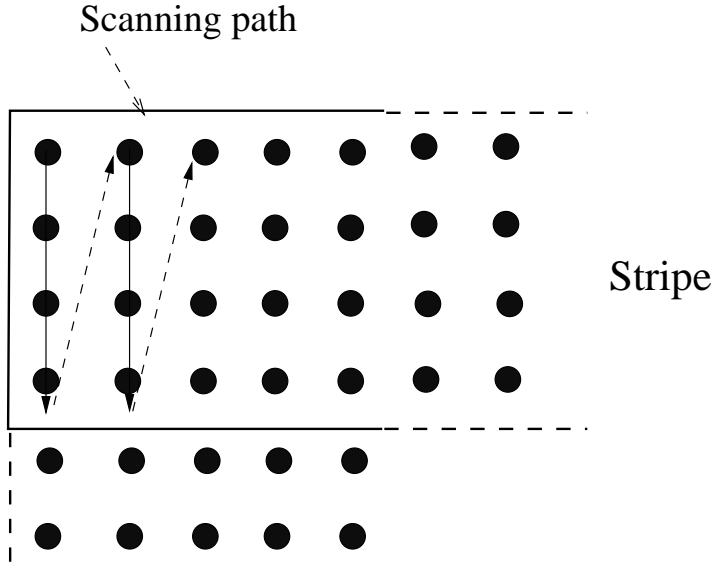}}
\caption{Stripe-oriented scanning of the quantization indices in order to induce a one-dimensional ordering.} \label{stripe}
\end{figure}

The two-dimensional sequence of quantization indices is arranged into a one-dimensional ``temporal'' 
ordering according to the so called ``stripe-oriented'' scanning scheme illustrated in Fig.~\ref{stripe}.
Each stripe is formed by four rows of quantization indices \cite{TaMa} and a number of columns equal to the dimension 
of the subband. Then, each quantized coefficient $u_k$ is identified uniquely by its position $k$
in the resulting one-dimensional arrangement $\uv$.

JPEG2000 models each bit-plane $\uv^{(p)}$ as a binary correlated
source, where the probability distribution of a bit $u_{p,k}$
depends on the value of certain neighboring bits in the same plane and on certain
bits in the upper bit-planes $p+1,\ldots,P$. Without entering into the
fine details of the scheme, that is extremely tedious and can
be learned from standard references \cite{TaMa}, we concentrate on
the {\em qualitative} features of the probability model.

The local dependency of bit $u_{p,k}$ on its neighbors can be
formulated as a Markov chain, conditioned on the realization of
the symbols in the upper bit-planes $\uv^{(p+1)},\ldots,\uv^{(P)}$. 
Consider the conditional probability distribution
\begin{equation} \label{cond-Pk}
\PP^{(K)}_\theta(\uv^{(p)}|\uv^{(p+1)},\ldots,\uv^{(P)}) = \prod_{k=1}^K
\PP_\theta(u_{p,k}|u_{p,1},\ldots,u_{p,k-1},\uv^{(p+1)},\ldots,\uv^{(P)})
\end{equation}
The underlying conditional Markov model for the $p$-th bit-plane is represented by the block
diagram of Fig.~\ref{markov-model}.

\begin{figure}[thpb]
\centerline{\resizebox{0.8\textwidth}{!}{\input
source-markov-model.pstex_t}} 
\caption{Markov model underlying the
conditional bit-plane joint distribution
$\PP^{(K)}_\theta(\uv^{(p)}|\uv^{(p+1)},\ldots,\uv^{(P)})$.}
\label{markov-model}
\end{figure}

The Markov chain state $\pi_{p,k} = (u_{p,k-1},\ldots,u_{p,k-L})$ is formed by the content
of a {\em causal} sliding window of previous bits in the same bit-plane
(the content of the shift-register of Fig.~\ref{markov-model}), for some integer $L$.
Furthermore, the dependency of bit $u_{p,k}$ on the bits
in the upper bit-planes is also confined to a {\em local} collection of index pairs
$(p',k') \in \Sc_{p,k}$. In other words, the local dependency set $\Sc_{p,k}$ is defined as the set of
index pairs $(p',k')$ for $p' > p$ such that
\[ \{u_{p'',k''} : (p'',k'') \notin \Sc_{p,k}, p'' > p\} \rightarrow
\{\pi_{p,k}, \; \{u_{p',k'} : (p',k') \in \Sc_{p,k}\}\}
\rightarrow u_{p,k} \] 
is a Markov chain. 
Obviously, for the top bit-plane we have $\Sc_{P,k} = \emptyset$ for all $k$.

\begin{figure}
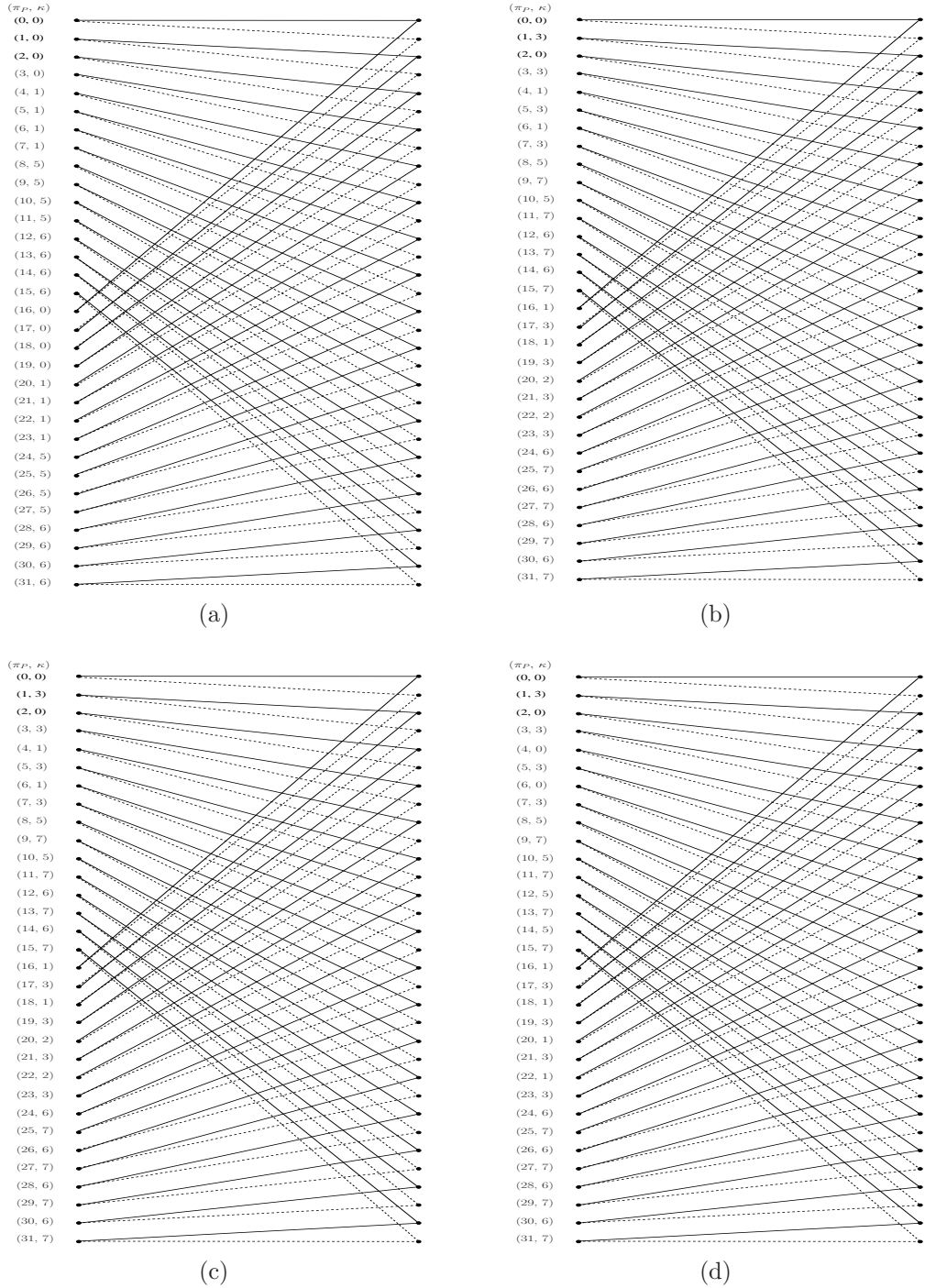

\label{trellis-diag}
 \subfloat[]
  {\resizebox{6 cm}{8.5 cm}{\input trellis1.pstex_t}}
  \hspace{1 cm}
  \subfloat[]
{\resizebox{6 cm}{8.5 cm}{\input trellis2.pstex_t}}\\
\subfloat[]
  {\resizebox{6 cm}{8.5 cm}{\input trellis3.pstex_t}}
   \hspace{1 cm}
    \subfloat[]
   {\resizebox{6cm}{8.5 cm}{\input trellis4.pstex_t}}
\caption{Trellis diagram for the Markov model of the $P$-th bit-plane: the Markov chain is 
time-variant, and the four sections (a), (b), (c) and (d) correspond to the four different positions on a bit
in the stripe (see Fig. \ref{stripe}).}
\end{figure}

The Markov model of Fig.~\ref{markov-model} has a {\em fixed} state diagram structure, and it is parameterized by the transition probabilities
\begin{equation} \label{param1}
\PP_\theta(u_{p,k}|u_{p,1},\ldots,u_{p,k-1},\uv^{(p+1)},\ldots,\uv^{(P)})
= \PP_\theta\left (u_{p,k}|
\underbrace{u_{p,k-L},\ldots,u_{p,k-1}}_{\pi_{p,k}}, \{u_{p',k'} :
(p',k') \in \Sc_{p,k}\} \right )
\end{equation}
Fortunately, these probabilities take on distinct values only for
a small number of equivalent configurations of the conditioning
bits. In the JPEG2000 parlance, we say that the conditioning bits
$(\pi_{p,k},\{u_{p',k'} : (p',k') \in \Sc_{p,k}\})$ define the
{\em context} of bit $u_{p,k}$. Despite the fact that we have
$2^{L + |\Sc_{p,k}|}$ possible configurations, many of them are
equivalent. We define the context function at bit-plane $p$ as
$\Kc_p : \FF_2^{L + |\Sc_{p,k}|} \rightarrow \{0,\ldots,M-1\}$, for
some integer $M$. We say that two configurations are equivalent if
their image under $\Kc_p$ (the associated ``context'') is the same. 
Then, for $\kappa \in  \{0,\ldots,M-1\}$ and $u_{p,k} \in \{0,1\}$ we have
\[ \PP_\theta\left (u_{p,k}| \pi_{p,k}, \{u_{p',k'} : (p',k') \in \Sc_{p,k}\} \right ) =
\PP_\theta(u_{p,k} | \kappa) \]
for all configuration of the conditioning bits such that
$\Kc(\pi_{p,k},\{u_{p',k'} : (p',k') \in
\Sc_{p,k}\}) = \kappa$.  
In JPEG2000 we have $M=17$, where 12 contexts are used for the magnitude bit-planes and 5
contexts are used for the sign bit-plane. 
More details about the Markov state diagram structure and on context definition in the notation of this paper (which is rather different from the 
current JPEG2000 descriptions available in the literature \cite{TaMa}) 
is available in \cite{maria-thesis}. Fig.~\ref{trellis-diag} shows the trellis diagram corresponding 
to the Markov chain of Fig.~\ref{markov-model} for the top ($P$-th) bit-plane. 
We have $L = 5$, which yields a 32-state trellis.
The states are enumerated from 0 to 31, the label next to each state contains the state number and the 
corresponding context $\kappa$. There are four non-equivalent trellis sections with the same state transition 
structure (defined by the shift-register, see footnote (11) in Section \ref{sec:Decoder}) but different correspondence
between state and context value, that depends on the position of the bit in the stripe.
As seen from Fig. \ref{stripe} there are four different positions correspoding to the four rows forming the stripe.
Here, trellis sections from (a) to (d) corresponds to the four positions from top to bottom.
Each state has two outgoing transitions corresponding to $u_{P,k}$ being 0 (solid) or 1 (dashed).
For example, for state 26 in section (a), the solid transition corresponds to $u_{P,k} = 0$ with probability $\PP_\theta(0|5)$ and
and the dashed transition corresponds to $u_{P,k} = 1$ with probability $1 - \PP_\theta(0|5)$. 
We showed the $P$-th bit-plane for the LL and LH subbands. The state-context corespondence for the other subbands is different, and it is
determined as explained in \cite{maria-thesis}. For the $p$-th bit-planes with $p < P$ the correspondence depends on the 
value of the conditioning bits in the set $\Sc_{p,k}$ and then
it varies with the time index $k$. It would be therefore very cumbersome to represent these trellises 
in this paper.

At this point, it should be clear that the Markov model is completely determined by the LLRs
\begin{equation} \label{a-priori}
\nu_{p,k}(\kappa) = \log \frac{\PP_\theta(u_{p,k} = 0| \kappa)}{\PP_\theta(u_{p,k} = 1| \kappa)}, \;\; \kappa = 0,\ldots,M-1
\end{equation}
In our scheme, we use a non-sequential ML estimator for the above probabilities, that is easily obtained by computing
the empirical frequency of zeros for each context $\kappa$, where separate bit-counts are maintained
for each data block with homogeneous local statistics as explained in the following.

As said before, the Markov model must be matched to the {\em local} statistics of $\uv$, that it is in general non-stationary. 
Different probability estimates are computed for each $(p,d)$-th segment. 
Furthermore, we found that it is convenient to estimate a local statistics for groups of 
adjacent stripes, generally smaller than a whole segment.

Suppose that segment $(p,d)$ is partitioned into $m_{p,d}$ regions 
(groups of adjacent stripes) and over each such region the model transition 
probabilities are locally estimated. For all positions $k$ in the same region, 
the LLRs defined in (\ref{a-priori}) take on the same value that depend only on the region, and not on the time index $k$.
We denote such parameters by $\nu(p,d,\ell,\kappa)$, where $p$ denotes the bit-plane, 
$d$ denotes the subband, $\ell = 1,\ldots,m_{p,d}$ denotes the region in the $(p,d)$ segment and 
$\kappa$ is the context. Then, $\nu_{p,k}(\kappa) = \nu(p,d,\ell,\kappa)$ for all positions $k$ corresponding to
the $\ell$-th region of the $(p,d)$-th segment. 
It follows that the overall Markov model is piecewise stationary, and it is defined by 
\begin{equation} \label{number-of-parameters}
\Mc =  M_0 \sum_{d=1}^{3D+1} m_{d,0}  + M_1 \sum_{d=1}^{3D+1} \sum_{p=1}^P m_{d,p}
\end{equation}
real parameters, where $M_1$ denotes the number of distinct contexts for the magnitude bit-planes and $M_0$
denotes the number of distinct contexts for the sign bit-plane, with $M_0 + M_1 = M$. 
The model parameter $\theta$ coincides with the collection of all the $\Mc$ LLRs $\{\nu(p,d,\ell,\kappa)\}$ defined above.

The estimated parameters $\widehat{\theta}$ must be sent to the decoder separately 
and must be highly protected against channel errors, since the decoder needs the probability model for reconstruction
(see Section \ref{sec:Decoder}). In the next section we shall discuss the compression-only performance of the scheme 
based on the above defined probability model. We discuss the optimization of the model parameter representation length and, 
as a sanity check, we compare the compression-only performance obtained by our Markovian probability modeler followed by 
arithmetic coding with that obtained by the standard JPEG2000. 


\subsection{Compression-only performance} \label{subsec:Comp}

If the probability model and estimator illustrated before is used
for compression only, the resulting output length (in bits) is
given by $B_{\rm tot} = B + B_{\rm model}$, where $B = \sum_{p=0}^P B_p$ with
$B_p$ given in (\ref{outputlength}) is the number of bits
necessary to represent the bit-planes using the estimated
probability model (this can be essentially achieved by using
arithmetic coding based on the estimated probability $\PP^{(K)}_{\widehat{\theta}}(\uv)$), 
and where $B_{\rm model}$ is the {\em model redundancy}, i.e., the number of bits necessary to represent the estimated probability 
model parameter $\widehat{\theta}$.

Thanks to the energy packing property of the DWT transform, the higher 
bit-planes are not identically zero only for ``low frequency'' subbands. 
We do not encode the identically zero subbands by adding a bit flag in
the model parameter to notify the receiver about all-zero subbands.


In order to minimize the total output length, $B_{\rm model}$ must be optimized.
In particular, we have to choose the number of regions $m_{p,d}$ to partition each segment, and the number of 
bits $q_{p,d}$ for the description of each LLR parameter.
To this regard, we have investigated a few possibilities. 
One option consists of defining regions as groups of $N_s$ adjacent stripes, where $N_s$ the same for all segments. 
Another option consists of having different grouping values $n_{p,d}$ in each segment $(p,d)$.
In this case, the values $n_{p,d}$ for each segment must be added to the model description.
As for the quantization bits for parameter representation, one option consists of using 
Rissanen's description length bound \cite{Rissanen1984Universal}. The achievability part of Theorem 1 in \cite{Rissanen1984Universal}
suggests to represent each parameter $\nu(p,d,\ell,\kappa)$ by $\left \lceil \frac{1}{2}  \log_2 K_{p,d} \right \rceil$ bits, 
where $K_{p,d}$ denotes the length of the groups in the $(p,d)$ segment. 

Since Rissanen's bound holds on average and it is an asymptotic result, 
it might not yield the best choice for a given source realization of
finite length. Hence, for the sake of comparison, we have also considered a brute-force 
bit-allocation algorithm to find the global optimum over all values of $n_{p,d}$ and $q_{p,d}$.
The brute-force search is initialized by letting $q_{p,d} = 0$ and $n_{p,d} = 1$ 
for all $p,q$, which yields equiprobable bits ($\nu(p,d,\ell,\kappa) = 0$ for all $p,d,\ell,\kappa$).
Therefore, the initial value of the total output length is $B_{\rm tot} = K (P + 1)$, i.e., 
the length of the original redundant sequence $\uv$. Then, we search over all $n_{p,d} = 1,2,3,\ldots$ and $q_{p,d} = 1,2,\ldots$ for 
the global minimum of the total description length of each $(p,d)$ segment.
The search over $q_{p,d}$ is stopped when increasing the model parameter quantization bits by one does not correspond 
to a decrease of the total description length. 
Even though this search might appear computationally heavy, it should be noticed that in practice only a few values of $n_{p,d}$ and
$q_{p,d}$ need to be considered. 
Also, since each segment $(p,d)$ is independently encoded, the global minimum of the output length is found by independently minimizing
the description length of each segment $(p,d)$. Hence, the brute-force bit-allocation is actually feasible in practice.
In this case, the pair of parameters $(n_{p,d}, q_{p,d})$  for each segment $(p,d)$ must be included in the model
description is to be added to the total output length. 

\begin{figure}[htbp]
\centerline{\includegraphics[scale= 0.6]{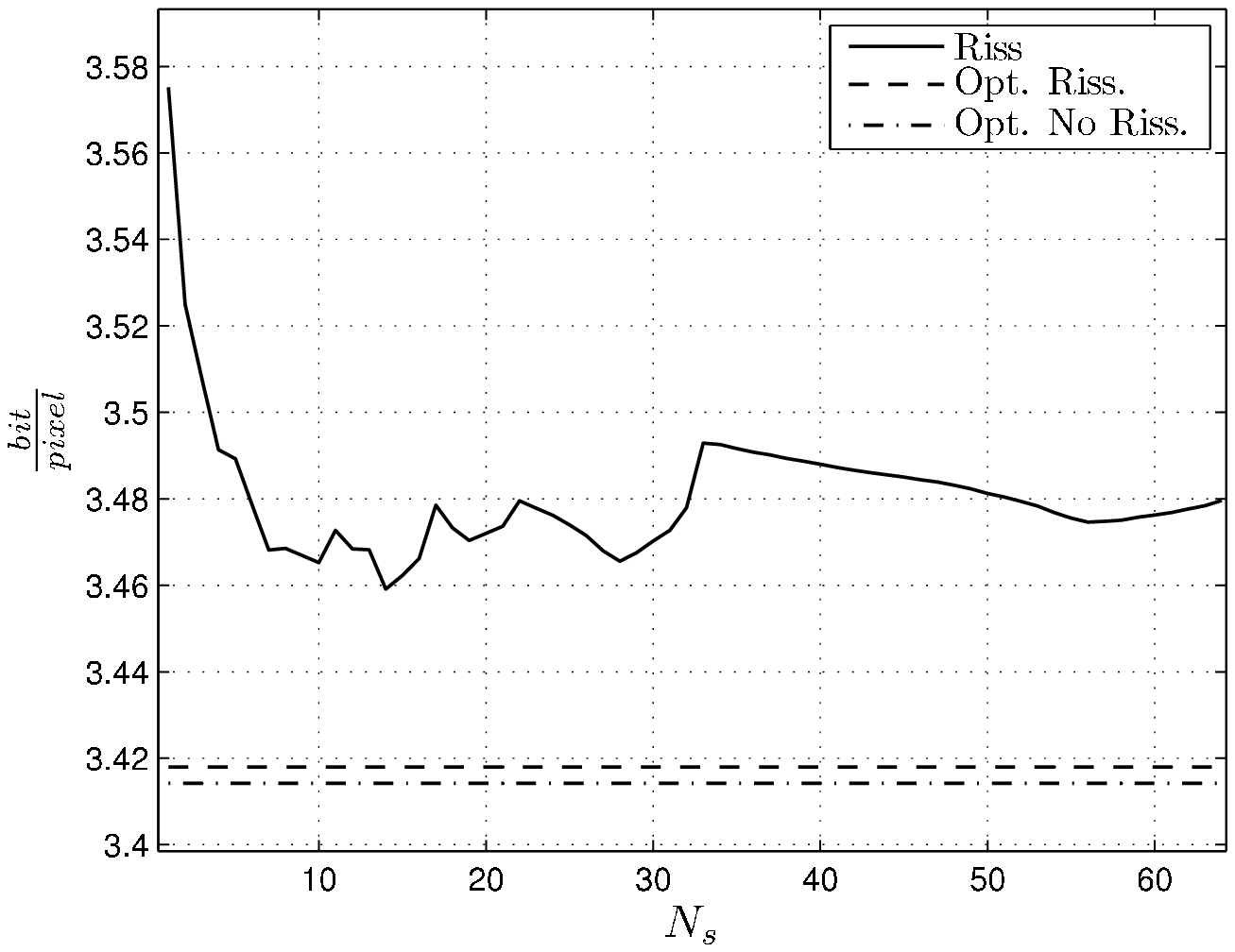}}
\caption{Total output length for various options of the model description for the Goldhill test image.}\label{fig:Goldhill_Stripes}
 \end{figure}

\begin{figure}[htbp]
\centerline{\includegraphics[scale= 0.6]{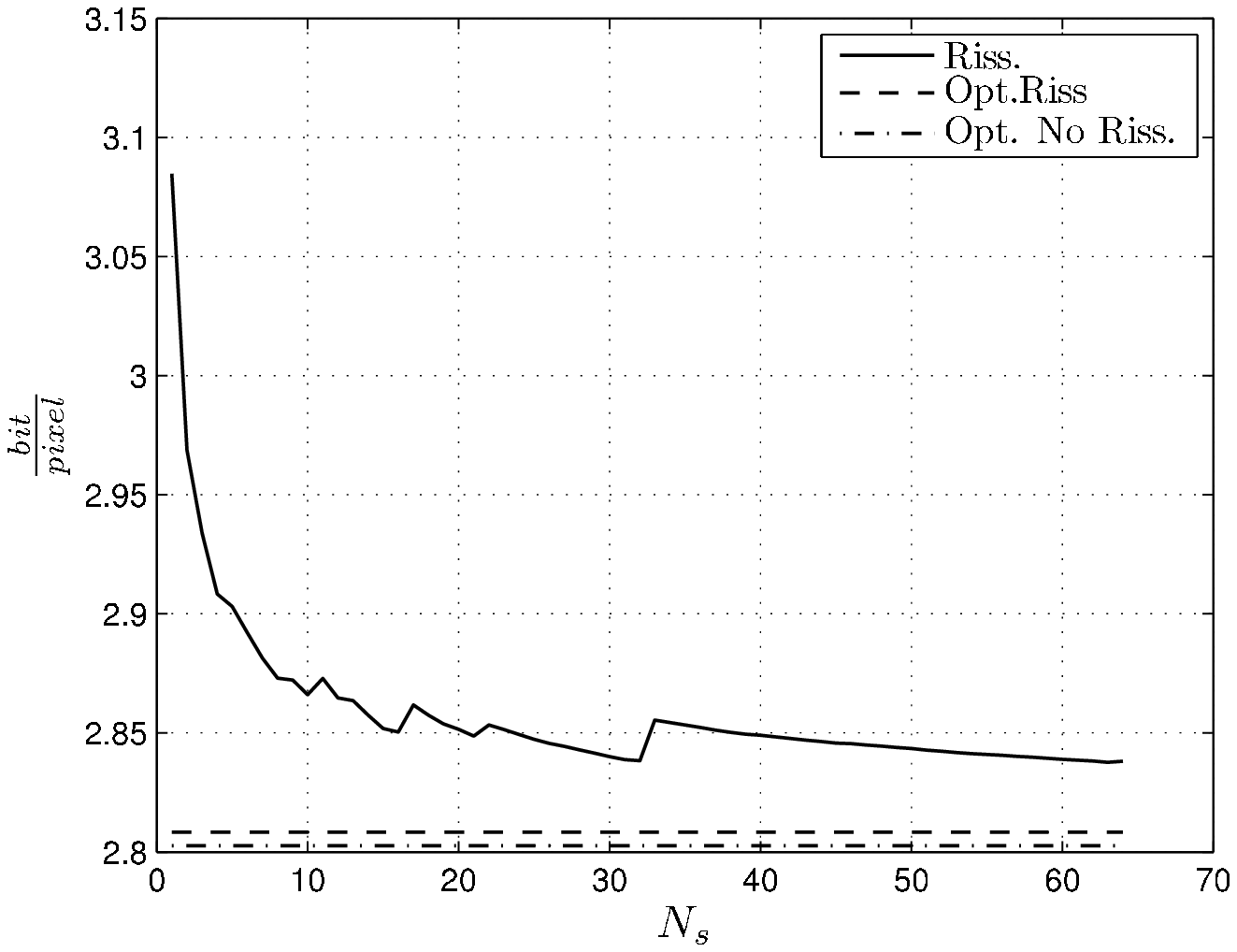}}
\caption{Total output length for various options of the model description for the Lena test image.}\label{fig:Lena_Stripes}
 \end{figure}

We run some tests and comparisons based on
the monochrome ``Goldhill'' $512\times512$ and the monochrome ``Lena''
$512\times512$ test images, after $D = 2$ stages of Daubechies DWT \cite{AnBa,TaMa}. 
Quantization is on 512 levels (corresponding to $P = 8$). 
In Fig. \ref{fig:Goldhill_Stripes} and \ref{fig:Lena_Stripes} we show the values of the 
total output length as a function of $N_s$ when the parameter quantization bits are set according to
Rissanen's bound. The curves have several local minima and maxima because of integer effects, since the segment lengths are generally not 
multiples of $N_s$ and for some values of $N_s$ we have spurious groups of stripes that cause the fluctuations. 
It is interesting to notice that the total output length is rather smooth with respect to $N_s$ and stays close 
to its global minimum for a wide range of values. Hence, optimization with respect to $N_s$ is not very critical in practice, 
provided that $N_s$ is chosen reasonably. This suggests that only a few values of $N_s$ should be tried in a practical implementation. 

The values obtained using optimized values $n_{p,d}$ different for each segment $(p,d)$ and Rissanen's model bit allocation
are reported as an a horizontal line, denoted by ``\textsl{Opt. Riss.}''.  Finally, the result of the brute-force bit-allocation is reported as a horizontal line, denoted by ``\textsl{Opt. No. Riss.}''. In both Fig. \ref{fig:Goldhill_Stripes} and \ref{fig:Lena_Stripes} it appears that optimizing 
with respect to stripe grouping values $n_{p,d}$ in each segment yields a significant advantage. 
On the contrary, the parameter quantization bit allocation given by Rissanen's bound is always very close to optimum. 
Since this allows a much faster optimization of the model description, this method is to be preferred and it is used for the rest of the results
presented in this paper.

Table \ref{tab:LengthComp} reports (first column) the total output length for
``Goldhill'' and ``Lena'' achieved by the proposed probability model using arithmetic coding to compress the data and the above
optimized model description. For comparison, the JPEG2000 output length (second column) is reported for the same quantization 
distortion $D_\Qc$. We notice that there is a remarkable agreement of the overall output length. 
This shows that the proposed Markov model is consistent with the probability model implicitly assumed by JPEG2000, and that the 
sequential adaptive entropy coder of JPEG2000 produces a redundancy very similar to the model redundancy obtained by our optimized model 
description. The slight advantage of our method is believed to be due to the fact that JPEG2000 resets the KT estimators 
on each code-block and inserts resynchronization symbols to limit error propagation. 

\begin{table}[tbp]
\begin{center}
\caption{Comparison between the JPEG2000 output length (in bit per pixel) and
the output length of an ideal entropy encoder based on the proposed probability modeler,
for $D=2$ stages of DWT, $8$ bit-planes and optimized number of quantization bits 
for the model parameters.}
\label{tab:LengthComp}
\vspace{0.7 cm}
\begin{tabular}{|c|c|c|}\hline
Image & Proposed Algorithm & JPEG2000 Encoder \\
\hline
Goldhill&3.45& 3.5\\
\hline
Lena &2.81& 2.84\\
 \hline
\end{tabular}
\end{center}
\end{table}

\section{Encoding and decoding} \label{sec:Decoder}

In this section we illustrate the linear joint source-channel encoding and obtain the Factor Graph (FG)
of the joint probability distribution of the source sequence $\uv$ and the channel output $\yv$.
The FG yields directly a low-complexity iterative joint source-channel decoder based on BP, 
by a completely standard application of the Sum-Product computation rules \cite{Ksch}.
Since the derivation of the BP algorithm is nothing more than an exercise, 
it will be omitted for the sake of conciseness.

\subsection{Linear encoding by punctured Turbo Codes} \label{sec:turbo}

In order to seek a good tradeoff between performance and complexity,
at possibly large but finite block length $K$, we resort to the use of well-known families
of linear binary codes for which efficient BP decoding is easily implementable.
In this paper we focus on the use of TCs \cite{BerGlav, ZhuAlaMitr, HAgWa06}.\footnote{LDPC codes \cite{RiU01, Gallager62, Gallager63, mncEL2}
and IRA codes \cite{Jin, Roumy}, as well as many variations thereof
proposed in the recent literature could be used here, after obvious modifications.}

As said in Section \ref{sec:proposed-approach}, we successively 
encode the bit-planes as $\xv^{(p)} = \uv^{(p)} \Gm_p$, for all
$p = 0,\ldots,P$. We focus on the encoder of a generic $p$-th
bit-plane and drop the index $p$ for notation simplicity. We
consider a TC family with two {\em identical} component binary Recursive Convolutional Codes (RCC) of rate 1. 
The RCC encoder is defined by the input output relation\footnote{$D$-transform notation: a sequence
$\ldots,x_-1,x_0,x_1,x_2,\ldots$ is represented by the Laurent series $x(D) = \sum_{\ell} x_\ell D^\ell$ in the indeterminate
$D$.} $x(D) = \frac{b(D)}{a(D)} u(D)$. As usual, the code generators $(a(D),b(D))$ are expressed by their 
coefficients in octal notation. For example, the RCC with generators $a(D) = 1 + D + D^2 + D^3 + D^4$ and $b(D) = 1 + D^4$ 
is indicated by $(37,21)_8$. We use a tail-biting encoder \cite{Ma86, Anderson98,Rosnes}. 
Hence, the the mapping $(u_1,\ldots,u_K) \mapsto (x_1,\ldots,x_K)$ is given by
\begin{equation} \label{tailbiting}
\xv = \uv \Am^{-1} \Bm
\end{equation}
where $\Am$ is the $K \times K$ circulant matrix  with first row
\[ (a_0,a_1,\ldots,a_\mu, \underbrace{0, 0 , \ldots, 0}_{K - \mu - 1}) \]
and $\Bm$ is the circulant matrix with first row
\[ (b_0,b_1,\ldots,b_\mu, \underbrace{0, 0 , \ldots, 0}_{K - \mu - 1}) \]
and $\mu$ denotes the RCC encoder memory.

The turbo encoder with puncturing is obtained as follows.  
Let $\Pim_1,\Pim_2$ denote two $K \times K$ permutation matrices (interleavers), 
and $\Rm_0, \Rm_1$ and $\Rm_2$ denote three {\em puncturing} matrices, 
of dimension $K \times n_0$, $K \times n_1$ and $K \times n_2$, respectively.
Notice that a $K \times n$ puncturing matrix is
a submatrix of a $K \times K$ permutation matrix
obtained by selecting $n$ out of $K$ columns. 
Then, a generator matrix for the TC with given RCC component, interleaver and
puncturing is given by
\begin{equation} \label{turbo-matrix}
\Gm = \left [ \Rm_0 | \Pim_1 \Am^{-1} \Bm \Rm_1 | \Pim_2 \Am^{-1} \Bm \Rm_2 \right ]
\end{equation}
The turbo-encoded output sequence is given by three blocks. The
first block (``systematic''), $\uv \Rm_0$ of length $n_0$
corresponds to a punctured version of the encoder input. The second and
the third blocks (``parity'') $\uv \Pim_1 \Am^{-1} \Bm \Rm_1$ and
$\uv \Pim_2 \Am^{-1} \Bm \Rm_2$ of length $n_1$ and $n_2$,
respectively, are obtained by permuting the input sequence, passing
it through the (tail-biting) RCC encoder, and puncturing its output.
The coding rate is given by $K /(n_0 + n_1 + n_2)$ \footnote{This encoder structure
cannot implement coding rates smaller than $1/3$. If coding rates
smaller than 1/3 are needed, we have to add RCC stages in
parallel. However, this was not needed in our simulations.}.

Next, we discuss the optimization of the TC encoder parameters, namely, 
the RCC generator polynomials, interleavers and puncturing matrices.
Intuitively, $\Gm$ in (\ref{turbo-matrix}) should {\em mimic} as closely as possible the generator matrix 
of a {\em random} linear code. In particular, a necessary (non-sufficient) condition for approaching the Shannon limit is that
the encoder output is {\em marginally} uniformly distributed \cite{verdu-shamai-output-statistics,ZhuAlaMitr}. 
In fact, $\Gm$ should map the statistically dependent and maginally non-uniform binary symbols of the input
$\uv$ into channel symbols $\xv$ with the required capacity-achieving uniform distribution. 
This problem is discussed in \cite{ZhuAlaMitr} for the case of non tail-biting TCs and in the limit of infinite block 
length (a ``convolutional coding'' framework). In particular, it is shown that the encoder output $x_k$
marginally uniform when $k$ is large, for a non-uniform i.i.d. encoder input $u(D)$ if and only if  $b(D)$ is not a multiple of $a(D)$.
Also, it is shown that under mild conditions on the generator polynomial $a(D)$ the state at time $k$ of the RCC encoders
is uniformly distributed over the encoder state space irrespectively of the input 
probability, for an i.i.d. input sequence and large $k$.

Our problem differs from \cite{ZhuAlaMitr} in the fact that we consider a block coding 
framework and tail-biting codes. Next, we shall show that by choosing a primitive polynomial $a(D)$, 
in the limit of large block length and large polynomial degree we obtain
both a marginally uniformly distributed encoder output and an encoder state uniformly disributed over the 
state space for all positions $k = 1,\ldots, K$ in the block. We have:

{\bf Lemma A.} The mapping $f(D) \mapsto \Fm$, that maps polynomials 
$f(D) = f_0 + f_1 D + \ldots, f_{K-1} D^{K-1}$ in the ring $[\FF_2[D]]_{(1 + D^K)}$ of polynomials residues 
modulo $1 + D^K$ into the $K \times K$ circulant matrix whose first row is $[f_0, f_1, \ldots, f_{K-1}]$ 
is a ring isomorphism. 

{\bf Proof.} Clearly, the zero polynomial is mapped into the zero matrix, the polynomial 1 is mapped into the identity matrix, 
the mapping is linear and preserves the product, i.e., $f(D) h(D)$ modulo $1+D^K$ (a cyclic convolution of the polynomial coefficients) 
is mapped into the product $\Fm \Hm$ of the corresponding matrices. \hfill $\square$

As a consequence, we have:

{\bf Lemma B.} $\Am$ defined above is non-singular if and only if $a(D)$ is not a divisor of zero in the ring $[\FF_2[D]]_{(1 + D^K)}$. 
In particular, $\Am$ is invertible if and only if $a(D)$  and $1 + D^K$ are relatively primes.
\hfill $\square$

Since $\Am^{-1}$ is a circulant matrix and, for what said before, we wish that its rows look as random as possible, 
we shall choose $a(D)$ to be a primitive polynomial of degree $\mu$. The existence of $\Am^{-1}$ is guaranteed by the following

{\bf Lemma C.} For $a(D)$ primitive, $\Am$ is invertible if and only if $2^\mu-1$ does not divide $K$.

{\bf Proof.} For Lemma B, we need that $a(D)$ and $1 + D^K$ are relatively prime. Since $a(D)$ is irreducible over $\FF_2$, 
it follows that the condition holds if and only if $a(D)$ does not divide $1 + D^K$.
The roots of $a(D)$ are primitive elements of the field $\FF_{2^\mu}$, the extension of degree $\mu$ over $\FF_2$. 
All the non-zero elements of $\FF_{2^\mu}$ are roots of the polynomial $1 + D^{2^\mu-1}$. 
Hence, $a(D)$ is a factor of $1 + D^{2^\mu-1}$. Finally, $1 + D^{2^\mu-1}$ does not divide $1 + D^K$ if and only if
$2^\mu-1$ does not divide $K$. \hfill $\square$

The condition that $K$ is not a multiple of $2^\mu-1$ is easily satisfied in practice, since typically $K$ is a power of two.
By choosing $a(D)$ primitive we have that the feedback shift register with coefficients given by $a(D)$ in the RCC encoder 
generates an $m$-sequence of period $2^\mu-1$, with Hamming weight $2^{\mu-1}$. This has the following nice consequence:

{\bf Lemma D.} 
If $2^\mu-1$ does not divide $K$, the circulant matrix $\Am^{-1}$ has first row $\tauv$ formed by the concatenation of
$\left \lfloor \frac{K}{2^\mu-1} \right \rfloor$ periods of the $m$-sequence generated by $a(D)$, 
plus $K$ modulo $2^\mu-1$ extra symbols. In particular, for $K \gg 2^\mu-1 \gg 1$ 
we have that the {\em normalized} Hamming weight of $\tauv$ is close to $1/2$. 

{\bf Proof.} Since $\Am^{-1}$ exists, the first row $\tauv$ is obtained by loading the feedback shif register by some non-zero 
configuration of the memory elements and feeding as input the sequence $u(D) = 1$. This clearly generates 
$\left \lfloor \frac{K}{2^\mu-1} \right \rfloor$ periods of the corresponding $m$-sequence, plus some tail symbols to arrive at the 
total length $K$. 

The Hamming weight of $\tauv$ is given by 
\[ \left \lfloor \frac{K}{2^\mu-1} \right \rfloor 2^{\mu-1} + \Delta \]
where $0 \leq \Delta \leq 2^\mu-2$ is the Hamming weight of the tail symbols. For large $\mu$ and $K \gg 2^{\mu}-1$ the above 
Hamming weight normalized by the block length $K$ is close to $\frac{2^{\mu-1}}{2^\mu-1} + \Delta/K \approx 1/2$. 
\hfill $\square$

Lemma D implies that we can construct a {\em structured} generator matrix $\Gm$ as in (\ref{turbo-matrix}) having a marginal 
empirical distribution of its entries close to $1/2$, i.e., close to that of the capacity-achieving linear coding ensemble. 
For example, consider $a(D) = 1 + D^3 + D^4$ (or $(23)_8$) and $K = 16$. The corresponding first row of $\Am^{-1}$ is equal to
\begin{equation} \label{tau-example} 
\tauv = [1, 0, 0, 0, 1, 0, 0, 1 ,    1 ,    0 ,    1 ,    0 ,    1 ,    1 ,    1 ,    1] 
\end{equation}
and has Hamming weight 9, so that $9/16 = 0.5625$. 
If we consider for example length $K = 64$, we would obtain $\tauv$ with Hamming weight 33, so that $33/64 = 0.5156$, 
that is already quite close to $1/2$. 

By choosing $b(D)$ of degree $\leq \mu$ and $a(D)$ primitive, it follows that $a(D)$ and $b(D)$ are relatively prime.
From \cite{ZhuAlaMitr} we have that the encoder output $x_k$ is asymptotically marginally uniformly distributed. However, since
we consider a tail-biting code, a cyclic shift of the RCC encoder output corresponds to a cyclic shift of the input and has the same
marginal distribution. It follows that, for large block length $K$, the output $x_k$ is marginally uniformly distributed for all
posititions $k = 1,\ldots,K$. 

Next, we turn our attention to the encoder state. The encoder state space coincides with $\FF_2^\mu$, which is an Abelian group 
(with respect to vector addition). In a tail-biting convolutional code, the {\em circulation state} is defined as the initial encoder state $\sv_0$ such that, for a given input $u(D)$, the final state is $\sv_K = \sv_0$, i.e., the path in the trellis corresponding to the initial state $\sv_0$ and input $u(D)$ closes onto itself (tail-biting condition). The circulation state $\sv$ is a {\em linear} function of the input $u(D)$. 
In particular, let $\Am'$ denote the $K \times \mu$ matrix obtained by taking the last $\mu$ columns of $\Am^{-1}$. 
Then, it is straightforward to see that $\sv = \uv \Am'$. For example, for 
the input $u(D) = 1$ the circulation state corresponding to $a(D) = 1 + D^3 + D^4$ and $K = 16$ is given by $\sv = (1,1,1,1)$, i.e., 
the last 4 positions of the sequence $\tauv$ in (\ref{tau-example}). 
Consider an input sequence $u(D)$ with i.i.d. symbols, such that $\PP(u_k = 1) = \rho$.
As argued before, a good joint source-channel code should have an encoder state distribution that is uniform 
over the state space irrespectively of the input bias probability $\rho$. This is given by the following:

{\bf Lemma E.} For $K \gg 2^\mu \gg 1$ and non-uniform i.i.d. encoder input $u(D)$ with 
$\PP(u_k =1) = \rho \in (0,1)$,\footnote{If $\rho = 0$ or $\rho = 1$ the entropy of $u(D)$ is equal to zero, and
we shall not encode constant all-zero or all-ones inputs. Hence, this restriction does not involve any loss of generality in our context.}  
the circulation state is almost uniformly distributed over the RCC encoder state space. 
Furthermore, the encoder state at any position in the trellis is also almost uniformly distributed.

{\bf Proof (sketch).}
First, we prove the following general simple result. \\
Fact: Let $\Ac$ denote a finite Abelian group of size $q$.
Consider $g = \sum_{i=1}^m a_i$ where the elements $a_i$ are independently selected according to the probability that puts
mass $1/(q-1)$ on all non-zero elements of $\Ac$ and probability 0 on the zero element (additive indentity of the group). 
Then, the distribution of $g$ converges to the uniform distribution on $\Ac$ for large $m$. 

Proof of the Fact: 
Consider $m = 2$, and without loss of generality denote the elements of $\Ac$ by the integers $0,1,\ldots,q-1$, where 0 denotes the  
additive identity and the $+$ rule is given by the addition rule of the group. Then, 
\begin{equation} \label{convolution} 
\PP(g = i) = \sum_{j=0}^{q-1} \PP(i - j) \PP(j), \;\; i = 0,\ldots,q-1 
\end{equation}
By defining the matrix 
\[ \Pm = \frac{1}{q-1} \left [ \onev \onev^T - \Id \right ] \]
(where $\onev$ denotes the all-ones vector of length $q$) 
and the probability vectors  $\pv_0 = (1,0,\ldots,0)^T$, $\pv_1 = (0,1/(q-1),\ldots,1/(q-1))^T$
and $\pv_2 = (\PP(g=0),\ldots,\PP(g=q-1))^T$ we find that $\pv_1 = \Pm \pv_0$ and that
(\ref{convolution}) can be written as
\[ \pv_2 = \Pm \pv_1 = \Pm^2 \pv_0 \]
Extending this to the case of general $m$, we find that the distribution of $g = \sum_{i=1}^m a_i$ is given by 
$\pv_m = \Pm^m \pv_0$. We can write
\begin{eqnarray} \label{powerP}
\Pm^m & = & \frac{1}{(q-1)^m} \sum_{i=0}^m {m \choose i} (-1)^{m-i} \left ( \onev\onev^T \right )^i \nonumber \\
& = & \frac{1}{(q-1)^m} \left [ (-1)^m \Id + \frac{1}{q} \sum_{i=1}^m {m \choose i} (-1)^{m-i} q^{i} \onev\onev^T + 
(-1)^m \frac{1}{q} \onev\onev^T - (-1)^m \frac{1}{q} \onev\onev^T \right ] \nonumber \\
& = & \frac{1}{(q-1)^m} \left [ \frac{(q-1)^m}{q} \onev\onev^T + (-1)^m \Id - (-1)^m \frac{1}{q} \onev\onev^T \right ]
\end{eqnarray}
From (\ref{powerP}) it is evident that $\Pm^m \rightarrow \frac{1}{q} \onev\onev^T$ for $m \rightarrow \infty$.
Then, we conclude that $\pv^m = \Pm^m \pv_0 \rightarrow (1/q) \onev$, as we wanted to show.

In order to show Lemma E, consider a typical realization of the input sequence $u(D)$. This has Hamming weight
$m \approx \rho K$. Furthermore, these can be located in any position of the input sequence with the same probability.
The matrix $\Am'$ defined before contains $\approx K/(2^\mu-1) \gg 1$ periodic repetitions of the $2^\mu - 1$ non-zero states
plus a segment not longer than $2^\mu - 2$ states. Each non-zero state appears essentially the same number of times in the rows of $\Am'$.
It follows that a ``one'' symbol in $u(D)$ uniformly distributed over
all positions $1,\ldots,K$ selects a non-zero state with $\approx$ uniform probability, 
equal to $1/(2^\mu-1)$. A typical input selects $m \approx \rho  K$ non-zero elements of the state space with 
uniform probability. Since the circulation state $\sv = \uv \Am'$ is given by the sum of these $m$ random non-zero elements, 
we apply the above Fact and conclude that, for large $K$, $\sv$ is uniformly distributed over the whole state space 
$\FF_2^\mu$ (including the zero state). 

Furthermore, the RCC encoder state when the encoder is driven by an i.i.d. sequence $u(D)$ defines a Markov chain. 
Since $a(D)$ is primitive, it is easy to see that the Markov chain is indecomposable and aperiodic. 
By construction, the transition matrix of the Markov chain corresponding to the RCC state 
has exactly two non-zero elements in each row and column, equal to $\rho$ and $1-\rho$.
It follows that the (unique) stationary distribution is given by the uniform distribution. 
Since the circulation state is (almost) uniformly distributed, any state at any given section of the trellis 
is also (almost) uniformly distributed, and it is exactly uniformly distributed as $K \rightarrow \infty$.
\hfill $\square$

Notice that the input sequence to the RCC encoder in the tail-biting turbo encoder defined before is
an interleaved version of the bit-plane bit sequence. Hence, even if the latter is not i.i.d. (e.g., in our case the bit-plane is a non-stationary Markov chain), after interleaving it will be close to i.i.d., with a probability $\widehat{\rho}$ equal to the fraction of ``ones'' in the bit-plane.
Notice also that, although we interleave the bit-plane sequence for the purpose of turbo encoding and BP decoding, 
the bit-plane entropy rate is computed taking into account the Markov memory structure, as said in Section \ref{sec:Encoder}. 
This has nothing to do with the entropy rate of the bit-plane {\em after} interleaving, given by 
$H_2(\widehat{\rho})$ where $H_2(\cdot)$ denotes the binary entropy function. 
By the concavity of entropy the latter may be much larger than the actual bit-plane entropy rate.


For the choice of $b(D)$ we follow the theory developed in \cite{ZhuAlaMitr}. In particular, letting
$b(D) = 1$ yields that the RCC encoder $\mu$-th order empirical distribution, as defined in \cite{verdu-shamai-output-statistics},
is uniformly distributed over $\FF_2^\mu$. This follows immediately from the fact that in this case the encoder output 
$x(D)$ coincides with the state bits and for Lemma E we have that the empirical distribution of any block $(x_{k-\mu},\ldots,x_{k-1})$ 
state bits is uniformly distributed over $\FF_2^\mu$. However, $b(D) = 1$ does not necessarily yield the best coding performance.
Hence, we have searched semi-exhaustively for generator polynomials $b(D)$ with $b_0 = b_\mu = 1$. 

For given RCC generators, the permutations $\Pim_1$ and $\Pim_2$
are chosen at random, by trial and error. For $K$ not too small, the effect of optimizing the permutations on the end-to-end
distortion of the proposed scheme is minimal. In fact, one
significant advantage of the proposed JSCC scheme is that its
performance is not dominated by the ``error floor'' region of the
BER \cite{BerGlav}: as soon as the BER drops to small values 
(waterfall region) the reconstruction quality of our JSCC scheme rapidly
improves and attains satisfactory distortion levels even though the
residual BER is not as small as it would be required by (lossless) data applications. 
In this respect, the proposed JSCC scheme puts much less stress on the code design 
than a conventional SSCC scheme!

More care must be deserved to the optimization of the puncturing
matrices. When the bits in a given bit-plane are not i.i.d. uniformly distributed, sending the systematic
part over the BIOS channel results in a suboptimal code. 
Hence, unless $\widehat{H}(U_p|U_{p+1},\ldots,U_P)$ is very close to 
1 bit/symbol, we let $n_0 = 0$. For $\Rm_1$ and $\Rm_2$, we have constructed (off-line) a library of 
puncturing matrices by following the incremental redundancy approach advocated in 
\cite{Hagenauer04,Hagenauer05,CaShVe_dimacs2,CaShVe_FnT}.
We stress the fact that the codes are constructed off-line, and do not depend on the actual source 
sequence to be encoded. By using a greedy pseudorandom progressive growth search (i.e., adding columns to the puncturing matrices), 
we have designed a library of nested puncturing matrices in order to cover all 
the rates $C/H - \delta$, for all quantized values $H$ in pre-determined fine grid in $[0,1]$. 
The library was designed assuming a Bernoulli source with entropy $H$ 
for all the quantized values of $H$. 
Further discussion on the design of the puncturing matrices is provided in \cite{maria-thesis} (see also \cite{HAgWa06}).

\begin{figure}[hbtp]
\centerline{\includegraphics[scale= 0.6]{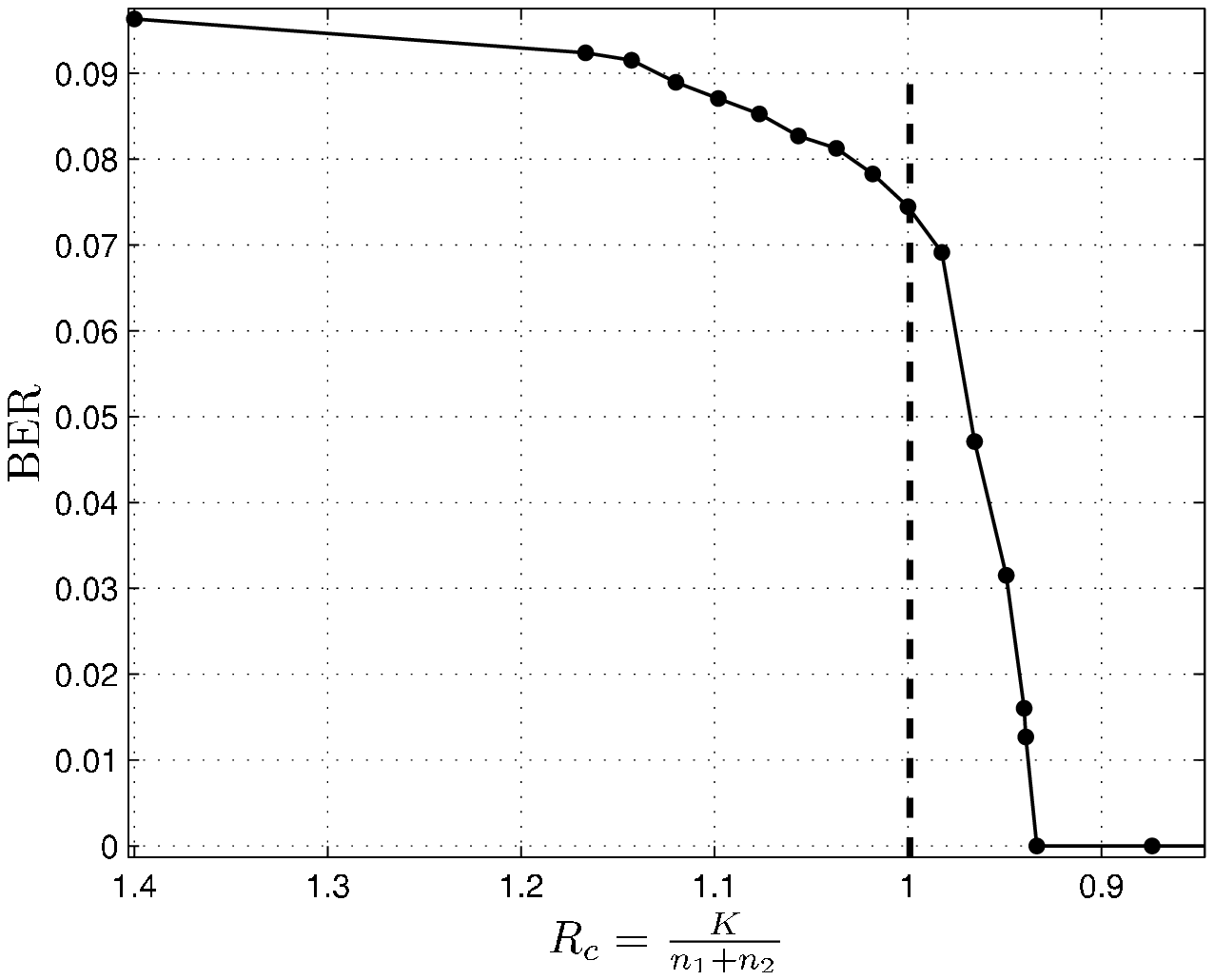}}
\caption{Threshold effect for the case of a Bernoulli
source $H(U)=0.5$, transmitted over a BSC $C=0.5$, for a family of progressively punctured TCs.}
\label{fig:transition}
\end{figure}

Fig.~\ref{fig:transition} shows the threshold effect for the case
of a binary Bernoulli source of length $K = 512 \times 512$, with entropy 
rate $H(U) = 0.5$, and a Binary Symmetric Channel (BSC) with capacity $C = 0.5$.
Notice that each point of Fig.~\ref{fig:transition} corresponds to a different puncturing matrix, generated by 
incremental redundancy as explained above.
The BER is averaged over a random choice of the interleavers. 
The transition of the family of progressively punctured TCs generated in this way is
remarkably close to the Shannon limit $\eta = \frac{C}{H(U)} = 1$. Again, we stress the fact that our scheme does 
not need very small BER in order to provide good reproduction results. BERs of the order to $10^{-2}$ are sufficient, 
as we will see in the examples of Section \ref{sec:Results}.

\subsection{Belief Propagation decoding} \label{sec:bp}

\begin{figure}[thpb]
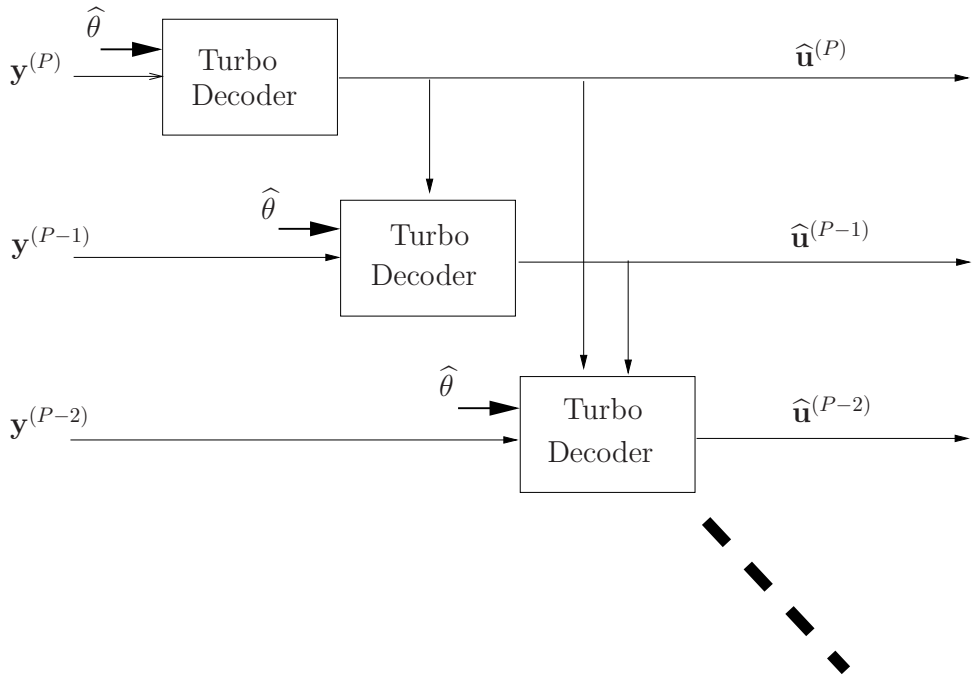

\centerline{\resizebox{0.8\textwidth}{!}{\input multistage-decoder.pstex_t}}
\caption{Structure of the successive bit-plane joint source-channel decoder.}
\label{multistage}
\end{figure}

The decoder is based on the successive decoding structure of
Fig.~\ref{multistage}, where the bit-planes are decoded in
sequence, and the decoder at level $p$ makes use of the hard
decisions made at the upper levels. These hard decisions determine the value of the conditioning bits needed for the conditional Markov model
at each level. An iterative version of the multistage decoder where soft information is exchanged between the
bit-plane decoders was also considered and simulated, but we found that it does
not provide any substantial improvements and therefore the additional (significant) 
complexity is not justified. 

It is assumed that the decoder knows the probability model $\PP^{(K)}_{\widehat{\theta}}(\uv)$ defined in Section
\ref{sec:Encoder} with the LLRs given in (\ref{a-priori}) as model parameters. As already said, the model parameters have to be sent separately, 
and need to be decoded with very high reliability. Fortunately, the model redundancy is very small in comparison with the data length and therefore
we can afford to use a low-rate code to protect the model parameters against channel errors. 
This is just a standard channel coding problem that we do not address here. However, we would like to stress that the simulation results
and the comparison with conventional SSCC is made by {\em including} the transmission of the model parameters, 
both in terms of redundancy and in terms of errors, i.e., the comparison is {\em fair} and was made without any optimistic assumption on perfect reception of the side information $\widehat{\theta}$.

Let's focus on the $p$-th component decoder and
let $\yv^{(p)}$ denote the BIOS channel output corresponding to the transmission 
of the coded $p$-th bit-plane
$\xv^{(p)} = \uv^{(p)}\Gm_p$. The goal of the $p$-th decoder is to compute the symbol-by-symbol posterior marginal probability
\begin{equation} \label{posterior-marg}
\PP_{\widehat{\theta}}(u_{p,k} | \yv^{(p)}, \widehat{\uv}^{(p+1)},\ldots,\widehat{\uv}^{(P)})
\end{equation}
where $\widehat{\uv}^{(p+1)},\ldots,\widehat{\uv}^{(P)}$ denote
the hard decisions on the upper bit-planes made at previous
decoding stages. 
This can be accomplished (approximately) by the BP algorithm applied to the FG of the underlying
joint probability distribution. In the case of Markovian sources and TCs, the 
BP takes on the particularly appealing form of three BCJR forward-backward
algorithms \cite{bah74} exchanging soft information in the form of ``extrinsic'' log-likelihood ratios (LLRs) \cite{Ksch}.
This follows by standard application of the Sum-Product computation rules
\cite{Ksch} to the resulting FG. Hence, this section is devoted to illuminating the FG structure.

Recall the ``marginalization'' notation of \cite{Ksch} for which
$\sum_{\sim v}$ denotes the sum over all variables in the
summation argument while keeping $v$ fixed. Furthermore, a
probability distribution $\PP(v)$ needs to be determined only up
to a proportionality constant (denoted by $\propto$), that can be
obtained by imposing the normalization $\sum_v \PP(v) = 1$. Using
Bayes rule and neglecting irrelevant proportionality terms, we obtained the factorization
\begin{eqnarray} \label{posterior-marg1}
\PP_{\widehat{\theta}}(u_{p,k} | \yv^{(p)}, \widehat{\uv}^{(p+1)},\ldots,\widehat{\uv}^{(P)}) & = & \sum_{\sim u_{p,k}}
\PP^{(K)}_{\widehat{\theta}}(\uv^{(p)} | \yv^{(p)}, \widehat{\uv}^{(p+1)},\ldots,\widehat{\uv}^{(P)}) \nonumber \\
& \propto &
\sum_{\sim u_{p,k}} \PP(\yv^{(p)}, \uv^{(p)} | \widehat{\uv}^{(p+1)},\ldots,\widehat{\uv}^{(P)},\widehat{\theta}) \nonumber \\
& = &
\sum_{\sim u_{p,k}} \sum_{\xv^{(p)}}
\PP(\yv^{(p)}, \xv^{(p)}, \uv^{(p)} | \widehat{\uv}^{(p+1)},\ldots,\widehat{\uv}^{(P)},\widehat{\theta}) \nonumber \\
& = &
\sum_{\sim u_{p,k}} \sum_{\xv^{(p)}}
\PP(\yv^{(p)} |\xv^{(p)}) \PP(\xv^{(p)}|\uv^{(p)}) \PP^{(K)}_{\widehat{\theta}}(\uv^{(p)}| \widehat{\uv}^{(p+1)},\ldots,\widehat{\uv}^{(P)}) \nonumber \\
& = & \sum_{\sim u_{p,k}} \sum_{\xv^{(p)}}
\left [ \prod_{j=1}^{N_p} \PP(y_{p,j} |x_{p,j} ) \right ]
1\{ \xv^{(p)} = \uv^{(p)} \Gm_p \} \cdot \nonumber \\
& & \cdot \PP^{(K)}_{\widehat{\theta}}(\uv^{(p)}|\widehat{\uv}^{(p+1)},\ldots,\widehat{\uv}^{(P)})
\end{eqnarray}
where we have used the fact that the BIOS channel is memoryless and where
$\PP(\xv^{(p)}|\uv^{(p)}) = 1\{\xv^{(p)} = \uv^{(p)}\Gm_p\}$ (an indicator function) 
since $\xv^{(p)}$ is a deterministic function of $\uv^{(p)}$.

Eventually, the desired posterior symbol-by-symbol probabilities are obtained by
marginalizing the joint distribution in the last line of (\ref{posterior-marg1}). 
From (\ref{cond-Pk}) and (\ref{param1}), the source probability term factors as
\begin{equation} \label{fact-p}
\PP^{(K)}_{\widehat{\theta}}(\uv^{(p)}|\widehat{\uv}^{(p+1)},\ldots,\widehat{\uv}^{(P)}) = \PP(\pi_{p,1})
\prod_{k=1}^K \PP_{\widehat{\theta}}(\pi_{p,k+1}|\pi_{p,k},\{\widehat{u}_{p',k'}: (p',k')\in \Sc_{p,k}\})
\end{equation}
where the a priori probability of $\pi_{p,1}$ is given by
\[ \PP(\pi_{p,1}) = \left \{ \begin{array}{ll}
1 & \pi_{p,1} = 0\\
0 & \mbox{otherwise}
\end{array} \right . \]
and where the state transition probability is given by\footnote{The update of the shift register in the 
block diagram of Fig. \ref{markov-model} is given by 
\[ \pi_{p,k+1} = (u_{p,k},\overrightarrow{\pi}_{p,k}) \]
where $\overrightarrow{\pi}_{p,k}$ denotes a right shift by one position of the state register.}
\[ \PP_{\widehat{\theta}}(\pi_{p,k+1}|\pi_{p,k},\{\widehat{u}_{p',k'}: (p',k')\in \Sc_{p,k}\}) = \left \{ \begin{array}{ll}
\PP_{\widehat{\theta}}(u_{p,k} = u |\kappa) & \begin{array}{ll}
\mbox{if} \; \pi_{p,k+1} = (u,\overrightarrow{\pi}_{p,k}) \;\; \mbox{and} \\
\Kc_p(\pi_{p,k},\{\widehat{u}_{p',k'}: (p',k')\in \Sc_{p,k}\}) = \kappa \end{array} \\
0 & \mbox{otherwise}  \end{array} \right . \]
The corresponding FG takes on the form of a trellis \cite{Ksch}, 
reflecting the state transition diagram of the Markov probability model.

The other non-trivial component of the overall FG is given by the factorization of the 
code indicator function $1\{\xv^{(p)} = \uv^{(p)}\Gm_p\}$. It is well-known that for TCs this takes the form of
two trellises, one for each RCC component, interconnected by interleavers.

\begin{figure}[thpb]
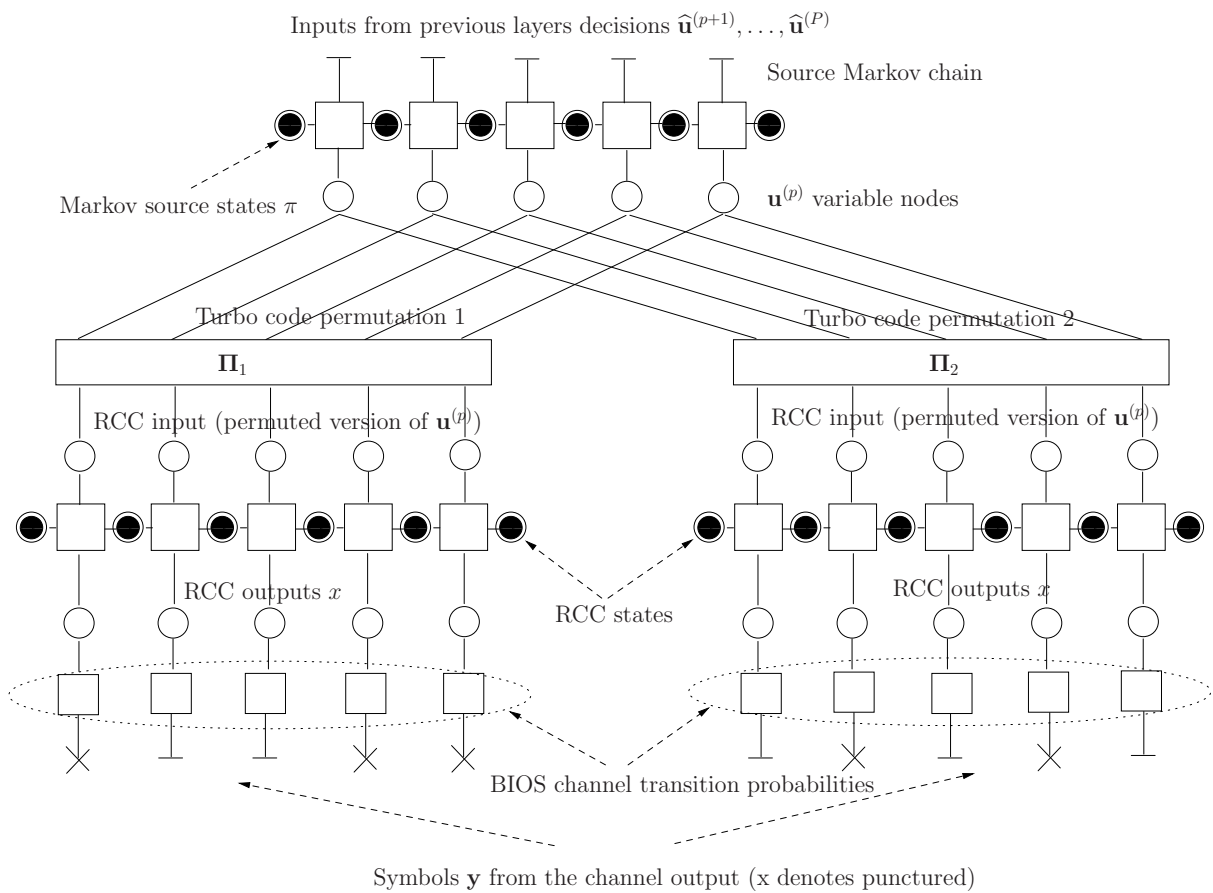

\centerline{\resizebox{1.0\textwidth}{!}{\input fg.pstex_t}}
\caption{Factor graph underlying the $p$-th component BP decoder.}
\label{FG}
\end{figure}

The factor graph (FG) corresponding to the factorization in the last line of (\ref{posterior-marg1}) is
given in Fig.~\ref{FG}. We use Wiberg's notation (see \cite{Ksch}), for which the FG
is a bipartite graph with variable nodes (circles) and function
nodes (boxes). State nodes are denoted by filled circles. 
A variable node is connected to a function node if the
corresponding variable is an argument of the
corresponding factor \cite{Ksch}.
In our case, the variable nodes correspond
to the bit-plane bits $u_{p,k}$, to the Markov source states $\pi_{p,k}$ and to the RCC encoder states.
The function nodes correspond to the state transition probabilities and to the BIOS channel transition probabilities.
The channel output symbols $y_{p,j}$ and the conditioning bits from the uper bit-planes appears as
``dongles''. The channel outputs corresponding to punctured coded symbols are represented as crossed dongles.
These symbols are treated as {\em erasures} by the decoder.

Standard application of Sum-Product computation rules to this FG yields the
iterative BP decoder, where one BCJR algorithm for each trellis
subgraph in the FG is used. The three BCJR component decoders
exchange messages in the form of LLRs \cite{Ksch} via the degree-3 
variable nodes $u_{p,1},\ldots,u_{p,K}$.
In our simulations, we used a BP scheduling where the BCJRs are activated
in round-robin (order is basically irrelevant). It should be noticed that symbol-by-symbol hard decisions are made 
for each stage $p$ and are used in subsequent stages as a priori information. However, in the soft-reconstruction scheme, 
the symbol-by-symbol a posteriori probabilities for each bit-plane are used in order to compute the reconstructed source 
via (\ref{soft-estimate1}).

\section{Results}\label{sec:Results}

In this section we present some results in terms of
Peak Signal to Noise Ratio (PSNR)\footnote{$PSNR\triangleq 10 \mbox{log}_{10}
\frac{A^2}{{\rm MSE}}$ is a measure of quality of image reconstruction
where MSE$= \frac{1}{K} \EE[|\sv - \widetilde{\sv}|^2]$ is the Mean Squared distortion of the 
reconstructed source and $A$ is the peak-to-peak signal amplitude. 
In our case, $A = 255$, since we are working with 8-bit images. 
We fix the source realization to be one of the test images 
considered and we take expectation $\EE[\cdot]$ with respect to the channel noise.}
versus efficiency $\eta$, measured as the number of source samples (pixels) per channel use.
We considered two 8-bit gray scale test images of size $512 \times 512$ pixels, referred to as
``Goldhill'' and ``Lena''. We run simulations over a Binary Symmetric Channel (BSC) 
and we designed our schemes for a BSC cross-over probability $\rho = 0.05$ 
(corresponding to $C = 1 - H_2(\rho) = 0.7136$ bits per channel use).

For comparison, we have considered a SSCC scheme obtained by
concatenating the output of the JPEG2000 encoder with a standard
punctured TC with rate optimized in order to work as close as
possible to the channel capacity. The simulation were performed by
using the OpenJPEG library, an open-source JPEG2000 codec written
in C language \cite{openjpeg}. We consider the case in which the
error resilience tools provided by JPEG2000 standard (i.e SOP marker,
SEGMARK marker and  ERTEM strategy \cite{TaMa}) are enabled. 
In order to compute the performances of the JPEG2000 compressed images, we
transmit and protect separately the header and the markers: they are received error free by the decoder.

As for the SSCC, several generator polynomials have been tested.
Here we report only the case of $(37,21)_8$ \cite{BerGlav}, 
that yields the best results. 
Since the output of JPEG2000 can be considered as i.i.d. uniform bits, the TC are conventional and 
we used standard regular puncturing patterns available in the literature. 

We would like to stress the fact that these results are presented
here for the purpose of a {\em proof of concept}: no particular
effort has been made to optimize the codes beyond the simple
incremental redundancy greedy selection of the puncturing patters
described before. We hasten to say that the result might be probably
improved by better code design. Furthermore, as mentioned in the literature review of Section \ref{sec:Intro}, 
a whole range of schemes between the basic SSCC scheme considered here and a fully JSCC are available, 
based on tuning the channel coding redundancy for each JPEG2000 encoded block 
\cite{Nosra, Chande, ChFa, Stan, Stankov, Hama05, maria_icc07}.
We did not consider comparisons with these schemes since they are {\em considerably} more complicated 
than the proposed JSCC, and they are actually {\em not yet used} in today's applications. 
Given the number and the variety of schemes proposed in the literature, a full comparison is well beyond 
the scope of this paper.

Tables \ref{tab:tab2} and \ref{tab:tab3} summarize the main
characteristics of the JSCC schemes for the two test images
considered, respectively. The number of encoded bits 
in each bit-plane is indicated by $K_p$. This may be less than $512^2 = 262144$ since
some segments, especially for the top bit-planes, might be identically zero. 
The third column shows the empirical entropy rate  
$\widehat{H}_p \eqdef \widehat{H}(U_p|U_{p+1},\ldots,U_P)$ as defined in (\ref{empirical-entropy}).
The fourth column shows the bound on coding efficiency for each bit-plane
given by $\bar{\eta}_p = C/\widehat{H}_p$. 
We used generators $(23,35)_8$ for all bit-planes.

\begin{table}[htbp]
\begin{center}
\caption{Goldhill} \label{tab:tab2}
\begin{tabular}{|c|c|c|c|}\hline

Bit-plane $p$ & length $K_p$ & empirical entropy $\widehat{H}_p$ & efficiency $\bar{\eta}_p$  \\
\hline
8 & $16384$ & $0.2966$ &2.4063 \\
\hline
7 & $29184$ & $0.3158$ &2.2595 \\
 \hline
6 & $71680$ & $0.2468$ &2.8906\\
 \hline
5 & $142336$ & $0.2341$ &3.0474\\
 \hline
4 & $223232$ & $0.3124$ &2.2837\\
  \hline
3 & $253440$ & $0.5431$ &1.3138 \\
  \hline
2 & $262144$ & $0.8172$ &0.8731\\
  \hline
1 & $262144$ &$0.9586$ &0.7444 \\
  \hline
0 (Sign) & $262144$ & $0.6138$ &1.1626\\
\hline
\end{tabular}
\end{center}
\end{table}

\begin{table}[htbp]
\begin{center}
 \caption{Lena} \label{tab:tab3}
\begin{tabular}{|c|c|c|c|}\hline

Bit-plane $p$ & length $K_p$ & empirical entropy $\bar{H}_p$ & efficiency $\bar{\eta}_p$  \\
 \hline
8 & $16384$ & $0.345$ &2.0691 \\
\hline
7 & $29184$ & $0.258$ &2.7598 \\
 \hline
6 & $71680$ & $0.277$ &2.5764\\
 \hline
5 & $142336$ & $0.191$ &3.7315\\
 \hline
4 & $223232$ & $0.190$ &3.7527\\
  \hline
3 & $253440$ & $0.324$ &2.1972 \\
  \hline
2 & $262144$ & $0.672$ &1.0607\\
  \hline
1 & $262144$ &$0.925$ &0.7713\\
  \hline
0 (Sign) & $262144$ & $0.4977$ &1.4337\\
\hline
\end{tabular}
\end{center}
\end{table}

In order to deliver the model parameters to the decoder with high reliability we have
separately encoded them using a conventional TC (generators $(37,21)_8$, rate 1/3), that achieves practically error-free 
performance for the BSC channel with capacity $0.7136$.
For reconstruction we have used the soft-bit decoding strategy given at the end of Section \ref{sec:proposed-approach}.

\begin{figure}[hbtp]
\centerline{\includegraphics[scale= 0.6]{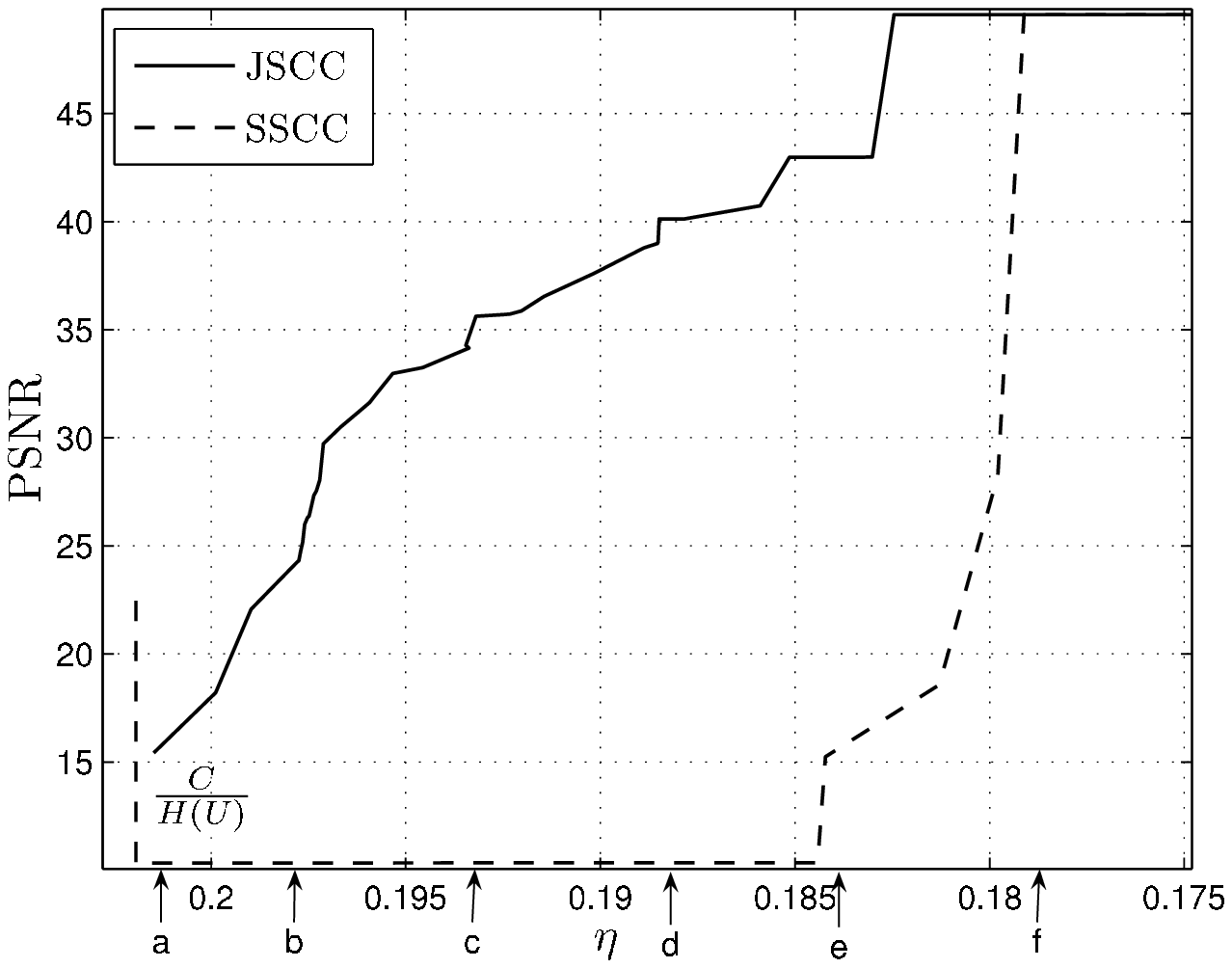}}
\caption{PSNR comparison between SSCC schemes and JSSC schemes for ``Goldhill''.}
\label{fig:rho005gold}
\end{figure}

\begin{figure}[hbtp]
\centerline{\includegraphics[scale= 0.6]{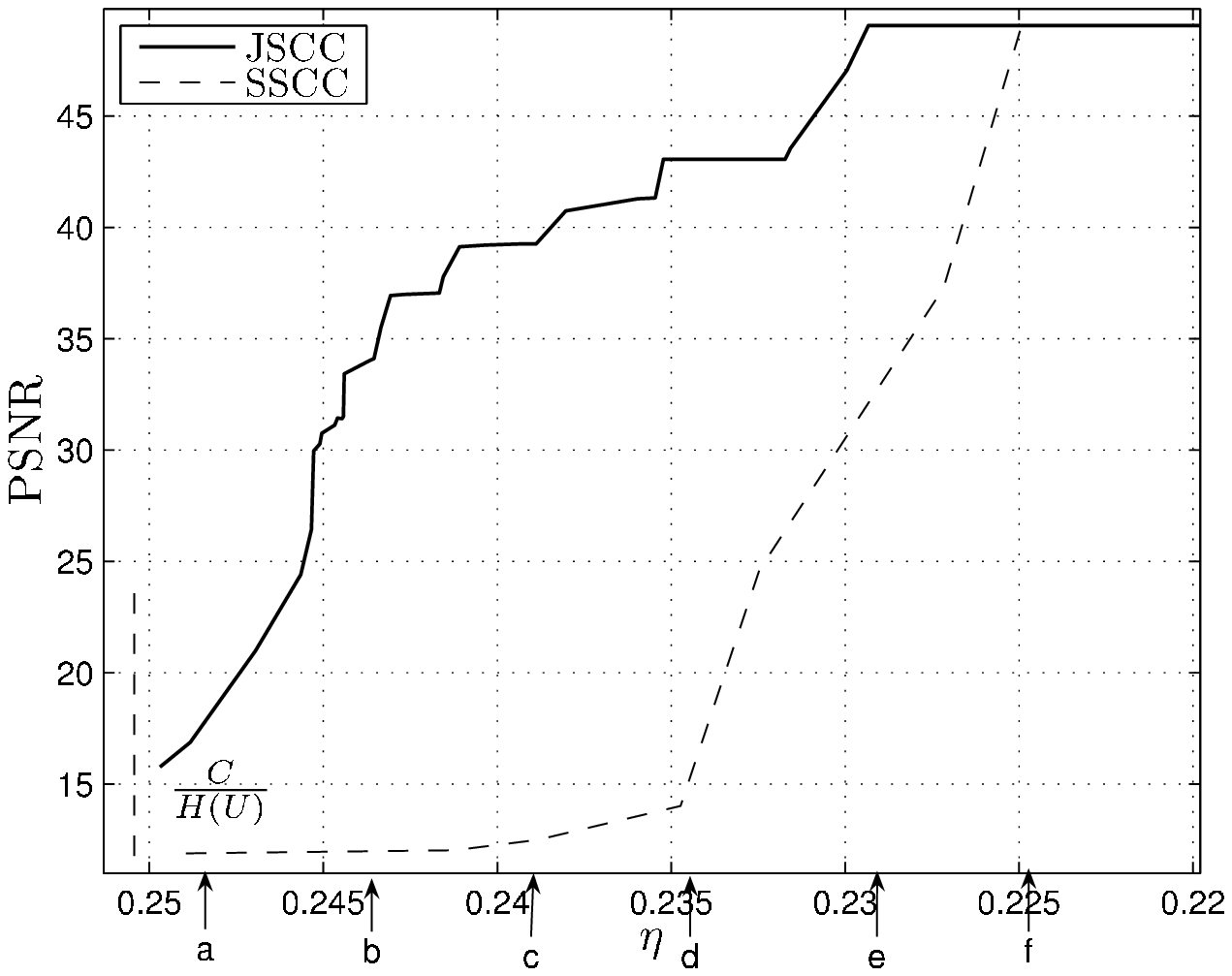}}
\caption{PSNR comparison between SSCC schemes and JSSC schemes for
``Lena''.}
\label{fig:rho005lena}
\end{figure}

Figs. \ref{fig:rho005gold} and \ref{fig:rho005lena} show the PSNR
performance of the proposed JSCC scheme and the conventional SSCC.
The resulting PSNR is plotted versus the coding efficiency, in order to put in
evidence the gap of the actual coding scheme with respect to the
Shannon limit defined as in Section \ref{sec:separated-scheme}. Notice that the coding efficiency decreases while moving towards the 
right of the horizontal axis. As the coding efficiency decreases, the PSNR reaches its maximum
value which corresponds to the quantization distortion $D_{\Qc}$, equal to $PSNR_{Goldhill} =49.57 $ and 
$PSNR_{Lena}= 49.08$ dB for ``Goldhill'' and ``Lena'', respectively.

PSNR curves may not tell much about the actual quality of the reconstructed image.
Hence, we include also {\em snapshots} of the reconstructed images for both the JSCC and the SSCC schemes 
corresponding to the efficiencies marked on Figs. \ref{fig:rho005gold} and \ref{fig:rho005lena}
as ``a,b,c,d,e,f''. The corresponding values $\eta_a,\eta_b,\ldots,\eta_f$ have increasing gap from the 
Shannon limit.

\begin{figure}[htbp]
\centerline{
a) \includegraphics[scale=0.23]{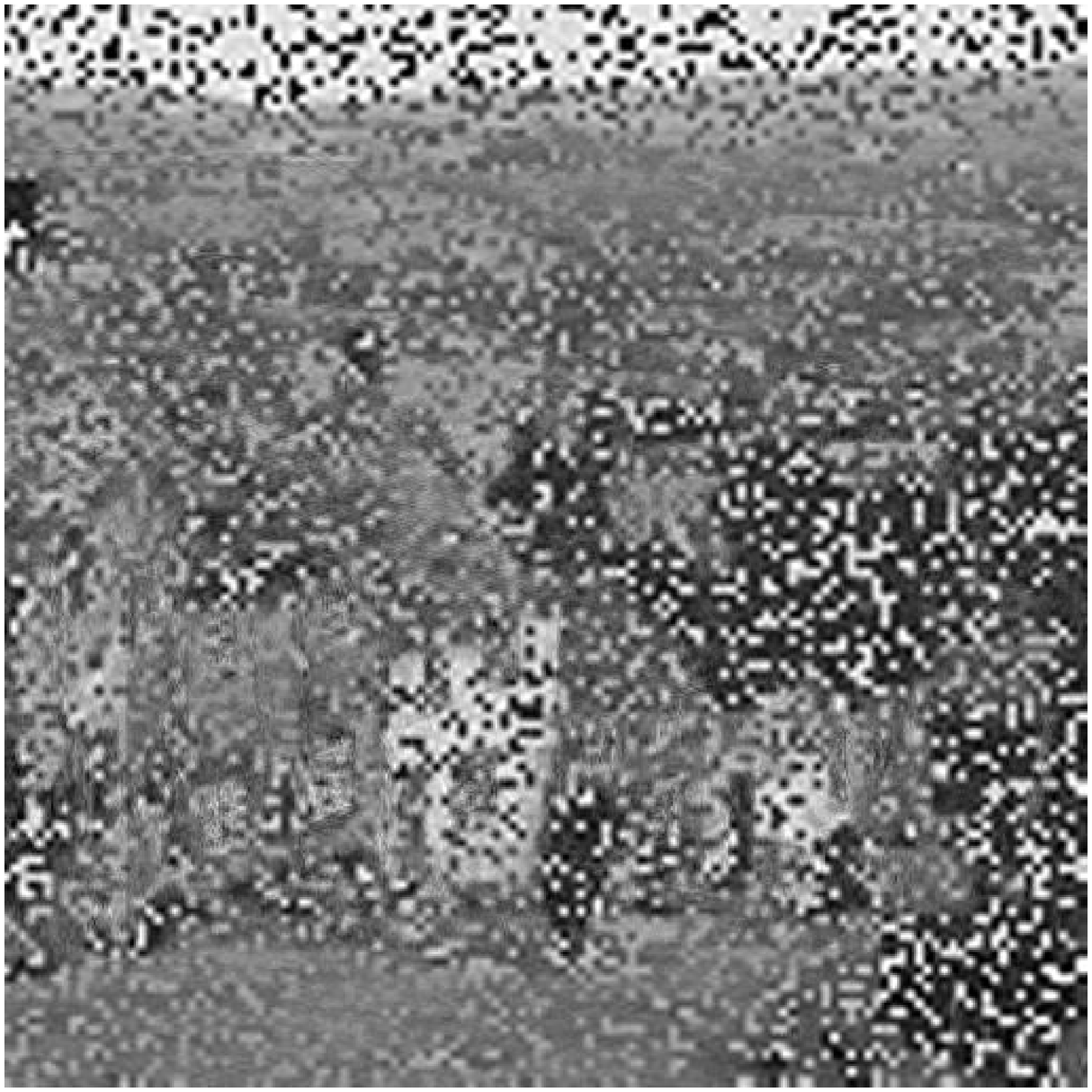} \hspace{3mm}
b) \includegraphics[scale=0.23]{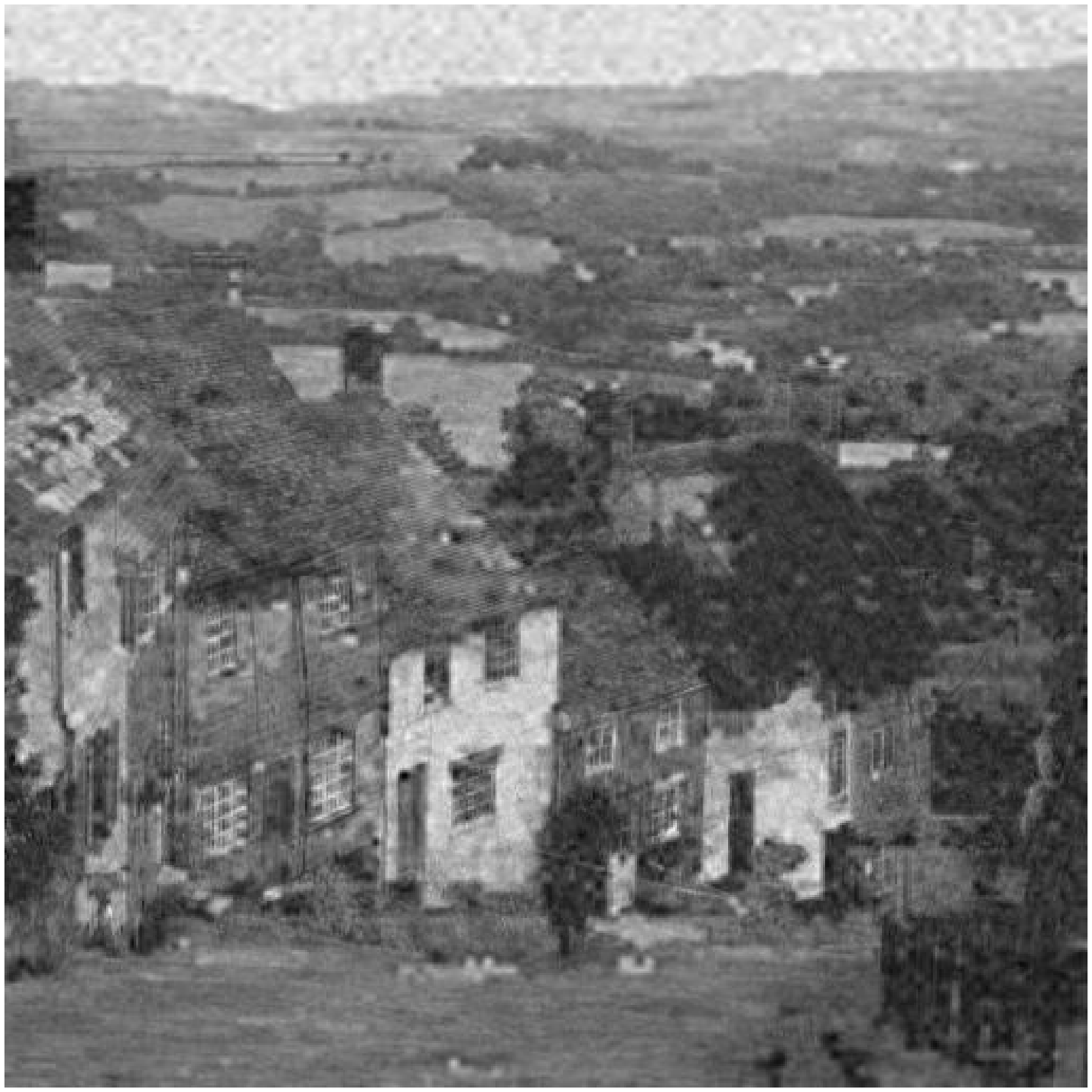} \hspace{3mm}
c) \includegraphics[scale=0.23]{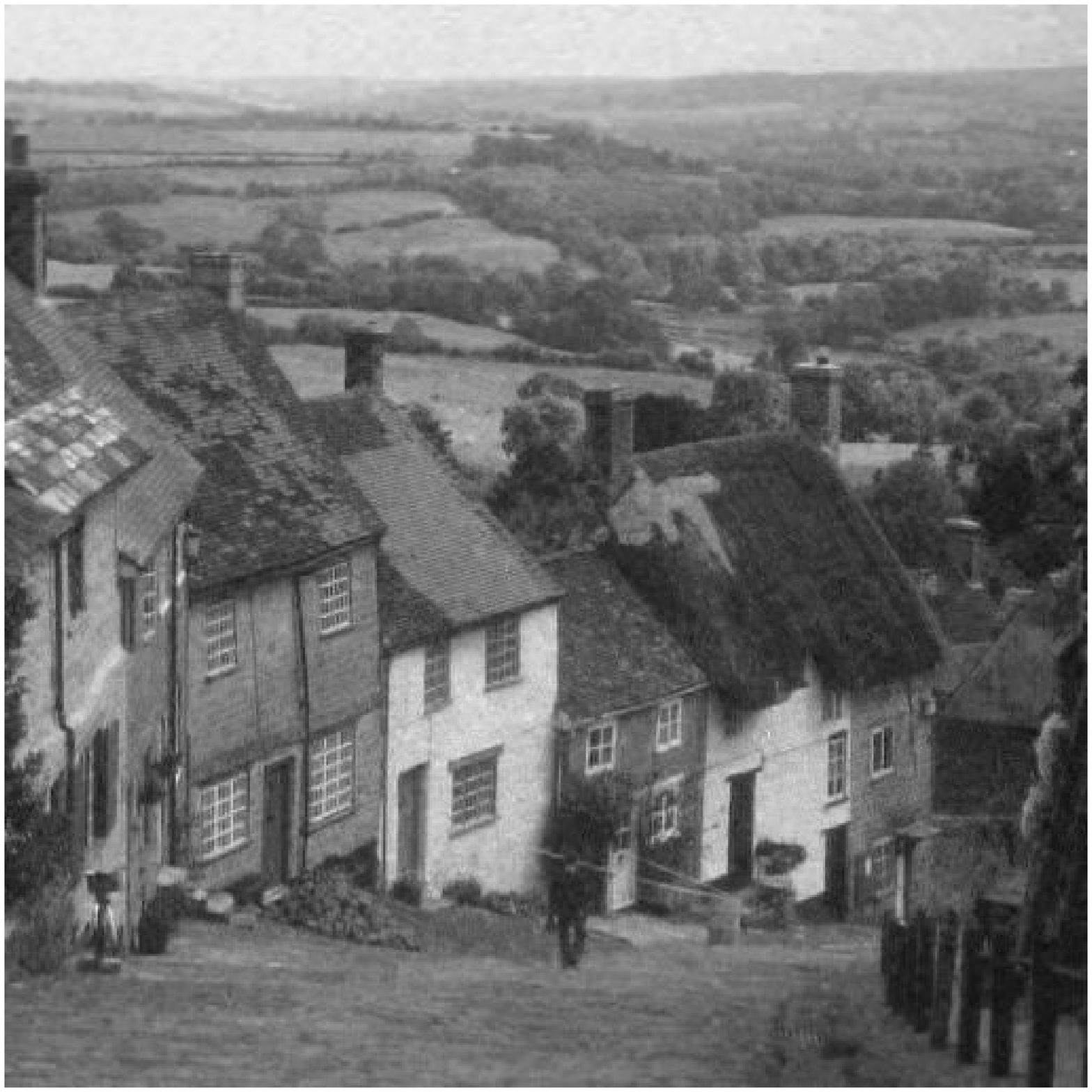}
} 
\centerline{
d) \includegraphics[scale=0.23]{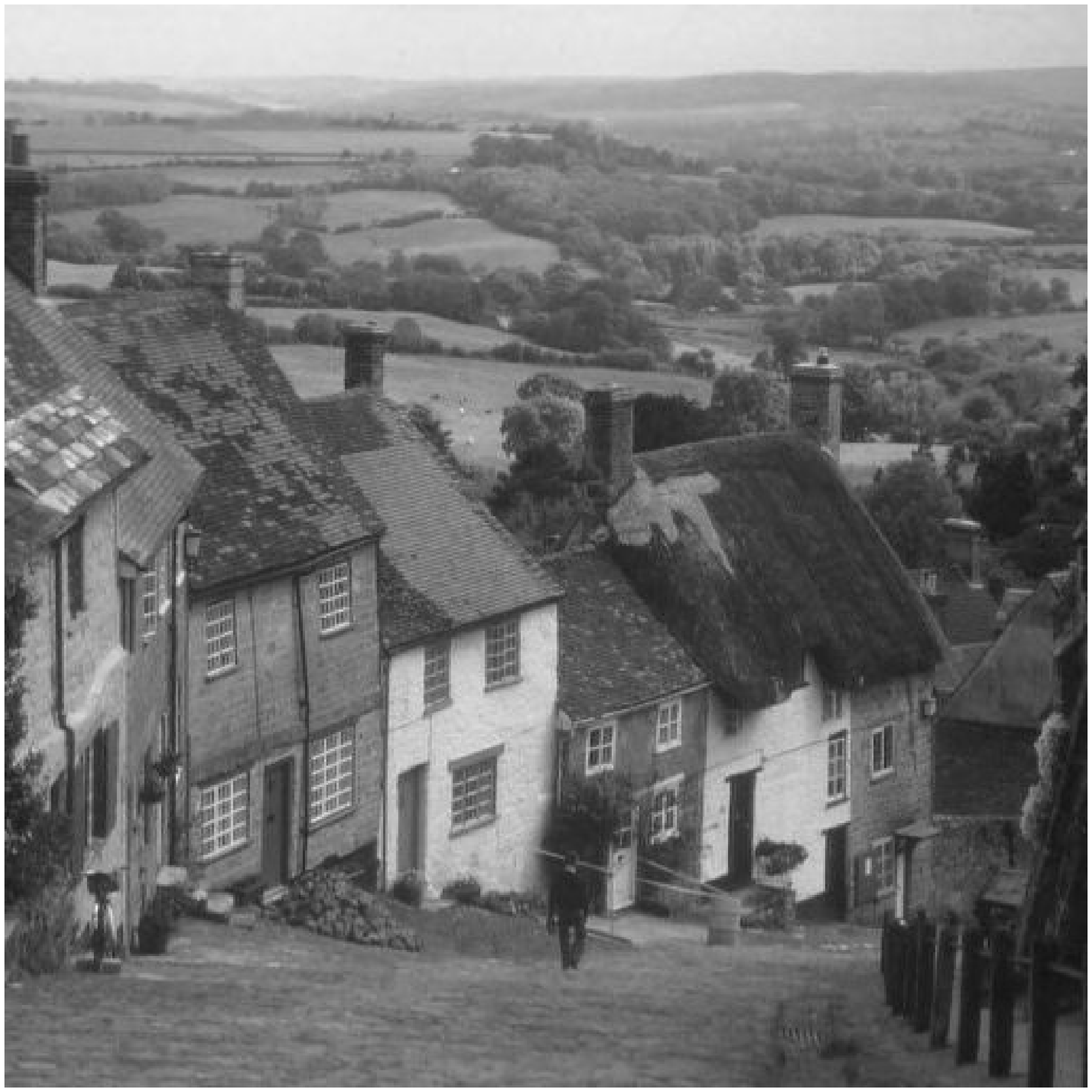} \hspace{3mm}
e) \includegraphics[scale=0.23]{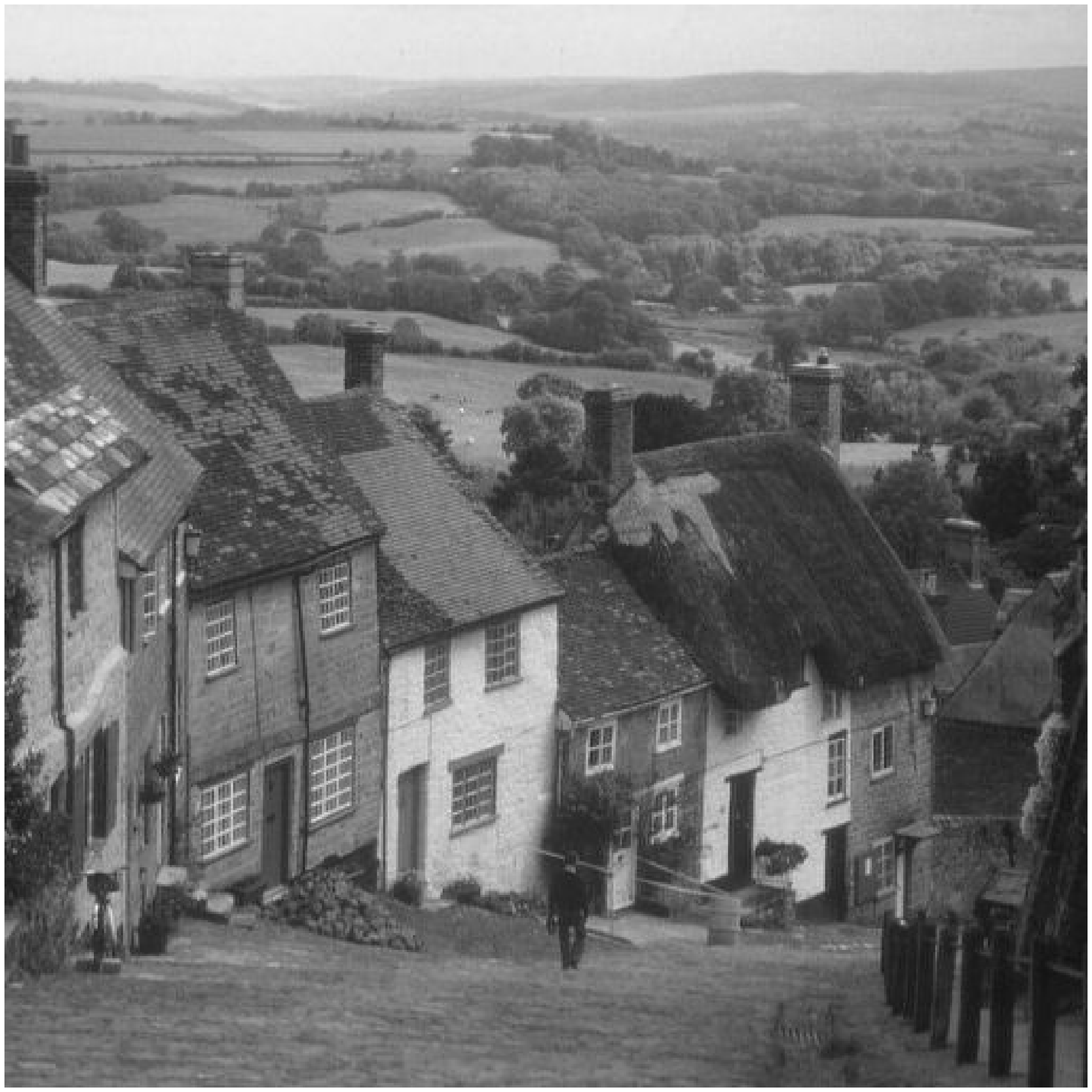} \hspace{3mm}
f) \includegraphics[scale=0.23]{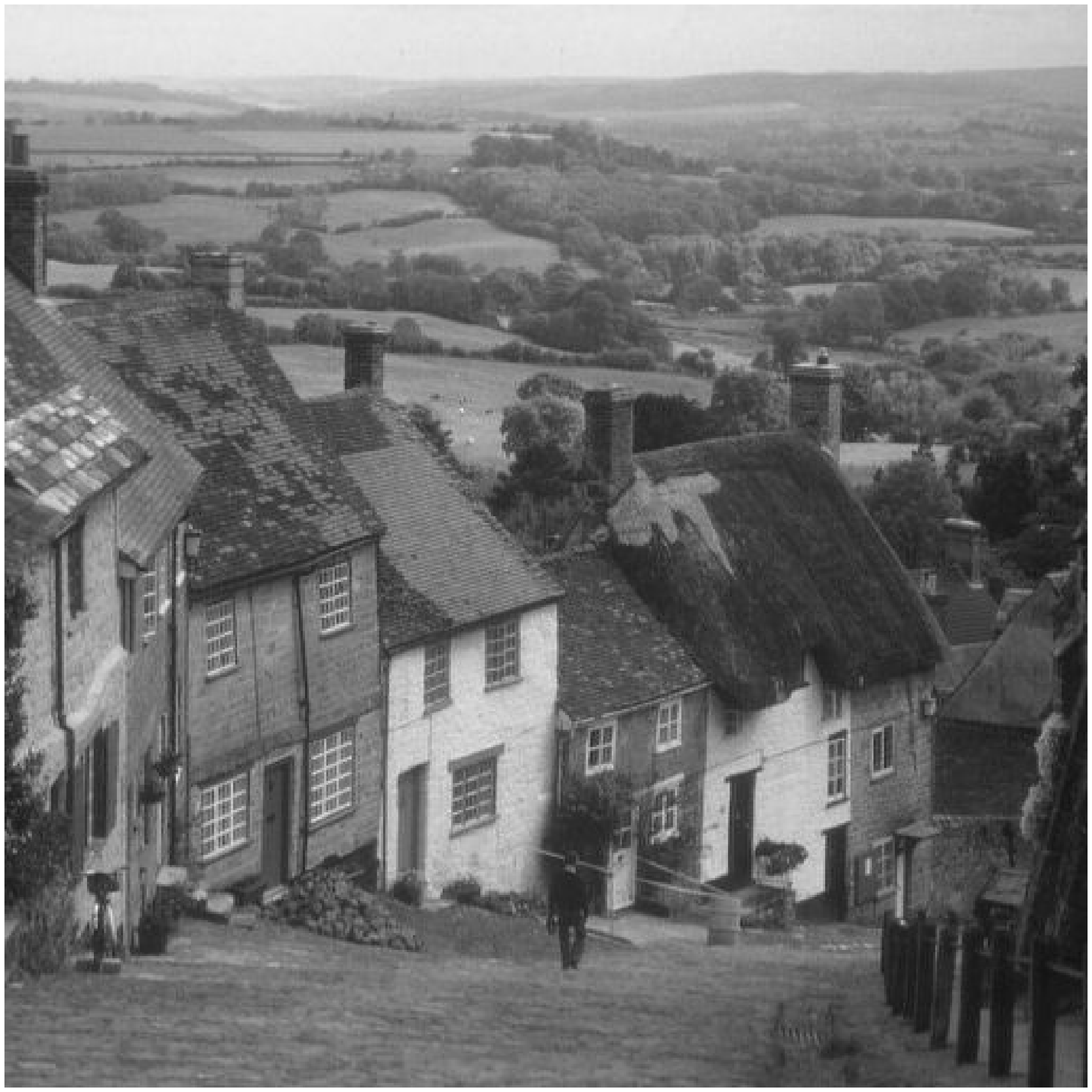}
} 
\centerline{
a) \includegraphics[scale=0.23]{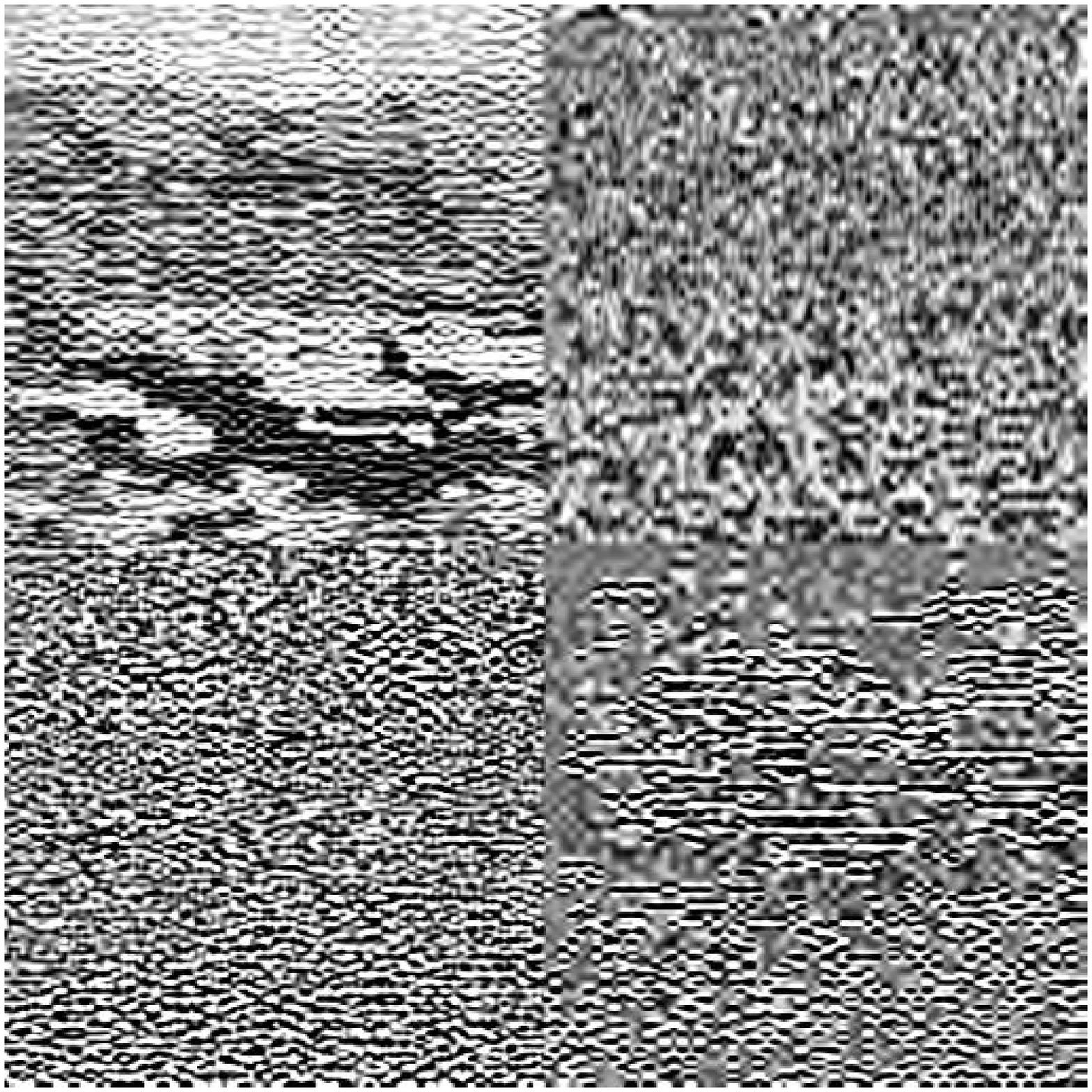} \hspace{3mm}
b) \includegraphics[scale=0.23]{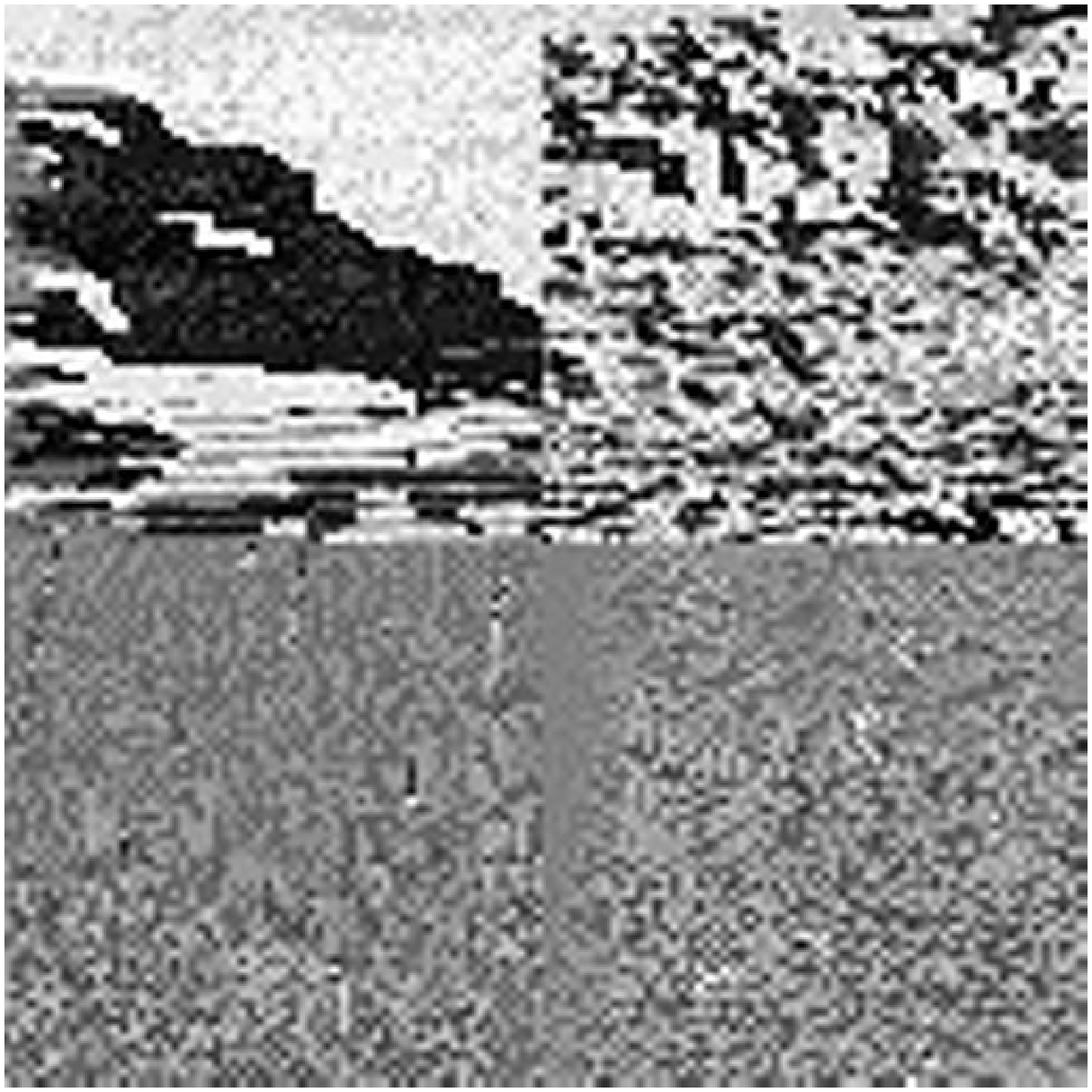} \hspace{3mm}
c) \includegraphics[scale=0.23]{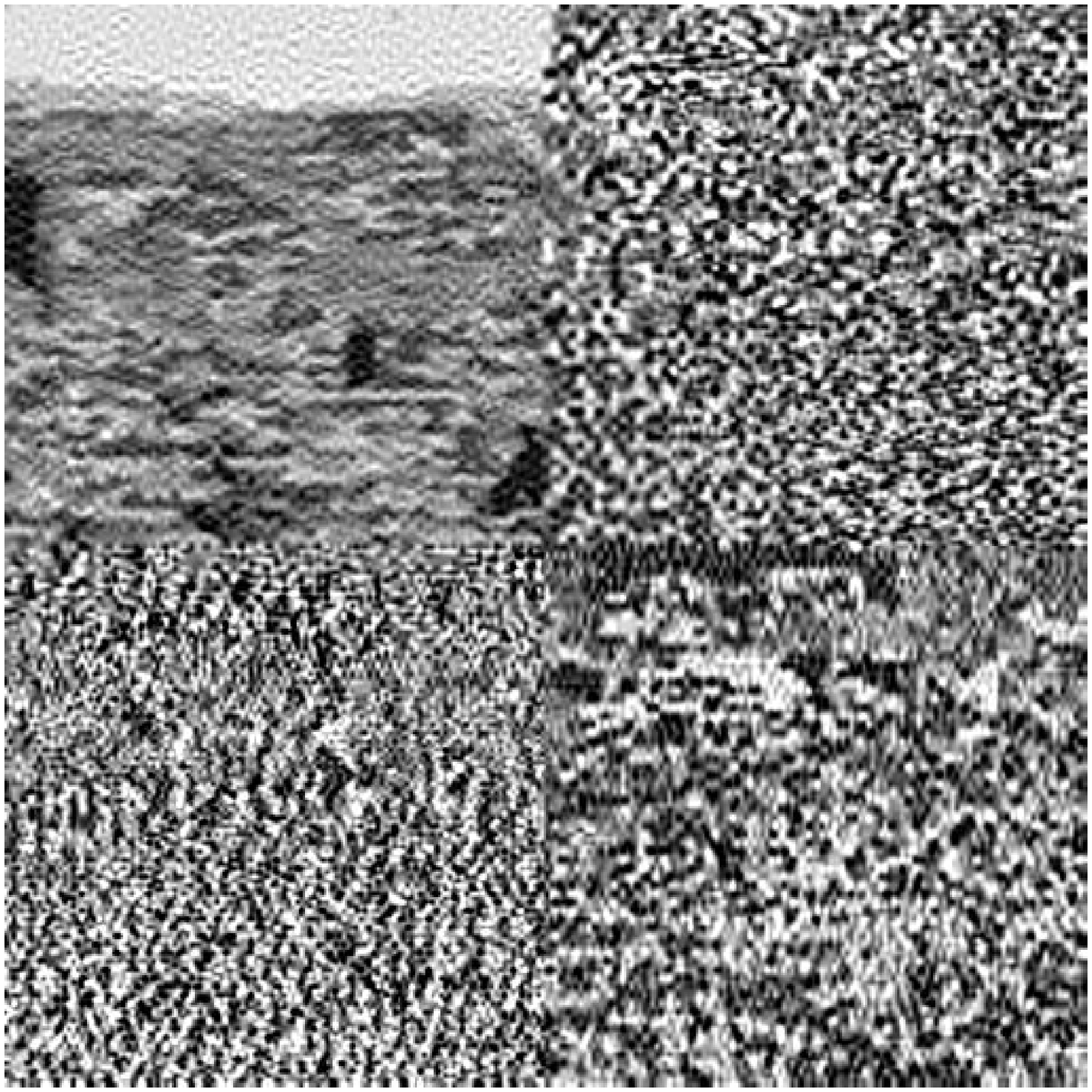}
}
\centerline{
d) \includegraphics[scale=0.23]{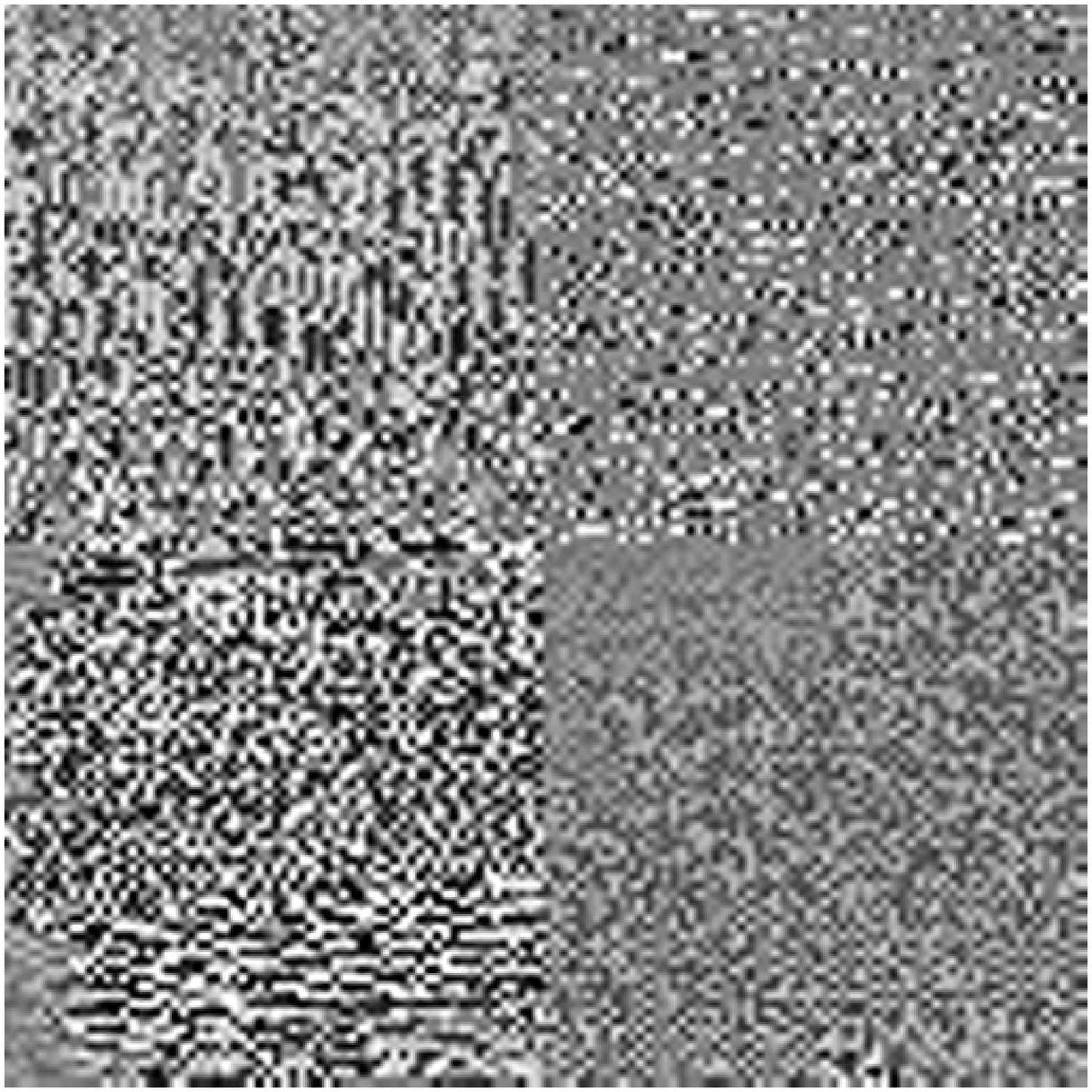} \hspace{3mm}
e) \includegraphics[scale=0.23]{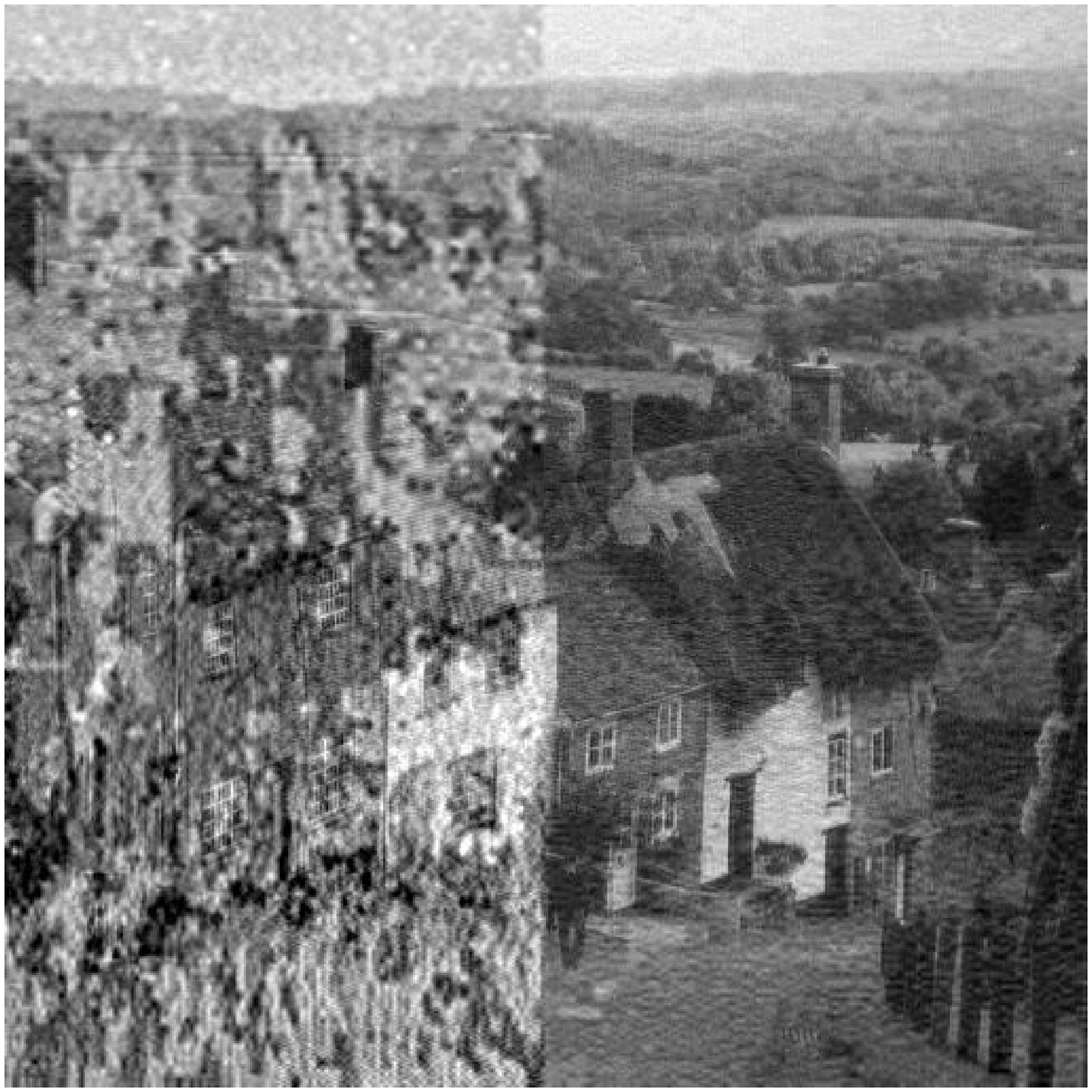} \hspace{3mm}
f) \includegraphics[scale=0.23]{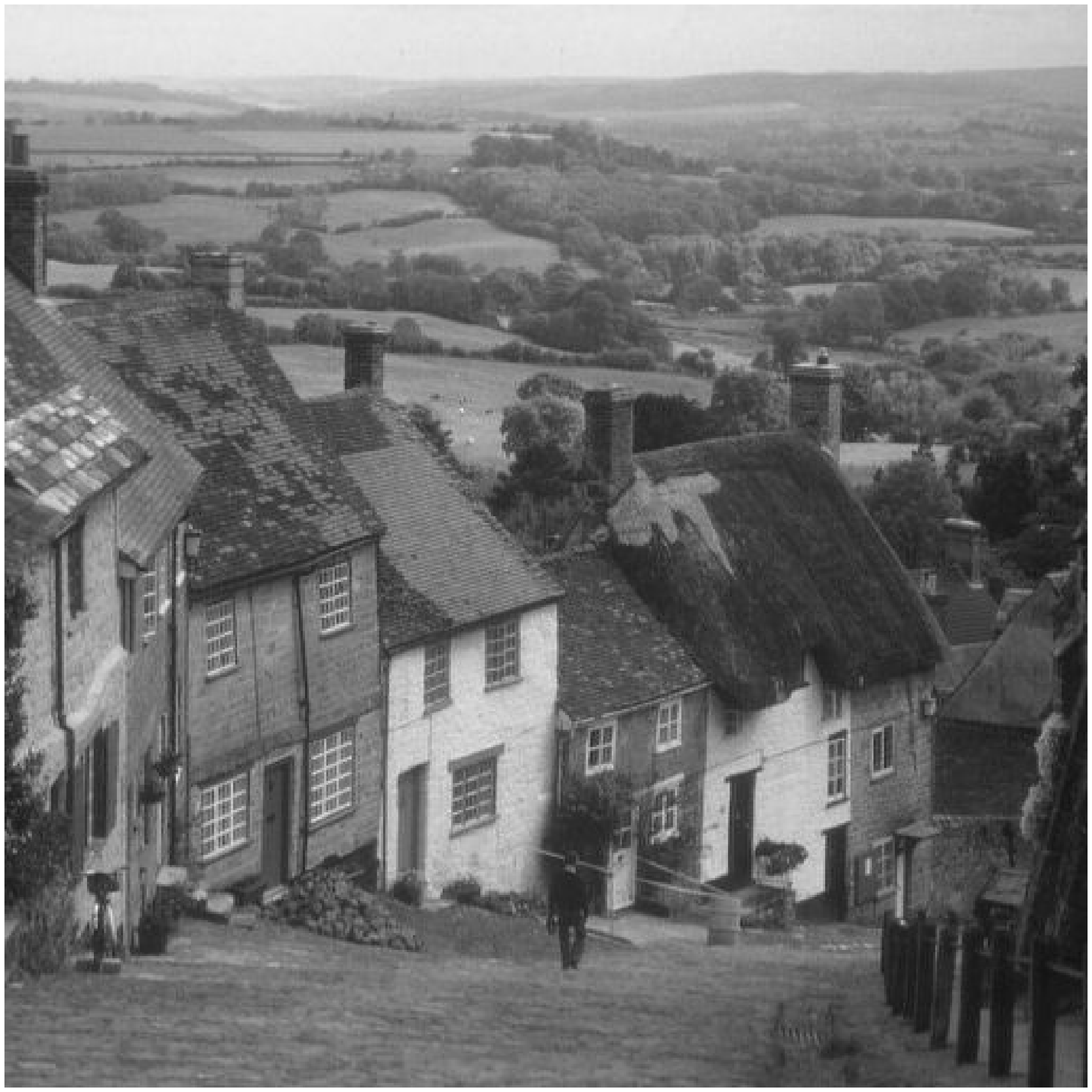}
}
\caption{Visual comparison for JSSC (first set) and the SSCC (second set) 
and the Goldhill image, for decreasing values of $\eta_a,\ldots,\eta_f$ indicated in 
Fig. \ref{fig:rho005gold} (from left to right).}
\label{Comparison_Gold}
\end{figure}

\begin{figure}[htbp]
\centerline{
a) \includegraphics[scale=0.23]{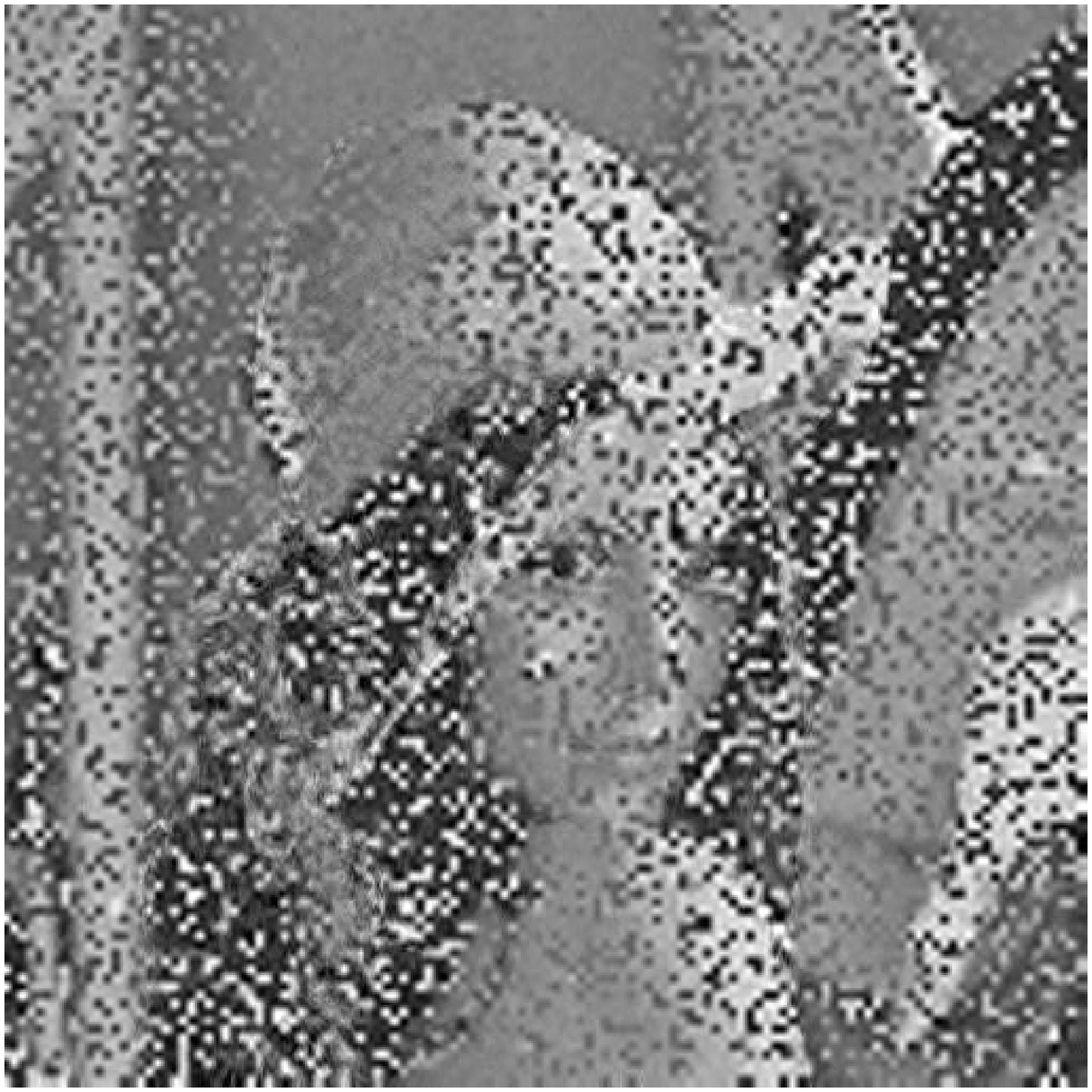} \hspace{3mm}
b) \includegraphics[scale=0.23]{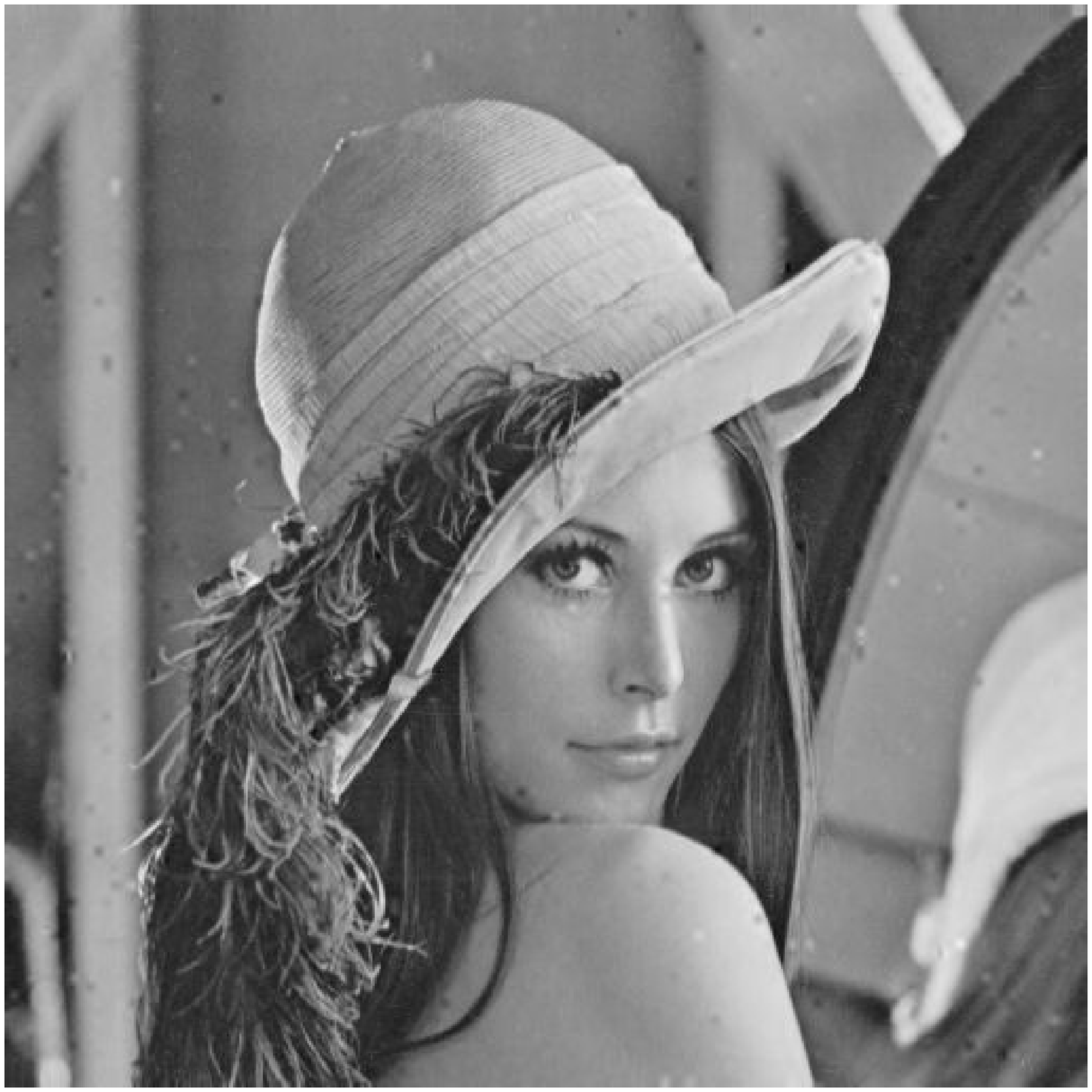} \hspace{3mm}
c) \includegraphics[scale=0.23]{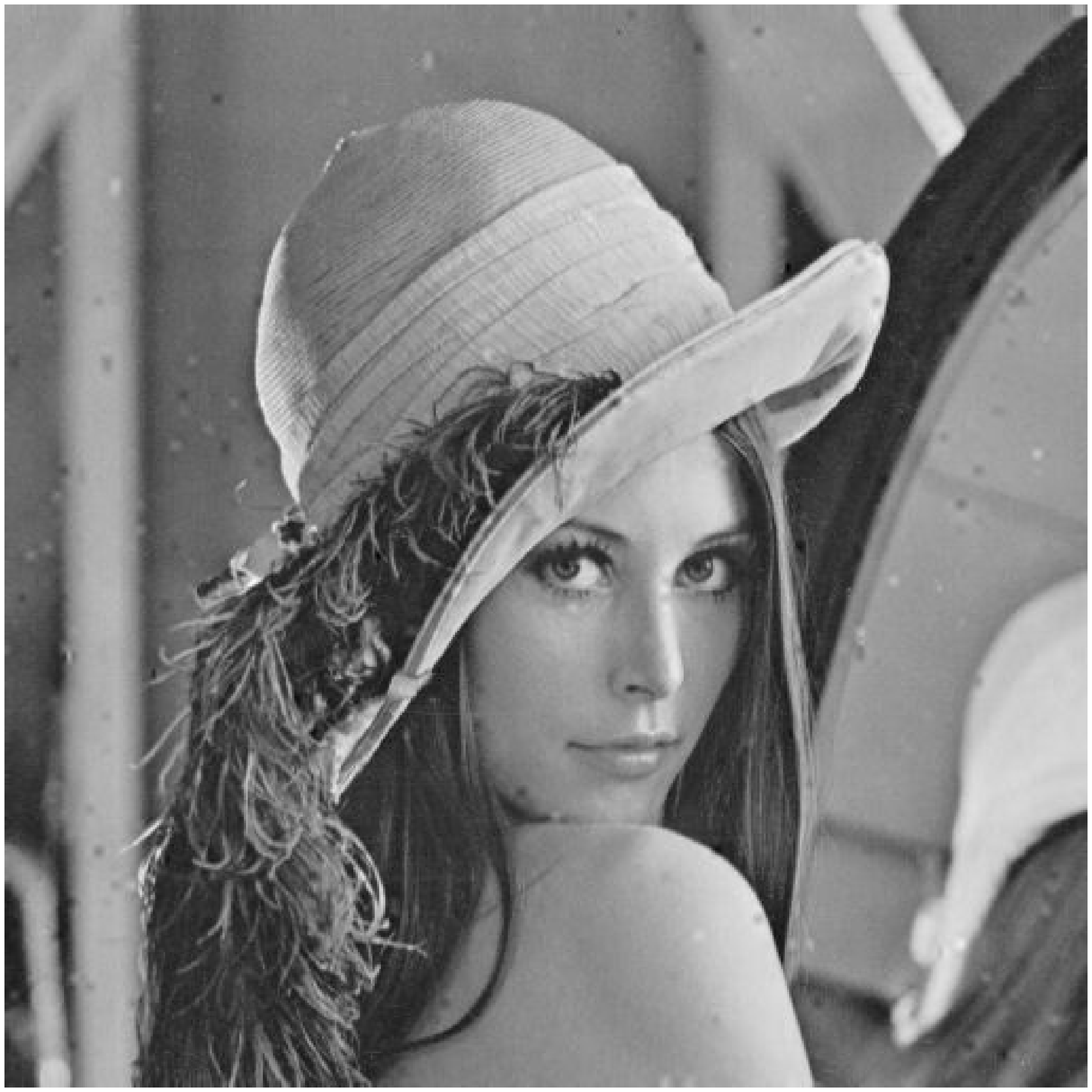}
} 
\centerline{
d) \includegraphics[scale=0.23]{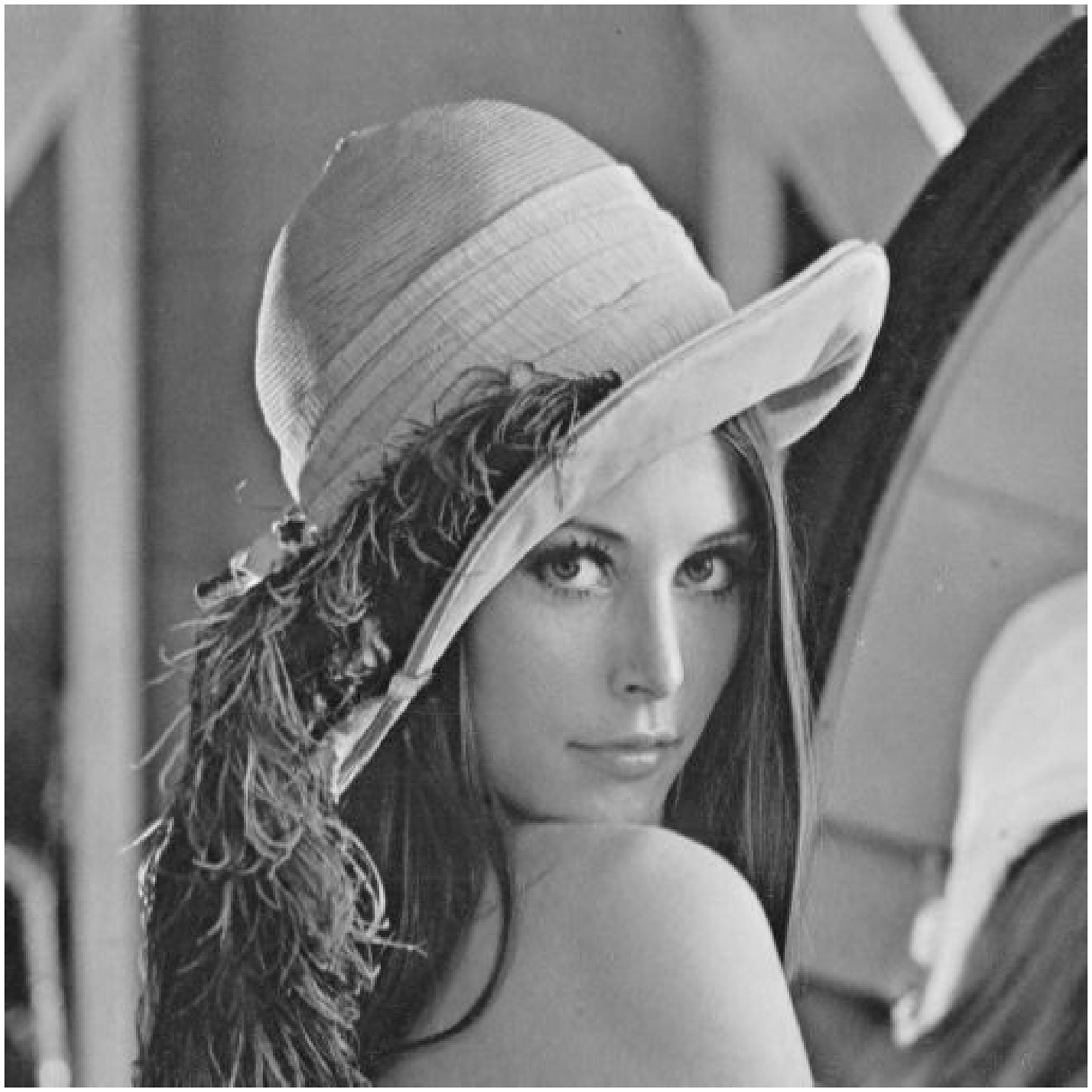} \hspace{3mm}
e) \includegraphics[scale=0.23]{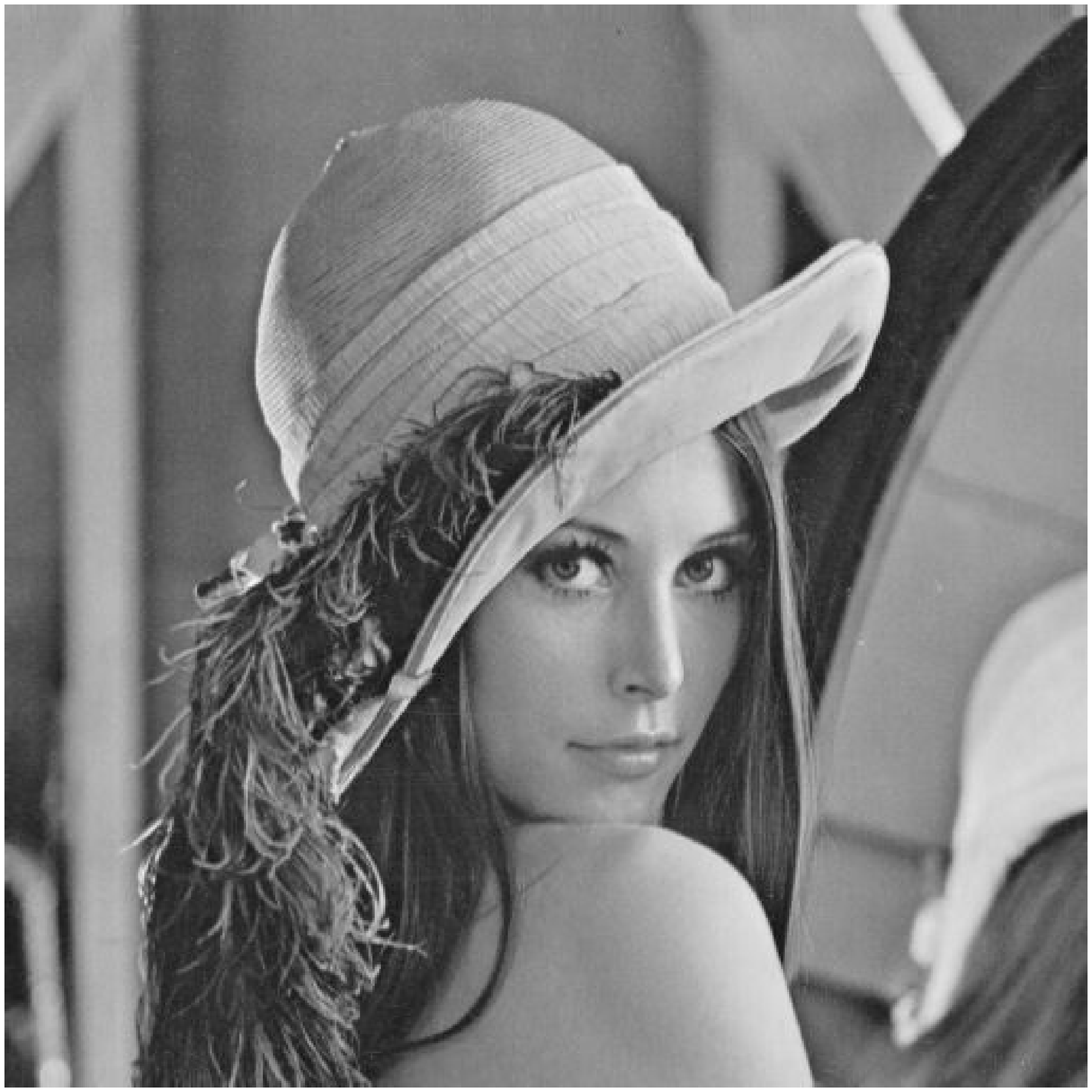} \hspace{3mm}
f) \includegraphics[scale=0.23]{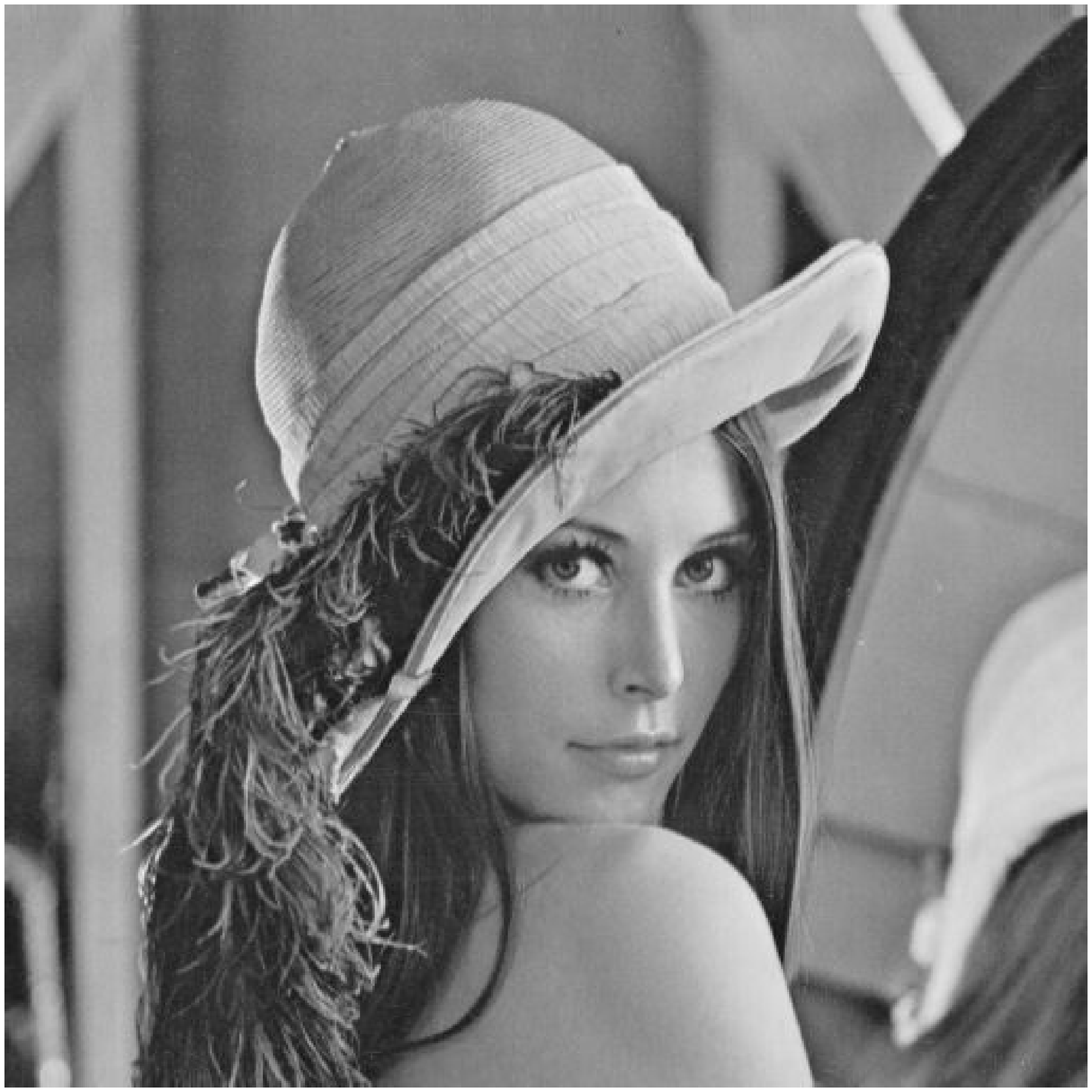}
} 
\centerline{
a) \includegraphics[scale=0.23]{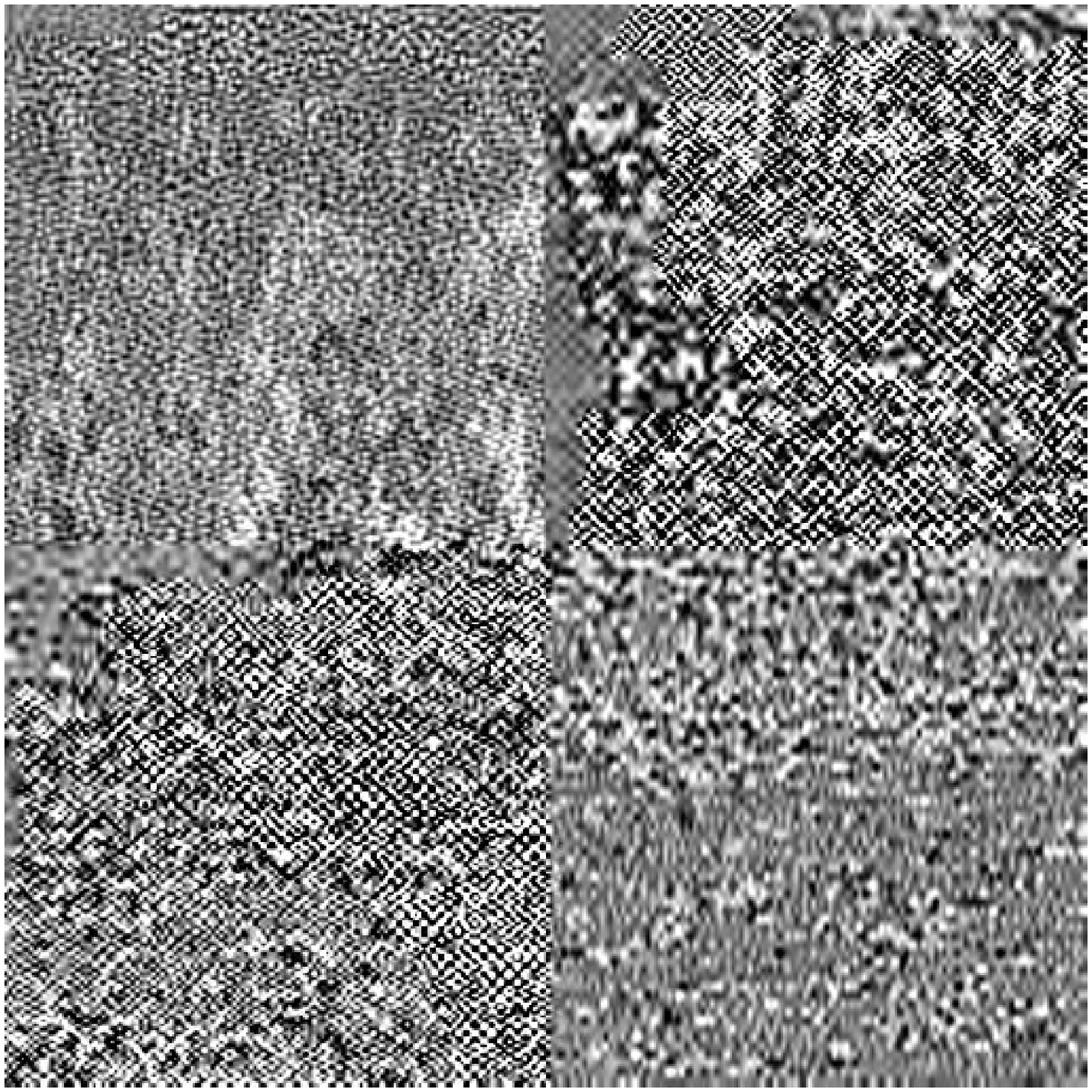} \hspace{3mm}
b) \includegraphics[scale=0.23]{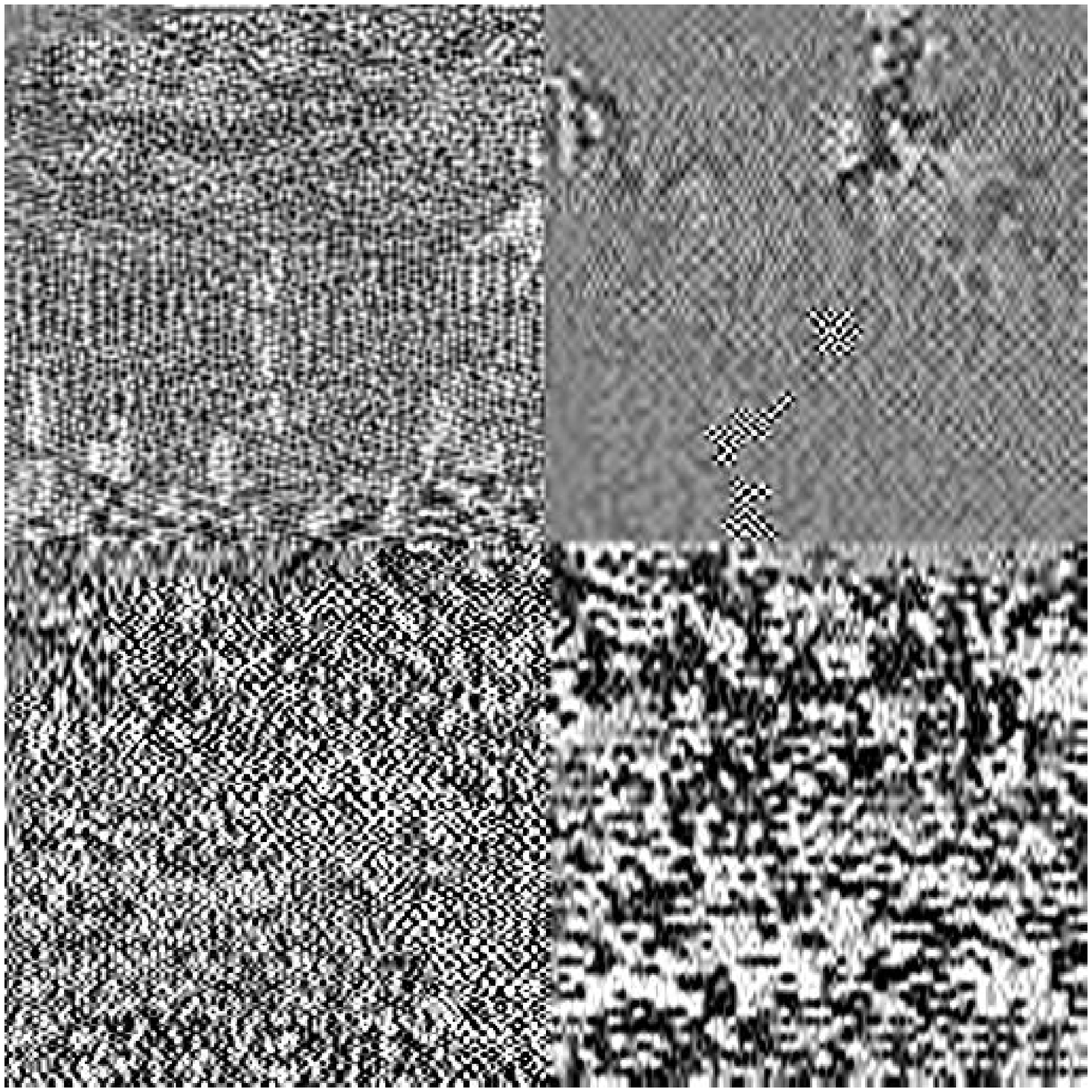} \hspace{3mm}
c) \includegraphics[scale=0.23]{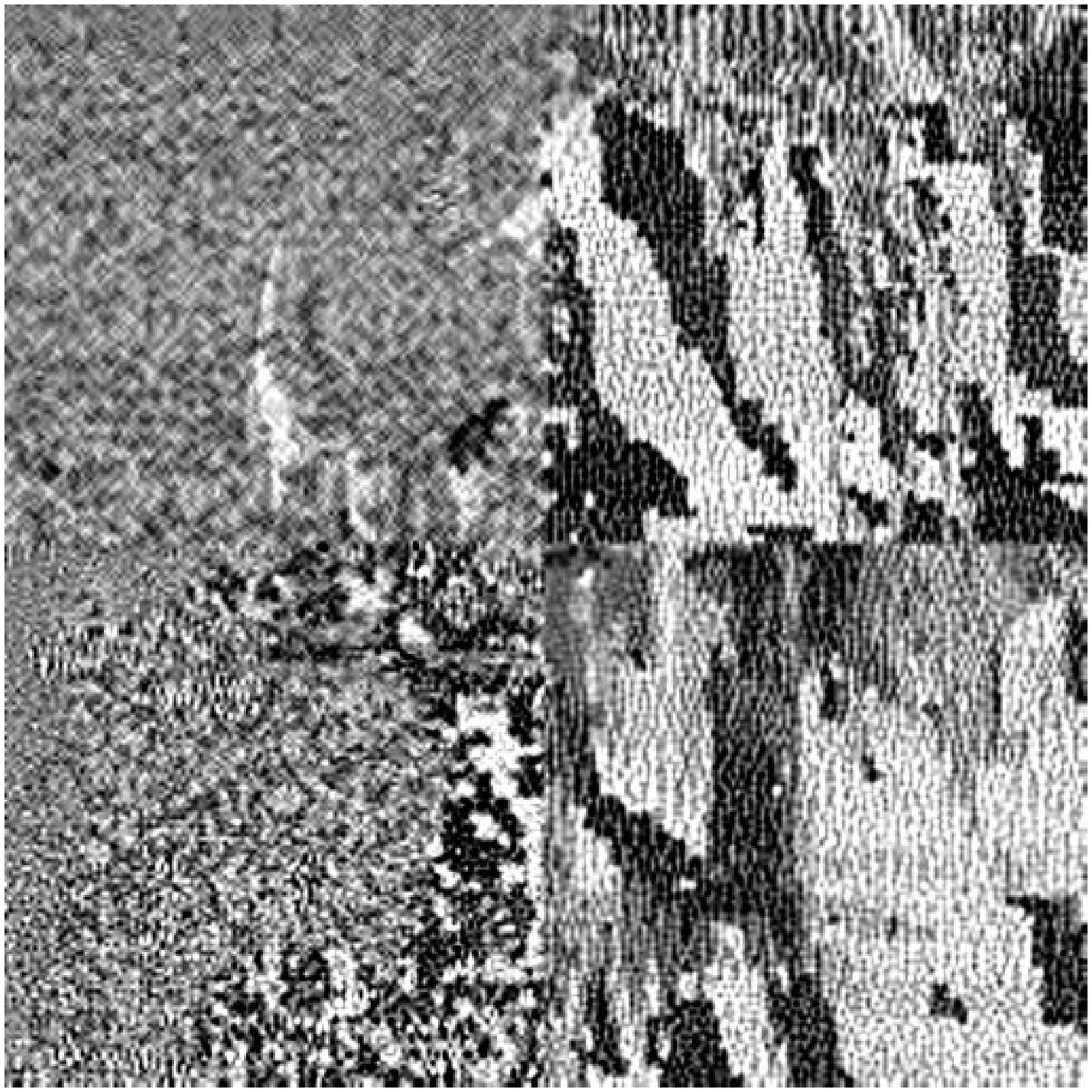}
}
\centerline{
d) \includegraphics[scale=0.23]{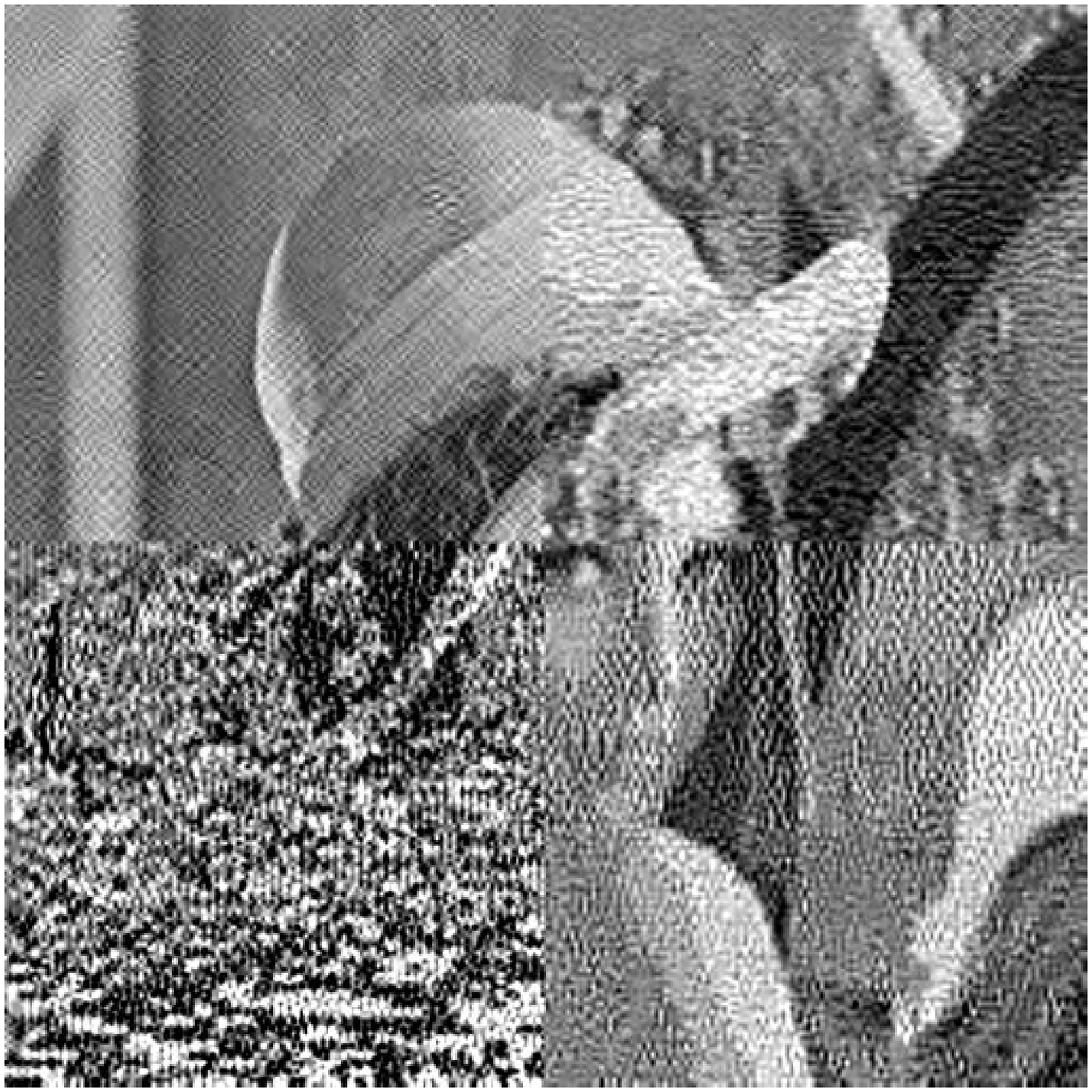} \hspace{3mm}
e) \includegraphics[scale=0.23]{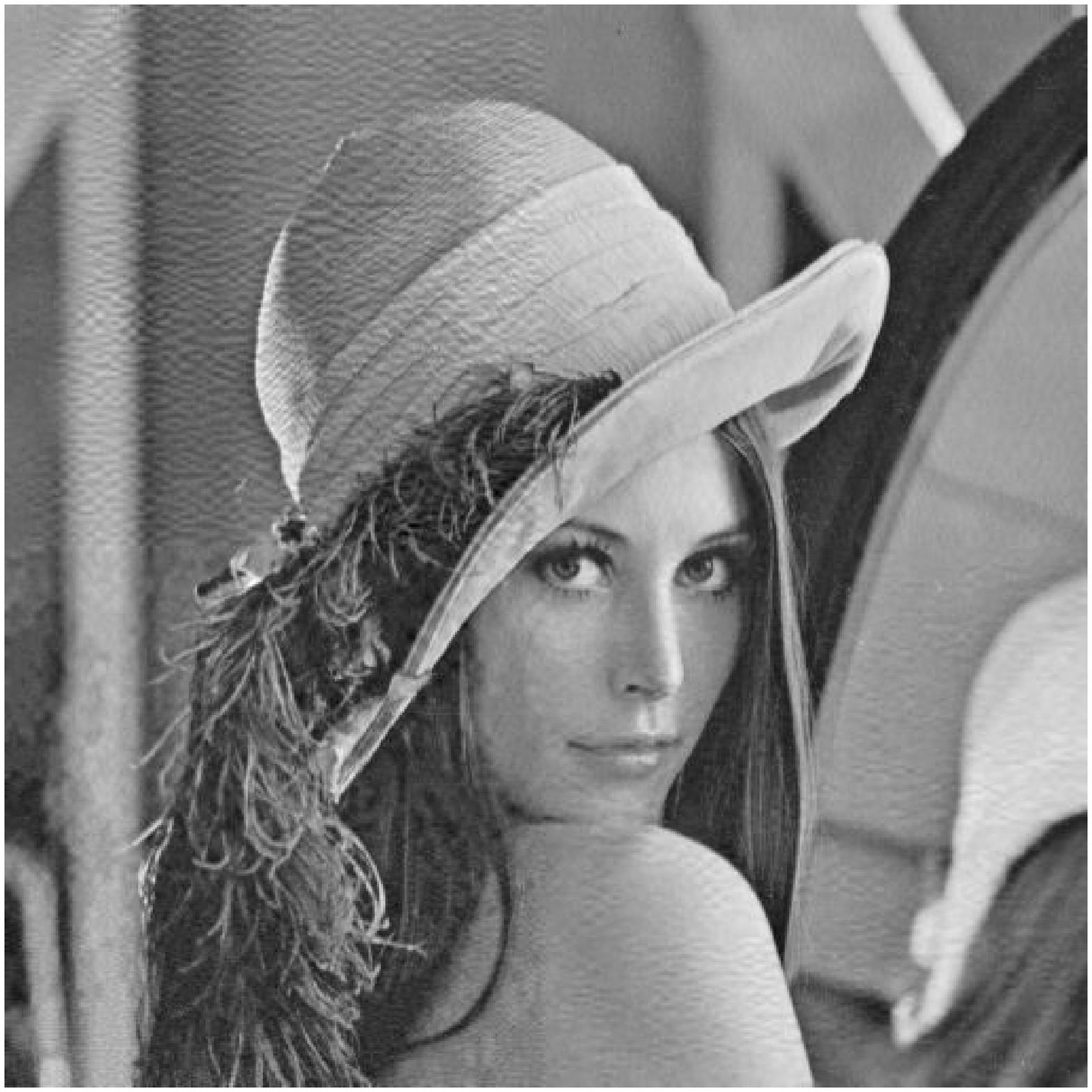} \hspace{3mm}
f) \includegraphics[scale=0.23]{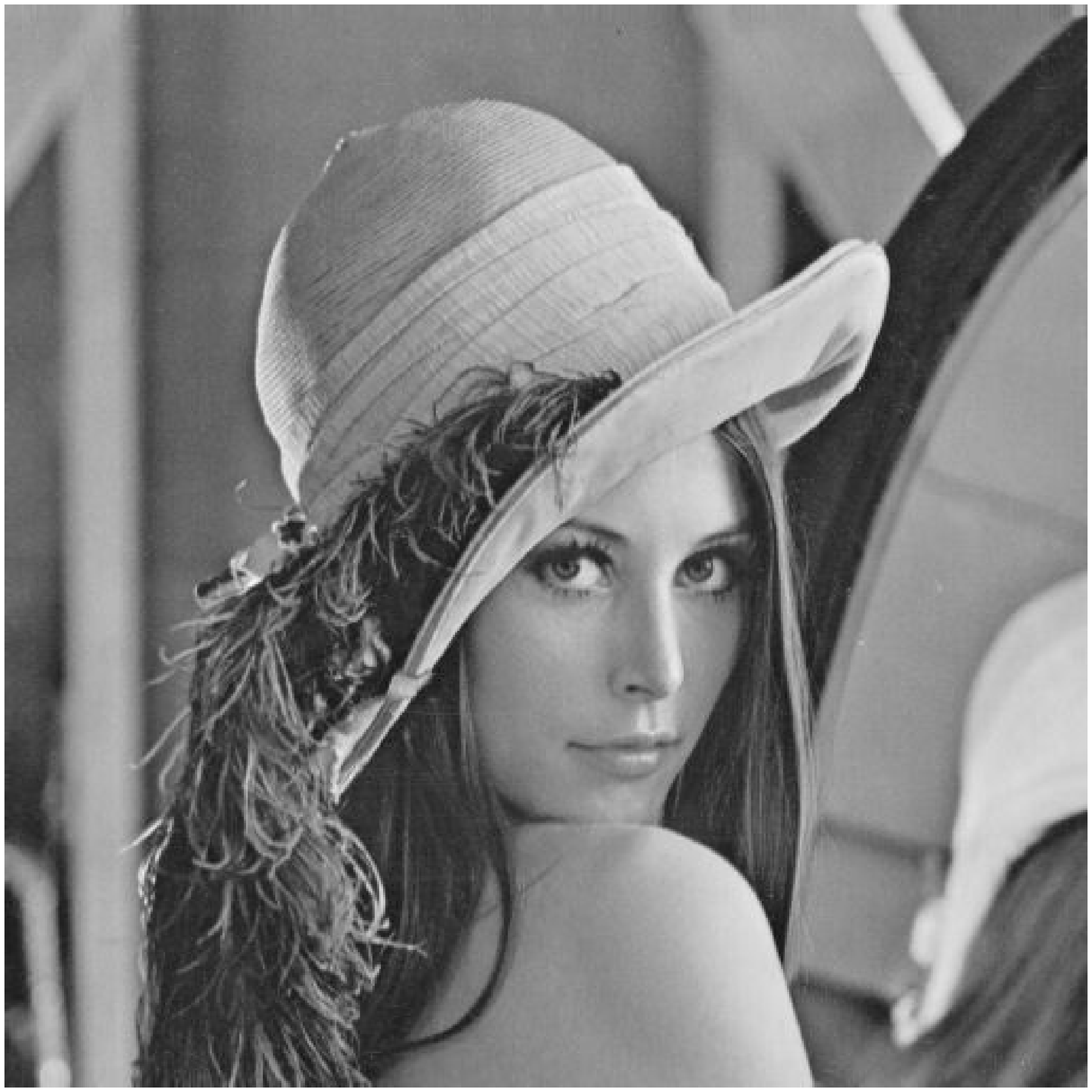}}
\caption{Visual comparison for JSSC (first set) and the SSCC (second set) 
and the Lena image, for decreasing values of $\eta_a,\ldots,\eta_f$ indicated in 
Fig. \ref{fig:rho005lena} (from left to right).}
\label{Comparison_Lena}
\end{figure}

These snapshots, shown in Figs. \ref{Comparison_Gold} and \ref{Comparison_Lena}, 
illustrate the main claims of this paper and provide experimental evidence about the effectiveness of the
proposed JSCC scheme: it is clearly visible that as soon as $\eta$ is slightly above the Shannon limit, our JSCC scheme
achieves an acceptable reconstruction quality. On the contrary, the conventional SSCC  achieves very poor quality for a much wider range of the gap from Shannon limit, and suddenly achieves the quantization distortion when the channel code is able to eliminate completely the 
channel errors (with very high probability). This behavior reflects the catastrophicity of the entropy coding stage of the 
conventional scheme, which is essentially {\em eliminated} by our approach.
Moreover, in the SSCC the reconstructed images clearly show the square patterns due to the 
catastrophic error propagation inside code-blocks, because of residual bit errors after channel decoding.
Even when the PSNR is not so low, these patterns reduce significantly the perceived image quality. 
On the other hand, the reconstructed images generated by the proposed JSCC method present a ``salt and pepper'' noise, 
that is less annoying for the perceived image quality.
This nature of the generated noise opens the door to the study of the concatenation of our approach with the recently proposed
discrete universal denoising algorithm (DUDE) \cite{dude}, that can take advantage of the apparently independent memoryless residual noise
after JSCC decoding in order to further improve the image visual quality. This interesting approach is not pursued in this paper and it is 
mentioned in Section \ref{sec:Conclusions} among the directions for future work.

\section{Conclusion} \label{sec:Conclusions}

We presented a new scheme for pragmatic Joint Source-Channel Coding
that builds over two information theoretic key ideas: a) the optimal 
rate-distortion performance can be univerally approached by low-dimensional 
quantization followed by entropy coding; b) the optimal performance of (almost) lossless joint-source channel coding
for a general discrete source and a wide class of channels (BIOS channels in our case) can be attained by 
linear coding. Our method is {\em general} in the sense that allows to take advantage of the vast experience 
developed in lossy compression of practically relevant sources. In fact, the scheme keeps the front-end of 
state-of-the art source transform coders, comprising a linear projection on a convenient orthonormal basis (such as
DWT, short-time DFT, DCT) and scalar quantization,  unchanged. Also, the probability models developed in state-of-the art source coders
can be reused, provided that these models can be represented in some easily factored form, for the purpose of the joint source-channel Belief Propagation decoder. We developed an example based on JPEG2000 and Turbo Codes. In our case, the probability model reduces to
a binary Markov chain for each bit-plane, conditionally on the bits from upper bit-planes. 
The Markov chains are generally non-stationary. Nevertheless, we showed that there exist binary linear codes that operate arbitrarily close to
the Shannon limit for any arbistrary source: a single codebook can handle all sources with given sup-entropy. This prompts for a scheme
that selects the coding rate for each bit-plane by measuring the {\em empirical entropy} of the bit-plane itself, based on the probability model
used by the decoder, and transmitted to the decoder separately. 

The proposed scheme was simulated for some classical test images over a BSC, and compared with a standard separated approach based on JPEG2000
and state-of-the art Turbo Codes. We achieve much better performance in terms of PSNR and, most importantly, in terms of subjective image reconstruction quality. Remarkably, our scheme puts very little stress on the channel coding scheme, and does not require to attain
very low BER in order to achieve good end-to-end distortion. We interpret this fact by noticing that 
using linear codes we eliminate the catastrophic behavior of traditional entropy coding based on Huffman or arithmetic coding, by producing a
mapping of the redundant quantization bits over the channel symbols that has a well-conditioned ``soft-inverse''.
Furthermore, the use of BP decoding allows to exploit the decoder soft-output in order to produce (approximated) MMSE estimates of the 
transform coefficients using soft-bits.

We would like to conclude by enumerating some topics for further research:
\begin{itemize}

\item Several families of linear codes can be used instead of punctured TCs. In particular, instead of using a library of 
codes to cover a quantized range of rates, we can use {\em fountain codes} \cite{luby,raptor} in order to produce 
the required amount of redundancy ``on the fly'', depending on the empirical entropy of the source and, in the case of 
a compound of BIOS channels whose capacity is known to the transmitter, depending on the capacity of the channel. 
The use of fountain codes for {\em universal} lossless compression was investigated in \cite{CaShVe_dimacs2}, and a similar encoding scheme
can be used in this context. Furthermore, the same approach can be used in multicast applications, where the same source must be sent to
several users and each user may have a different channel capacity, as in the classical ``fountain coding'' setting.

\item The representation of the probability models implicitly assumed in various state-of-the art source coders in terms of 
(conditional) Markov chains is a problem of independent interest. We did not investigate in the details other image coders or coders for 
different kind of sources, but we believe that an approach similar to that taken here can be applied to a variety of source coders.
In particular, Markovian models for state-of-the art speech/audio and video coders may lead to very interesting 
practical applications.

\item The restriction to BIOS channels is made here essentially for simplicity. It is clear that the output of the binary joint-source channel encoder can be concatenated with any suitable modulator in order to drive a non-binary input channel, as commonly done in a ``coded modulation''
approach.  In particular, the binary multistage encoding advocated here to encode the bit-planes can be married to 
{\em superposition coding}: instead of sending the encoded bit-planes in sequence, these can be modulated and superimposed. 
This approach may lead to efficient schemes for multicasting a common source to several users over Gaussian channels 
in different SNR (or fading) conditions \cite{phamdo,skoglund,krishna-it,krishna-asilomar,steiner-shamai}.

\item From our simulations it appears that the type of reconstruction ``noise'' achieved by the proposed scheme is very different from 
the typical effect of non-perfect reconstruction at the output of standard source coders. In fact, in our case it appears to
be much closer to some additive independent noise. Hence, our scheme appears to be ideally suited to be concatenated with the DUDE algorithm for denoising \cite{dude}.
 
\end{itemize}


\newpage
\clearpage

\appendix

\centerline{APPENDIX}

\section{Proof of Theorem 1} \label{app:proof}

Consider a memoryless stationary BIOS channel with input alphabet $\FF_2$, output alphabet $\Yc$, transition
probability $P_{Y|X}(y|x)$, and capacity $C$. Consider also a binary arbitrary source $V$, defined by the sequence of
$K$-th order joint probability distributions $P^{(K)}_V(\vv)$ over $\FF_2^K$, for $K = 1,2,\ldots$.
Notice that these probabilities need not have any structure, 
and the source need not be stationary or ergodic. The sup-entropy rate $\overline{H}(V)$ is defined in \cite{HV93} as the limsup in  
probability of the sequence of random variables
$-\frac{1}{K} \log_2 P^{(K)}_V(\vv)$, for $K \rightarrow \infty$. 
Theorem 1 is a simple corollary of the following results.

Let $\bv \in \FF_2^B$ be a binary vector and $\Gm_1 \in \FF_2^{B \times N}$ a binary matrix.
Consider the linear encoding rule $\cv = \bv \Gm_1$ and some decoding rule $\psi_1 : \Yc^N \rightarrow \FF_2^B$.
We define the conditional decoding error probability as
\begin{equation} \label{conditional-error}
\PP(e_1|\bv) = \PP(\psi_1(\yv) \neq \bv | \xv = \bv\Gm_1)
\end{equation}
Then, we have \cite[Problem 11, p. 114]{CK}

{\bf Lemma 1.} For every $\epsilon,\delta > 0$ and sufficiently large $N$ there exist $\Gm_1$ and 
$\psi_1$ such that $\PP(e_1|\bv) \leq \epsilon, \forall \; \bv \in \FF_2^B$, with $\frac{B}{N} \geq C - \delta$.
\hfill $\square$ 

We stated Lemma 1 in terms of the conditional error probability that, due to uniform error property of linear codes, 
is independent of the information message and therefore coincides with the average and maximal error probabilites. 
In particular, we let $\psi_1$ be the Maximum Likelihood decoding rule, defined by 
\begin{equation} \label{ML}
\psi_1(\yv) = {\rm argmax}_{\bv' \in \FF_2^B} \; \PP(\yv|\bv'\Gm_1) 
\end{equation}
Notice that this rule {\em ignores} the a priori probability of the information messages. Furthermore,
the resulting average error probability does not depend on this a priori probability.

Let $\vv \sim P^{(K)}_V$ be a vector of length $K$ generated by the arbitrary
binary source $V$, and let $\Gm_2 \in \FF_2^{K \times B}$ be a binary matrix.
Consider the linear encoding rule $\bv = \vv \Gm_2$ and some suitable decoding rule
$\psi_2 : \FF_2^B \rightarrow \FF_2^K$.
We define the average decoding error probability as
\begin{equation} \label{average-pe}
\PP(e_2) = \sum_{\vv \in \FF_2^K} \PP(\psi_2(\bv) \neq \vv | \bv = \vv \Gm_2) P^{(K)}_V(\vv)
\end{equation}
The following result, which is the key for the proof of Theorem 1, is due to Verd\'u and Shamai \cite{CaShVe_FnT} and generalizes 
a well-known result for memoryless sources \cite[Problem 7, p. 24]{CK} to sources with an arbitrary statistics:

{\bf Lemma 2.}
For any $\epsilon,\delta > 0$ and sufficiently large $K$ there exist
$\Gm_2$ and $\psi_2$ such that $\PP(e_2) \leq \epsilon$ with $\frac{B}{K} \leq \overline{H}(V) + \delta$.
\hfill $\square$

Putting together Lemma 1 and 2, consider the following linear coding scheme.
Fix $\epsilon_1,\epsilon_2,\delta_1,\delta_2 > 0$ and choose $K,B$ and $N$ such 
that there exist
$\Gm_1$ satisfying Lemma 1 with $\epsilon_1,\delta_1$ and
$\Gm_2$ satisfying Lemma 2 with $\epsilon_2,\delta_2$.
We map linearly the source sequence $\vv$ of length $K$ into the channel codeword
$\xv = \vv \Gm_2\Gm_1$ of length $N$. At the receiver, we use a concatenated decoder 
based on $\psi_1$ and $\psi_2$, that is, we let $\widehat{\vv} = \psi_2(\psi_1(\yv))$.
Notice that $\psi_1(\cdot)$ is applied to $\yv$ by ignoring the prior probability on $\vv \Gm_2$ 
induced by the source distribution, and $\psi_2(\cdot)$ is applied on the inner decoding 
output $\psi_1(\yv)$ by ignoring the actual channel output $\yv$. 
This corresponds to {\em separated} decoding and it is, evidently, suboptimal.
The error event $e$ of the concatenated decoder is contained in $e_1 \cup e_2$, where $e_1$ and $e_2$ are the error events of the inner 
and outer decoders, respectively. By the union bound we have
\begin{equation} \label{Pebound}
\PP(e) \leq \PP(e_1) + \PP(e_2) \leq \epsilon_1 + \epsilon_2 = \epsilon
\end{equation}
The resulting efficiency is given by 
\begin{equation}
\eta = \frac{K}{N} \geq \frac{\frac{B}{\overline{H}(V) + \delta_2}}{\frac{B}{C - \delta_1}} \geq \frac{C}{\overline{H}(V)} - \delta
\end{equation}
for some $\delta > 0$. Clearly, both $\epsilon$ and $\delta$ vanish as $\epsilon_1,\epsilon_2,\delta_1$ and $\delta_2$
vanish. This proves Theorem 1.

Finally, it is clear that the suboptimal concatenated linear scheme with separated two-stage decoding given above
cannot perform better than an optimal linear scheme where we design directly an encoding matrix $\Gm \in \FF_2^{K \times N}$ (not necessarily as given by the product of a ``compression'' matrix $\Gm_2$ and a ``channel coding'' matrix $\Gm_1$), 
and where we use the optimal Maximum a Posteriori (MAP) decoder instead of the concatenation of $\psi_1$ and $\psi_2$.
For the sake of completeness, the MAP decoder is given by
\begin{eqnarray} \label{MAP-joint}
\widehat{\vv} & = & {\rm argmax}_{\uv \in \FF_2^K} \;\; \PP(\uv|\yv) \nonumber \\
& = & {\rm argmax}_{\uv \in \FF_2^K} \;\; \PP(\yv|\uv \Gm) P_V^{(K)}(\uv)
\end{eqnarray}

\bibliographystyle{IEEEtran}
\bibliography{biblio}
\end{document}